\documentclass[aps,prc,twocolumn,superscriptaddress,notitlepage]{revtex4-1}

\usepackage{geometry} 
\usepackage{graphicx} 
\usepackage{epsfig}
\usepackage{epstopdf}
\usepackage{amsmath}
\usepackage{amsfonts}
\usepackage{amssymb} 
\usepackage{color}
\usepackage{multirow}

\newcommand{\Real}{\Re\text{e}}
\newcommand{\Imag}{\Im\text{m}}

\begin{document}

\vspace*{-3cm}\hspace{4.5cm} IRFU-15-12

\title{The E00-110 experiment in Jefferson Lab's Hall A:\\Deeply Virtual Compton Scattering off the Proton at 6~GeV}
\author{M.~Defurne}
\affiliation{CEA, Centre de Saclay, Irfu/Service de Physique Nucl\'eaire, 91191 Gif-sur-Yvette, France}
\author{M.~Amaryan}
\affiliation{Old Dominion University, Norfolk, Virginia 23508, USA}
\author{K.~A.~Aniol}
\affiliation{California State University, Los Angeles, Los Angeles, California 90032, USA}
\author{M.~Beaumel}
\affiliation{CEA, Centre de Saclay, Irfu/Service de Physique Nucl\'eaire, 91191 Gif-sur-Yvette, France}
\author{H.~Benaoum}
\affiliation{Syracuse University, Syracuse, New York 13244, USA}
\affiliation{Department of Applied Physics, University of Sharjah, UAE}
\author{P.~Bertin}
\affiliation{Universit\'e Blaise Pascal/CNRS-IN2P3, F-63177 Aubi\`ere, France}
\affiliation{Thomas Jefferson National Accelerator Facility, Newport News, Virginia 23606, USA}
\author{M.~Brossard}
\affiliation{Universit\'e Blaise Pascal/CNRS-IN2P3, F-63177 Aubi\`ere, France}
\author{A.~Camsonne}
\affiliation{Universit\'e Blaise Pascal/CNRS-IN2P3, F-63177 Aubi\`ere, France}
\affiliation{Thomas Jefferson National Accelerator Facility, Newport News, Virginia 23606, USA}
\author{J.-P.~Chen}
\affiliation{Thomas Jefferson National Accelerator Facility, Newport News, Virginia 23606, USA}
\author{E.~Chudakov}
\affiliation{Thomas Jefferson National Accelerator Facility, Newport News, Virginia 23606, USA}
\author{B.~Craver}
\affiliation{University of Virginia, Charlottesville, Virginia 22904, USA}
\author{F.~Cusanno}
\affiliation{INFN/Sezione Sanit\`{a}, 00161 Roma, Italy}
\author{C.W.~de~Jager}
\affiliation{Thomas Jefferson National Accelerator Facility, Newport News, Virginia 23606, USA}
\affiliation{University of Virginia, Charlottesville, Virginia 22904, USA}
\author{A.~Deur}
\affiliation{Thomas Jefferson National Accelerator Facility, Newport News, Virginia 23606, USA}
\author{R.~Feuerbach}
\affiliation{Thomas Jefferson National Accelerator Facility, Newport News, Virginia 23606, USA}
\author{C.~Ferdi}
\affiliation{Universit\'e Blaise Pascal/CNRS-IN2P3, F-63177 Aubi\`ere, France}
\author{J.-M.~Fieschi}
\affiliation{Universit\'e Blaise Pascal/CNRS-IN2P3, F-63177 Aubi\`ere, France}
\author{S.~Frullani}
\affiliation{INFN/Sezione Sanit\`{a}, 00161 Roma, Italy}
\author{E.~Fuchey}
\affiliation{Universit\'e Blaise Pascal/CNRS-IN2P3, F-63177 Aubi\`ere, France}
\affiliation{CEA, Centre de Saclay, Irfu/Service de Physique Nucl\'eaire, 91191 Gif-sur-Yvette, France}
\author{M.~Gar\c con}
\affiliation{CEA, Centre de Saclay, Irfu/Service de Physique Nucl\'eaire, 91191 Gif-sur-Yvette, France}
\author{F.~Garibaldi}
\affiliation{INFN/Sezione Sanit\`{a}, 00161 Roma, Italy}
\author{O.~Gayou}
\affiliation{Massachusetts Institute of Technology,Cambridge, Massachusetts 02139, USA}
\author{G.~Gavalian}
\affiliation{Old Dominion University, Norfolk, Virginia 23508, USA}
\author{R.~Gilman}
\affiliation{Rutgers, The State University of New Jersey, Piscataway, New Jersey 08854, USA}
\author{J.~Gomez}
\affiliation{Thomas Jefferson National Accelerator Facility, Newport News, Virginia 23606, USA}
\author{P.~Gueye}
\affiliation{Hampton University, Hampton, Virginia 23668, USA}
\author{P.A.M.~Guichon}
\affiliation{CEA, Centre de Saclay, Irfu/Service de Physique Nucl\'eaire, 91191 Gif-sur-Yvette, France}
\author{B.~Guillon}
\affiliation{Laboratoire de Physique Subatomique et de Cosmologie, 38026 Grenoble, France}
\author{O.~Hansen}
\affiliation{Thomas Jefferson National Accelerator Facility, Newport News, Virginia 23606, USA}
\author{D.~Hayes}
\affiliation{Old Dominion University, Norfolk, Virginia 23508, USA}
\author{D.~Higinbotham}
\affiliation{Thomas Jefferson National Accelerator Facility, Newport News, Virginia 23606, USA}
\author{T.~Holmstrom}
\affiliation{College of William and Mary, Williamsburg, Virginia 23187, USA}
\author{C.E.~Hyde}
\affiliation{Old Dominion University, Norfolk, Virginia 23508, USA}
\author{H.~Ibrahim}
\affiliation{Old Dominion University, Norfolk, Virginia 23508, USA}
\affiliation{Physics Department, Cairo University, Giza 12613, Egypt}
\author{R.~Igarashi}
\affiliation{University of Saskatchewan, Saskatchewan, Saskatchewan, Canada, S7N 5C6}
\author{X.~Jiang}
\affiliation{Rutgers, The State University of New Jersey, Piscataway, New Jersey 08854, USA}
\author{H.S.~Jo}
\affiliation{Institut de Physique Nucl\'eaire CNRS-IN2P3, Orsay, France}
\author{L.J.~Kaufman}
\affiliation{University of Massachusetts Amherst, Amherst, Massachusetts 01003, USA}
\affiliation{Indiana University Department of Physics and CEEM Bloomington, IN 47405, USA}
\author{A.~Kelleher}
\affiliation{College of William and Mary, Williamsburg, Virginia 23187, USA}
\author{C.~Keppel}
\affiliation{Thomas Jefferson National Accelerator Facility, Newport News, Virginia 23606, USA}
\author{A.~Kolarkar}
\affiliation{CEA, Centre de Saclay, Irfu/Service de Physique Nucl\'eaire, 91191 Gif-sur-Yvette, France}
\author{E.~Kuchina}
\affiliation{University of Kentucky, Lexington, Kentucky 40506, USA}
\author{G.~Kumbartzki}
\affiliation{Rutgers, The State University of New Jersey, Piscataway, New Jersey 08854, USA}
\author{G.~Laveissi\`ere}
\affiliation{Universit\'e Blaise Pascal/CNRS-IN2P3, F-63177 Aubi\`ere, France}
\author{J.J.~LeRose}
\affiliation{Thomas Jefferson National Accelerator Facility, Newport News, Virginia 23606, USA}
\author{R.~Lindgren}
\affiliation{University of Virginia, Charlottesville, Virginia 22904, USA}
\author{N.~Liyanage}
\affiliation{University of Virginia, Charlottesville, Virginia 22904, USA}
\author{H.-J.~Lu}
\affiliation{Department of Modern Physics, University of Science and Technology of China, Hefei 230026, China}
\author{D.J.~Margaziotis}
\affiliation{California State University, Los Angeles, Los Angeles, California 90032, USA}
\author{M.~Mazouz}
\affiliation{Laboratoire de Physique Subatomique et de Cosmologie, 38026 Grenoble, France}
\affiliation{Facult\'e des sciences de Monastir, 5000 Tunisia}
\author{Z.-E.~Meziani}
\affiliation{Temple University, Philadelphia, Pennsylvania 19122, USA}
\author{K.~McCormick}
\affiliation{Rutgers, The State University of New Jersey, Piscataway, New Jersey 08854, USA}
\author{R.~Michaels}
\affiliation{Thomas Jefferson National Accelerator Facility, Newport News, Virginia 23606, USA}
\author{B.~Michel}
\affiliation{Universit\'e Blaise Pascal/CNRS-IN2P3, F-63177 Aubi\`ere, France}
\author{B.~Moffit}
\affiliation{College of William and Mary, Williamsburg, Virginia 23187, USA}
\author{P.~Monaghan}
\affiliation{Massachusetts Institute of Technology,Cambridge, Massachusetts 02139, USA}
\author{C.~Mu\~noz~Camacho}
\affiliation{CEA, Centre de Saclay, Irfu/Service de Physique Nucl\'eaire, 91191 Gif-sur-Yvette, France}
\affiliation{Institut de Physique Nucl\'eaire CNRS-IN2P3, Orsay, France}
\author{S.~Nanda}
\affiliation{Thomas Jefferson National Accelerator Facility, Newport News, Virginia 23606, USA}
\author{V.~Nelyubin}
\affiliation{University of Virginia, Charlottesville, Virginia 22904, USA}
\author{R.~Paremuzyan}
\affiliation{Institut de Physique Nucl\'eaire CNRS-IN2P3, Orsay, France}
\author{M.~Potokar}
\affiliation{Institut Jozef Stefan, University of Ljubljana, Ljubljana, Slovenia}
\author{Y.~Qiang}
\affiliation{Massachusetts Institute of Technology, Cambridge, Massachusetts 02139, USA}
\author{R.D.~Ransome}
\affiliation{Rutgers, The State University of New Jersey, Piscataway, New Jersey 08854, USA}
\author{J.-S.~R\'eal}
\affiliation{Laboratoire de Physique Subatomique et de Cosmologie, 38026 Grenoble, France}
\author{B.~Reitz}
\affiliation{Thomas Jefferson National Accelerator Facility, Newport News, Virginia 23606, USA}
\author{Y.~Roblin}
\affiliation{Thomas Jefferson National Accelerator Facility, Newport News, Virginia 23606, USA}
\author{J.~Roche}
\affiliation{Thomas Jefferson National Accelerator Facility, Newport News, Virginia 23606, USA}
\author{F.~Sabati\'e}
\affiliation{CEA, Centre de Saclay, Irfu/Service de Physique Nucl\'eaire, 91191 Gif-sur-Yvette, France}
\author{A.~Saha}
\affiliation{Thomas Jefferson National Accelerator Facility, Newport News, Virginia 23606, USA}
\author{S.~Sirca}
\affiliation{Institut Jozef Stefan, University of Ljubljana, Ljubljana, Slovenia}
\author{K.~Slifer}
\affiliation{University of Virginia, Charlottesville, Virginia 22904, USA}
\affiliation{University of New Hampshire, Durham, New Hampshire, 03824, USA}
\author{P.~Solvignon}
\affiliation{Temple University, Philadelphia, Pennsylvania 19122, USA}
\author{R.~Subedi}
\affiliation{Kent State University, Kent, Ohio 44242, USA}
\author{V.~Sulkosky}
\affiliation{College of William and Mary, Williamsburg, Virginia 23187, USA}
\author{P.E.~Ulmer}
\affiliation{Old Dominion University, Norfolk, Virginia 23508, USA}
\author{E.~Voutier}
\affiliation{Laboratoire de Physique Subatomique et de Cosmologie, 38026 Grenoble, France}
\author{K.~Wang}
\affiliation{University of Virginia, Charlottesville, Virginia 22904, USA}
\author{L.B.~Weinstein}
\affiliation{Old Dominion University, Norfolk, Virginia 23508, USA}
\author{B.~Wojtsekhowski}
\affiliation{Thomas Jefferson National Accelerator Facility, Newport News, Virginia 23606, USA}
\author{X.~Zheng}
\affiliation{Argonne National Laboratory, Argonne, Illinois, 60439, USA}
\author{L.~Zhu}
\affiliation{University of Illinois, Urbana, Illinois 61801, USA}
\collaboration{The Jefferson Lab Hall A Collaboration}

\date{\today}

\begin{abstract}
We present final results on the photon electroproduction ($\vec{e}p\rightarrow ep\gamma$) cross 
section in the deeply virtual Compton scattering (DVCS) regime and the 
valence quark region from Jefferson Lab experiment E00-110. Results from an analysis of a subset of these data were published before, but the analysis has been improved which is described here at length, together with details on the experimental setup. Furthermore, additional data have been analyzed resulting in photon electroproduction cross sections at new kinematic settings, for a total of 588 experimental bins. Results of the $Q^2$- and $x_B$-dependences of both the helicity-dependent and helicity-independent cross sections are discussed. The $Q^2$-dependence illustrates the dominance of the twist-2 handbag amplitude in the kinematics of the experiment, as previously noted. Thanks to the excellent accuracy of this high luminosity experiment, it becomes clear that the unpolarized cross section shows a significant deviation from the Bethe-Heitler process in our kinematics, compatible with a large contribution from the leading twist-2 DVCS$^2$ term to the photon electroproduction cross section. The necessity to include higher-twist corrections in order to fully reproduce the shape of the data is also discussed. The DVCS cross sections in this paper represent the final set of experimental results from  E00-110, superseding the previous publication.
\end{abstract}


\maketitle
\tableofcontents
\section{Introduction}
\label{sec:intro}
\begin{figure}[htbp]
\begin{center}
\includegraphics[width=\linewidth]{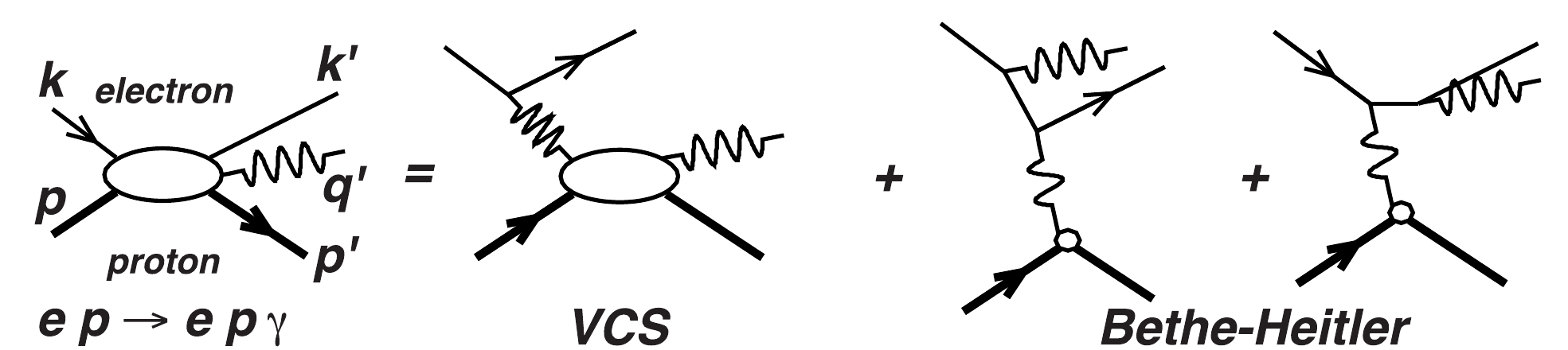}
\caption{
Lowest-order QED amplitude for the $ep\rightarrow ep\gamma$ reaction including its decomposition.
 The momentum four-vectors of all external particles are labeled at left.
The net four-momentum transfer to the proton is 
$\Delta_\mu=(q-q')_\mu=(p'-p)_\mu$. In the virtual Compton scattering
(VCS) amplitude, the (spacelike) virtuality of the incident photon is
$Q^2=-q^2=-(k-k')^2$.  In the Bethe-Heitler (BH) amplitude, the virtuality
of the incident photon is $-\Delta^2=-t$.  Standard $(e,e')$ invariants
are $s_e=(k+p)^2$, $x_B=Q^2/(2q\cdot p)$ and $W^2=(q+p)^2$.}
\label{fig:epepg}
\end{center}
\end{figure}

In the past two decades the Deeply Virtual Compton Scattering (DVCS) and Deep Virtual Meson Production (DVMP) 
reactions have emerged as powerful probes of the quark-gluon structure of the proton, neutron, and other atomic nuclei.
DVCS, specifically, refers to  the reaction
$\gamma^*p\rightarrow p\gamma$ in the Bjorken limit of Deep Inelastic
Scattering (DIS) $\gamma^* p$ kinematics, but at low net invariant momentum transfer $t$ to the target (in this case the proton).
Experimentally, we can access DVCS through exclusive electroproduction of real photons $ep\to ep\gamma$, where the DVCS amplitude interferes with the so-called Bethe-Heitler (BH) process (Figure~\ref{fig:epepg}). Given the knowledge of the proton elastic form factors, the BH contribution is calculable in Quantum Electrodynamics (QED) since it corresponds to the emission of the photon by the incoming or the outgoing electron. 

DVCS is the simplest probe of a new class
of light-cone  matrix elements, called 
Generalized Parton Distributions (GPDs) \cite{Mueller:1998fv}.  
These reactions offer the exciting prospect to obtain 3-dimensional tomographic images of the transverse spatial distributions of
partons (elementary quarks and gluons) as  functions of the parton light-cone momentum fraction
\cite{Mueller:1998fv, Ji:1996nm, Ji:1996ek, Ji:1997gm, Radyushkin:1997ki,Radyushkin:1996nd}.
In the kinematics of the present experiment, the GPDs are dominated by the quark light-cone matrix elements.
The correlation of transverse spatial and longitudinal momentum
information contained in the GPDs provides a new tool
to evaluate the contribution of quark orbital angular momentum
to the proton spin \cite{Ji:1996ek}.

\begin{figure}[htbp]
\begin{center}
\includegraphics[width=\linewidth]{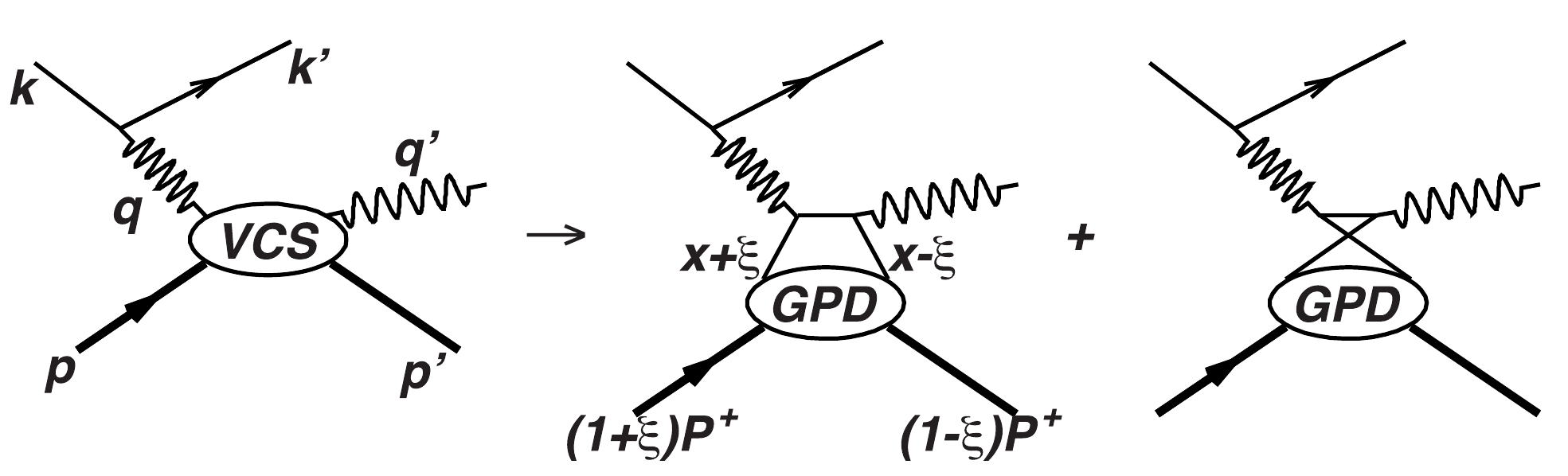}
\caption{The leading-order Virtual Compton Scattering (VCS) amplitude  in the limit of large $Q^2$, fixed $x_B$, and small
$t = \Delta^2$.  The kinematic variable $\xi = -(q+q')^2/(2(q+q')\cdot P)$, with $P = (p+p')/2$.  In the aforementioned limit
$\xi \rightarrow x_B/(2-x_B)$ and $2\xi \rightarrow \Delta^+/P^+=(\Delta^0+\Delta^z)/(P^0+P^z)$, with the $z$-direction parallel to
$\mathbf P$ in the $\mathbf q+ \mathbf P=0$ CM frame. Similarly, in the middle diagram, the quark and proton lines are labeled
by their `+' momentum fractions and '+' momentum components, respectively}
\label{fig:GPD_Full}
\end{center}
\end{figure}

The proof of Quantum Chromodynamics (QCD) factorization theorems \cite{Collins:1996fb,Ji:1998xh} established that in the Bjorken limit of high $Q^2$ at fixed $x_B$, the GPDs are the leading contribution to the 
$\gamma^* p \to \gamma p$ amplitude in an expansion in inverse powers of $Q^2$ (twist-expansion). 
Higher terms in the expansion are sensitive to more complicated correlation matrix elements ({\it e.g.} $qqg$ correlations).
The leading order DVCS amplitude is determined by four GPDs, which are defined in terms of vector 
$H$ and $E$ and axial vector $\tilde{H}$ and $\tilde{E}$ light-cone matrix elements.
The Generalized Parton Distributions enter the DVCS cross section through Compton Form Factors (CFFs),
which are integrals over the quark loops of the two diagrams of Fig.~\ref{fig:GPD_Full}.
For example, the CFF $\mathcal H$ corresponding to the GPD $H$ is defined through ($f\in\{u,d,s\}$) \cite{Belitsky:2001ns}:
\begin{multline}
\!\!{\mathcal H}(\xi,t)=\!\sum_{f}\frac{e_f^2}{e^2}
\Biggl\{
         i\pi     \left[H_f(\xi,\xi,t) \!-\! H_f(-\xi,\xi,t)\right]
  \\ +
 {\mathcal P}
    \int_{-1}^{+1} dx
\left[ \frac{1}{\xi-x} - \frac{1}{\xi+x} \right]   H_f(x,\xi,t)\Biggr\}.
\label{eq:CFF}
\end{multline}
Thus, the imaginary part accesses GPDs along the line $x=\pm\xi$,
whereas the real part probes GPD integrals over $x$. The `diagonal' GPD, $H(\xi,\xi,t=\Delta^2)$ is not a positive-definite probability
density, however it is a transition density with the momentum transfer $\Delta_\perp$ Fourier-conjugate to the transverse distance $r$
between the active parton and the center-of-momentum of the spectator partons in the target \cite{Burkardt:2007sc}.
  Furthermore, the real part of
the Compton Form Factor is determined by a dispersion integral over the diagonal $x=\pm \xi$ plus a $D$-term \cite{Teryaev:2005uj, Anikin:2007yh, Anikin:2007tx, Diehl:2007ru}.
This $D$-term~\cite{Polyakov:1999gs} only has support in the ERBL region $|x|<\xi$ in which the GPD is determined by $q\overline{q}$ exchange in the $t$-channel.

GPDs have generated an intense experimental activity. Beam spin asymmetries for DVCS in the valence region were first measured by the HERMES~\cite{Airapetian:2001yk} and CLAS~\cite{Stepanyan:2001sm} collaborations.
Cross sections were first measured at low-$x_B$ by the H1 \cite{Adloff:2001cn} and ZEUS collaborations \cite{Chekanov:2003ya}.
These results were  followed with more
detailed studies  of the $Q^2$-, $W^2$-, and $t$-dependence of the cross sections
\cite{Aktas:2005ty,Aaron:2007ab,Chekanov:2008vy,Aaron:2009ac}. 
The HERMES collaboration has measured a diverse range of  asymmetries on the proton, including longitudinal-spin 
\cite{Airapetian:2005jc}, transverse-spin \cite{Airapetian:2008aa,Airapetian:2011uq}, 
beam-charge    \cite{Airapetian:2006zr}, and 
kinematically complete beam-spin asymmetries \cite{Airapetian:2012pg}.
Detailed studies of the DVCS cross section as a function of $W^2$, $Q^2$, and $t$ by the ZEUS \cite{Chekanov:2003ya}
and H1 Collaborations \cite{Aaron:2007ab} demonstrated the factorization of the cross section, and the dominance of 
gluon GPDs at low $x_B$.  The first measurements of the DVCS cross section in the valence region were obtained by the
present experiment \cite{MunozCamacho:2006hx}, together with an extraction of DVCS off the neutron 
\cite{Mazouz:2007aa}.  A subsequent JLab Hall A experiment is analyzing the beam energy dependence of the DVCS 
cross section \cite{E07-007,E08-025}.
 Beam-spin and longitudinal target spin asymmetries in the valence
region were measured in     CLAS \cite{Girod:2007aa,Chen:2006na,Seder:2014oaa,Pisano:2015iqa}.
Extensive DVCS data taking has now started with the JLab upgrade \cite{Roche:2006ex}.  Over the next few years, a broad GPD program is planned
for the 12 GeV beams at Jefferson Lab and the high energy muon beams at CERN in the COMPASS experiment.

\section{Theoretical Framework}
\label{sec:theory}
 The photon electroproduction cross section of a polarized lepton beam of energy $k$  off
 an unpolarized target of mass $M$ is sensitive to the coherent interference of the DVCS amplitude
 with the Bethe-Heitler amplitude (see Fig. \ref{fig:epepg}). It can be written as:
\begin{multline}
\frac{d^5\sigma(\lambda,\pm e)}{dQ^2 dx_B dt d\phi d\phi_e} =
  \frac{d^2\sigma_0}{dQ^2 dx_B} \frac{1}{e^6}\times\\  \left[
       \left|\mathcal T^{BH} \right|^2 +
        \left| \mathcal T^{DVCS}\right|^2  \mp
        \mathcal I  \right] \label{eq:dsigDVCS}
\end{multline}
\begin{eqnarray}
 \frac{d^2\sigma_0}{dQ^2 dx_B} &=&
\frac{\alpha_{\rm QED}^3}{16\pi^2(s_e-M^2)^2 x_B} 
\frac{1}{\sqrt{1+\epsilon^2}}   \\
\epsilon^2 &=& 4 M^2 x_B^2/Q^2 \nonumber \\
s_e&=& 2 M k + M^2 \nonumber 
\label{eq:dsig0}
\end{eqnarray}
where $\phi_e$ is the azimuthal angle of the scattered electron around the beam axis in the laboratory frame, $\phi$ is the azimuthal angle between the leptonic and hadronic planes defined in the Trento convention
\cite{Bacchetta:2004jz}, $\lambda$ is the electron helicity and the $+$($-$) stands for the sign of the charge of the lepton beam. The cross section does not depend on $\phi_e$ and this angle is integrated over, leaving effectively a 4-differential cross section.
The BH contribution $\mathcal T^{BH}$ is calculable in QED, given the $\sim 1\%$  knowledge of the
proton elastic form factors in our range of $-t < 0.4$~GeV$^2$. The other two
contributions to the cross section, the interference term $\mathcal I$ and the DVCS squared term $\left| \mathcal T^{DVCS}\right|^2$,
provide complementary information on GPDs. It is possible to exploit the structure of the
cross section as a function of the angle $\phi$ to separate
up to a certain degree the different contributions to the total cross section~\cite{Diehl:1997bu}.
The BH term is given in \cite{Belitsky:2001ns},
Eq.~(25), and only its general form is reproduced here:

\begin{eqnarray}
 |\mathcal T^{BH}|^2 &=&
\frac{e^6\sum_{n=0}^2 c_n^{BH} \cos(n\phi)}{x_B^2 t y^2 (1+\epsilon^2)^2  \mathcal P_1(\phi) \mathcal P_2(\phi)}
\label{eq:BHPhi}
\end{eqnarray}

The harmonic terms $c_n^{BH}$ depend upon
 bilinear combinations of the ordinary elastic form factors
$F_1(t)$ and $F_2(t)$ of the proton.  The factors $\mathcal P_i$ are the electron
propagators in the BH amplitude~\cite{Belitsky:2001ns}.

 The interference term in Eq.~(\ref{eq:dsigDVCS}) is a
linear combination of GPDs, whereas the DVCS$^2$ term is a
bilinear combination of GPDs.  These terms have the following harmonic structure:
\begin{widetext}
\begin{equation}
\mathcal I =
\frac{e^6}{x_B y^3 \mathcal P_1(\phi) \mathcal P_2(\phi) t }
\left\{ c_0^{\mathcal I} + \sum_{n=1}^3
     \left[  
   c_n^{\mathcal I}\cos(n\phi)
        +    \lambda s_n^{\mathcal I}\sin(n\phi) \right] \right\}
\label{eq:IntPhi}
\end{equation}
\begin{equation}
\left| \mathcal T^{DVCS} \right|^2  =
\frac{e^6}{y^2 Q^2} \left\{
c_0^{DVCS} + \sum_{n=1}^2 
  \left[   c_n^{DVCS} \cos(n\phi) + \lambda s_1^{DVCS} \sin(n\phi) \right]\right\}
\label{eq:DVCSPhi}
\end{equation}
\end{widetext}
The $c_0^{DVCS, \mathcal I}$ and $(c,s)_1^{\mathcal I}$ harmonics are dominated by twist-two  GPD terms, although they do have higher-twist admixtures that must be quantified by
the $Q^2-$dependence of each harmonic.  The $(c,s)_1^{DVCS}$ and $(c,s)_2^{\mathcal I}$ harmonics are dominated by twist-three matrix elements, although the same twist-two GPD terms
also contribute (but with smaller kinematic coefficients than in the lower Fourier terms).  The $(c,s)_2^{DVCS}$ and $(c,s)_3^{\mathcal I}$ harmonics
stem only from  twist-two double helicity-flip gluonic GPDs. They are formally suppressed by $\alpha_s$ and will be neglected here, but they do not mix with the  twist-two quark amplitudes.

The bilinear DVCS term has a twist-2 contribution that reads:
\begin{equation}
c_0^{\text DVCS} =   2\frac{2-2y+y^2+\frac{\epsilon^2}{2}y^2}{1+\epsilon^2}    \mathcal C^{\text DVCS}(\mathcal F,\mathcal F^\ast)
\label{eq:cdvcs}
\end{equation}
where $\mathcal F$ represents the set $\{\mathcal H,\,\mathcal E,\,\widetilde{\mathcal H},\,\widetilde{\mathcal E}\}$ of
twist-2 CFFs. 
The Fourier coefficients $c^{\mathcal I}_n$ and
$s^{\mathcal I}_n$ of the interference term are:
\begin{eqnarray}
c_n^{\mathcal I} &=& C_{++}^n\Real\,\mathcal C^{\mathcal I,n}_{++}(\mathcal F)
+ C^n_{0+}\Real\,\mathcal C^{\mathcal I,n}_{0+}(\mathcal F_\text{eff})
\,,
\nonumber \\
s_n^{\mathcal I} &=& S^n_{++}\Imag\,\mathcal S^{\mathcal I,n}_{++}(\mathcal F)
+ S^n_{0+}\Imag\,\mathcal S^{\mathcal I,n}_{0+}(\mathcal F_\text{eff})\,.\nonumber\\
\end{eqnarray}

The above coefficients are defined in terms of the photon helicity-conserving
\begin{eqnarray}
\mathcal C^{\mathcal I,n}_{++}(\mathcal F) \!&=&\! \mathcal C^{\mathcal I}\!(\mathcal F)\!+\!\frac{C^{V,n}_{++}}{C^n_{++}}\mathcal C^{\mathcal I, V}\!(\mathcal F)\!+\!\frac{C^{A,n}_{++}}{C^n_{++}}\mathcal C^{\mathcal I,A}\!(\mathcal F)
\nonumber \\\label{eq:ci}\\
\mathcal S^{\mathcal I,n}_{++}(\mathcal F) \!&=&\! \mathcal C^{\mathcal I}\!(\mathcal F)\!+\!\frac{S^{V,n}_{++}}{S^n_{++}}\mathcal C^{\mathcal I, V}\!(\mathcal F)\!+\!\frac{S^{A,n}_{++}}{S^n_{++}}\mathcal C^{\mathcal I,A}\!(\mathcal F)\nonumber\\
\end{eqnarray}
and helicity-changing amplitudes
\begin{multline}
\mathcal C^{\mathcal I,n}_{0+}(\mathcal F_\text{eff}) =\frac{\sqrt{2}}{2-x_B}\frac{\widetilde K}{Q}\bigg [ \mathcal C^{\mathcal I}(\mathcal F_\text{eff})+\\\frac{C^{V,n}_{0+}}{C^n_{0+}}C^{\mathcal I, V}(\mathcal F_\text{eff})+\frac{C^{A,n}_{0+}}{C^n_{0+}}\mathcal C^{\mathcal I,A}(\mathcal F_\text{eff})\bigg ]
\label{eq:ceff}
\end{multline}
\begin{multline}
\mathcal S^{\mathcal I,n}_{0+}(\mathcal F_\text{eff}) = \frac{\sqrt{2}}{2-x_B}\frac{\widetilde K}{Q}\bigg [\mathcal C^{\mathcal I}(\mathcal F_\text{eff})+\\\frac{S^{V,n}_{0+}}{S^n_{0+}}C^{\mathcal I, V}(\mathcal F_\text{eff})+\frac{S^{A,n}_{0+}}{S^n_{0+}}\mathcal C^{\mathcal I,A}(\mathcal F_\text{eff})\bigg ]\,.
\end{multline}
The complete expressions of kinematic coefficients $C^n_{ab}$, $S^n_{ab}$ and $\widetilde K$ are given in~\cite{Belitsky:2010jw}. The $\mathcal C^{\mathcal I}$ and $\mathcal C^{DVCS}$~terms are respectively linear and bilinear combination of CFFs. For example:
\begin{equation}
\mathcal C^{\mathcal I}(\mathcal F)=F_1\mathcal H+\xi(F_1+F_2)\widetilde{\mathcal H}-\frac{t}{4M^2}F_2\mathcal E \,.
\end{equation}

\section{The E00-110 Experiment}
\label{sec:exp}

	The E00-110~\cite{E00110} experiment ran in Hall A at Jefferson Lab in the fall of 2004. Its goal was to measure the $Q^2-$dependence of the DVCS~\footnote{Formally, Deeply Virtual Compton Scattering or DVCS refers only
to the sub-process $\gamma^* p \to \gamma p$. However, DVCS is often used more loosely
in the literature to name the photon electroproduction process $e p \to e p \gamma$ used experimentally.} helicity-dependent cross sections at fixed value of $x_B$:
\begin{eqnarray}
	d^4 \sigma \!&=&\! \frac{1}{2}\left[ \frac{d^4\sigma(\lambda=+1)}{dQ^2 dx_B dt d\phi} + \frac{d^4\sigma(\lambda=-1)}{dQ^2 dx_B dt d\phi} \right] \\
	\Delta^4 \sigma \!&=&\! \frac{1}{2}\left[ \frac{d^4\sigma(\lambda=+1)}{dQ^2 dx_B dt d\phi} - \frac{d^4\sigma(\lambda=-1)}{dQ^2 dx_B dt d\phi} \right]  
\end{eqnarray}

\begin{figure}
\includegraphics[width=1\linewidth]{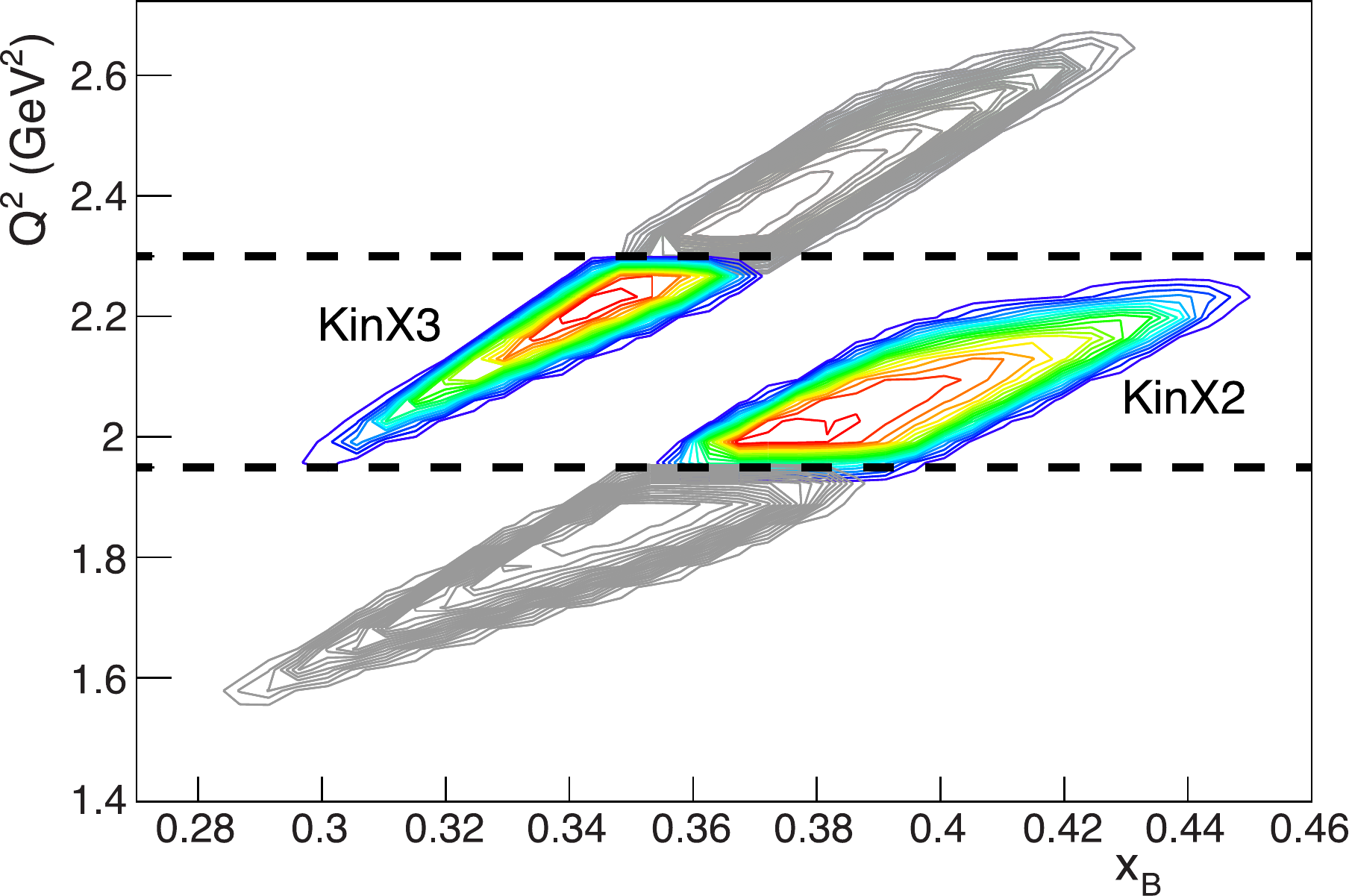}
\caption{(Color online) Distribution of $H(e,e^{\prime}\gamma)X$ events in the [$x_B$, $Q^2$] plane, for Kin2 ($x_B=0.36$, $Q^2=1.9\; {\rm GeV}^2$) and Kin3 ($x_B=0.36$, $Q^2=2.3\; {\rm GeV}^2$).
    Events for Kin$X$2 ($x_B=0.39$, $Q^2=2.06\; {\rm GeV}^2$) and Kin$X$3 ($x_B=0.34$, $Q^2=2.17\; {\rm GeV}^2$)  are bounded by the two horizontal lines at $Q^2=1.95$~GeV$^2$ and $Q^2=2.30$~GeV$^2$.} 
\label{fig:kin2x}
\end{figure}

Tab.~\ref{tab:DVCSkin} summarizes all the kinematic settings of this experiment. In addition to the $Q^2-$dependence at fixed $x_B$, we present here new results on the $x_B$-dependence of the DVCS cross section at fixed $Q^2$ by using a subset of the data from the Kin2 and Kin3 settings with $1.95<Q^2<2.30$~GeV$^2$, as illustrated in Fig.~\ref{fig:kin2x}. We labelled these new settings Kin~X2 and Kin~X3. They are centered at $x_B=0.40$ and $x_B=0.34$, respectively, for an averaged $Q^2=2.1$~GeV$^2$.
\begin{table*}
\begin{ruledtabular}
\begin{tabular}{lccccccc}
Setting & $k'$ (GeV/$c$) & $\theta_e$ $({}^\circ)$ & $Q^2$ (GeV$^2$) & $x_B$ & $\theta_q$ $({}^\circ)$ & $W$ (GeV) & $E_\gamma$ (GeV)\\
\hline
Kin1 & 3.53 & 15.6 & { 1.5} & 0.36 & $ -22.3$ & 1.9 & 2.14\\
Kin2 & 2.94  & 19.3 & { 1.9} & 0.36 & {$ -18.3$}& 2.0 & 2.73\\
Kin3 & 2.34 & 23.8 & { 2.3} & 0.36 & {$ -14.8$}& 2.2 & 3.32\\
\hline
KinX2 & 2.94 & 20.1 & 2.06 & 0.39 & $-18.6$ & 2.03 & 2.71\\
KinX3 & 2.36 & 23.1 & 2.17 & 0.34 & $-14.5$ & 2.26 & 3.33\\
\end{tabular}
\end{ruledtabular}
\caption{
Experimental $ep\rightarrow e p \gamma$ kinematics, for incident beam 
energy $E_b=5.7572$ GeV. $\theta_q$ is the central value of the {\bf q}-vector 
direction. $E_\gamma$ is the photon energy for $t=t_{\rm min}$. A subset of Kin2 and Kin3, with the cuts shown in Fig.~\ref{fig:kin2x}, provides the kinematic settings KinX2 and KinX3 at fixed $Q^2$ but varying $x_B$. Note that only the average kinematics for each setting are listed in this table : in order to minimize systematic bin centering effects, the results are presented or listed using the kinematics of each bin in $x_B$, $Q^2$ and $t$ according to their averaged experimental value in the bin. Our extraction procedure ensures that all $\phi$ bins are evaluated at the same kinematic setting, as explained in section~\ref{ssec:fit_proc}.}
\label{tab:DVCSkin}
\end{table*}

\begin{figure}[ht]
	\centerline{\includegraphics[width=\linewidth]{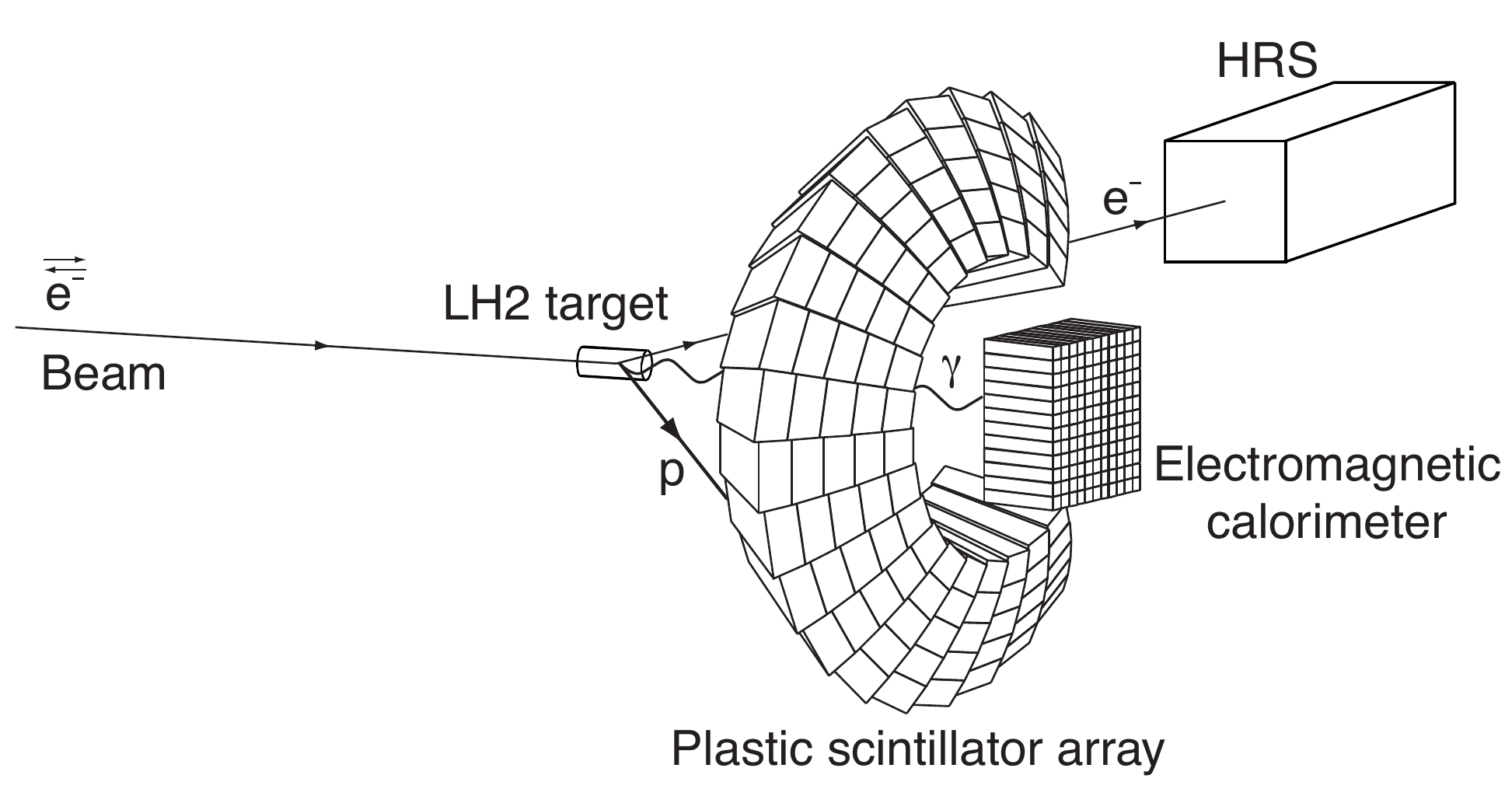}}
	\caption{\small Setup of the E00-110 experiment in Hall~A of Jefferson Lab. The photon calorimeter as well as the proton array were centered on the virtual photon direction ${\bf q}$, then shifted sidewise by half a calorimeter block away from the beam to limit the singles rate on the detector elements close to the beamline.} \label{fig:setuphallA}
	\end{figure}

\begin{figure}[ht]
	\centerline{\includegraphics[width=\linewidth]{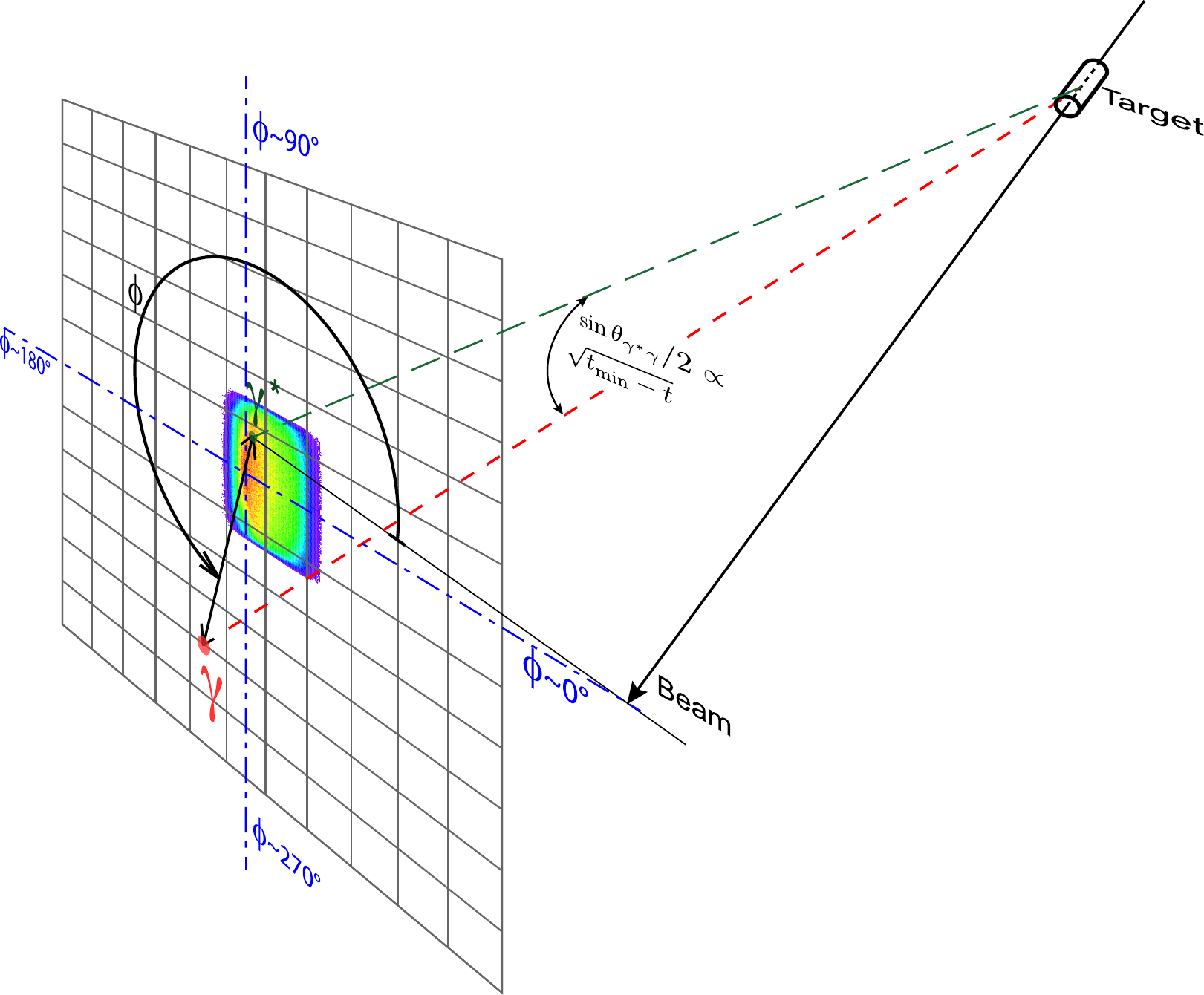}}
	\caption{(Color online) Perspective view of the downstream face of electromagnetic calorimeter. The virtual photon ($\gamma^\ast$) acceptance is shown projected
on the calorimeter plane. The distance (black line) between the "impact" of the virtual photon
and the detected real photon ($\gamma$) position is roughly
proportional to $\sqrt{t-t_{min}}$ for the small $t_{min}-t$ values of
this experiment. The angle $\phi$ is the azimuthal angle of the photon with respect to the
plane formed by the incident beam and the virtual photon.
For the central $(e,e')$ kinematics indicated by $\gamma^\ast$, $\phi$ is
measured counter-clockwise from the horizontal (to the right)
direction.} \label{fig:virtphot}
	\end{figure}

	The setup of the experiment is shown in Fig.~\ref{fig:setuphallA}.
	In order to counter the small cross section, this experiment used the high luminosity in Hall~A of Jefferson Lab, running at 
	$10^{37}$~cm$^{-2}$s$^{-1}$, which corresponds to 2.25~$\mu$A of electron beam on a 15-cm-long  liquid hydrogen target. The scattered electron was detected in the Hall~A left High Resolution Spectrometer (HRS), which provides a  momentum resolution $\delta p/p=2\cdot 10^{-4}$ and an angular resolution of 2~mrad in the horizontal plane~\cite{Alcorn:2004sb}. This pinpoints the electron kinematics ($x_B$ and $Q^2$), the electron scattering plane,
	 and the momentum direction ${\bf q}$ of the virtual photon of the virtual Compton amplitude. The emitted photon was detected in an electromagnetic calorimeter covering $\sim 0.1$~sr, with its front face 1.1~m from the target center, and centered in the direction of the virtual photon (shifted by half a calorimeter block). The spectrometer acceptance  of 6 msr and $\pm 4.5\%$ in momentum selects virtual photons in a small solid angle or $\sim 3$~msr, as illustrated in Fig.~\ref{fig:virtphot}. The detected photon direction (two angles) with respect to the virtual photon direction (calculated using the electron kinematics) determines the remaining two kinematic variables of the reaction: 
	 $t$ and $\phi$. The measurement of the detected photon energy 
	 allows for an exclusivity cut based on the squared missing mass of the recoil proton. As a cross-check on exclusivity, the recoil proton was detected in the Proton Array, a set of 100 blocks of plastic scintillator in a C-ring configuration around the virtual photon direction. This geometry was selected in order to have a simple azimuthal symmetry around the virtual photon direction, which is a key element for a smooth $\phi$ acceptance. 
	 
The basic equipment of Hall A, including the beamline, target system, and dual spectrometers is described in~\cite{Alcorn:2004sb}.
The following sections provide details specific to the present experiment.
\subsection{Electron Beam}
\label{ssec:beam}

	\subsubsection{Beam Energy}
The incident beam energy is measured by determining its bend in the arc section of the Hall A beamline~\cite{arcep}. Its deflection angle is computed from a set of wire scanners. The magnetic field integral of the eight dipoles of the beamline is compared to a reference magnet (9$^{\rm th}$ dipole). The measurement of the beam energy made during the experiment resulted in the value $E_b=5757.2\pm 0.1_\text{stat}\pm 0.1_\text{syst}$\,MeV.

	\subsubsection{Beam Current}
	\label{sssec:beamcurrent}

The beam current is measured using two resonant RF cavity monitors (Beam Current Monitors) tuned at the frequency of the accelerator (1.497\,GHz). The voltage at their outputs is proportional to the beam current and provides a continous monitoring of its value during the experiment. The absolute reference is provided by a separate monitor, a Parametric Current Transformer~\cite{Alcorn:2004sb}, which is calibrated by passing current of known value through a wire inside the beam pipe.

	\subsubsection{Beam Polarization}
	\label{sssec:beampol}
The electron beam polarization was measured concurrently with the regular data taking using the Hall A Compton polarimeter~\cite{Baylac:2002en}. At the entrance of the Hall, the beam is deflected by a chicane and interacts with a circularly polarized photon beam. The polarization of the electron beam can be obtained from the counting rate asymmetry from opposite beam helicities. The electrons that interact with the photon beam are detected by silicon micro-strips, while those that do not interact continue towards the experimental target. The photon beam is provided by a resonant Fabry-P\'erot cavity that amplifies a 230\,mW Nd:YaG laser ($\lambda=1064$\,nm) to 1200\,W.

The statistical error of a Compton measurement is inversely proportional to the square root of the
number of events and to the analyzing power of the polarimeter. In this experiment, a 1\%
statistical error could be achieved in 2.5\,h of data taking. However, this
is far from being the limiting factor. Since the Compton data was taken
during normal DVCS running, we can average over long periods of time in
order to make the statistical error negligible.

\begin{figure}
\centering\includegraphics[width=0.9\linewidth]{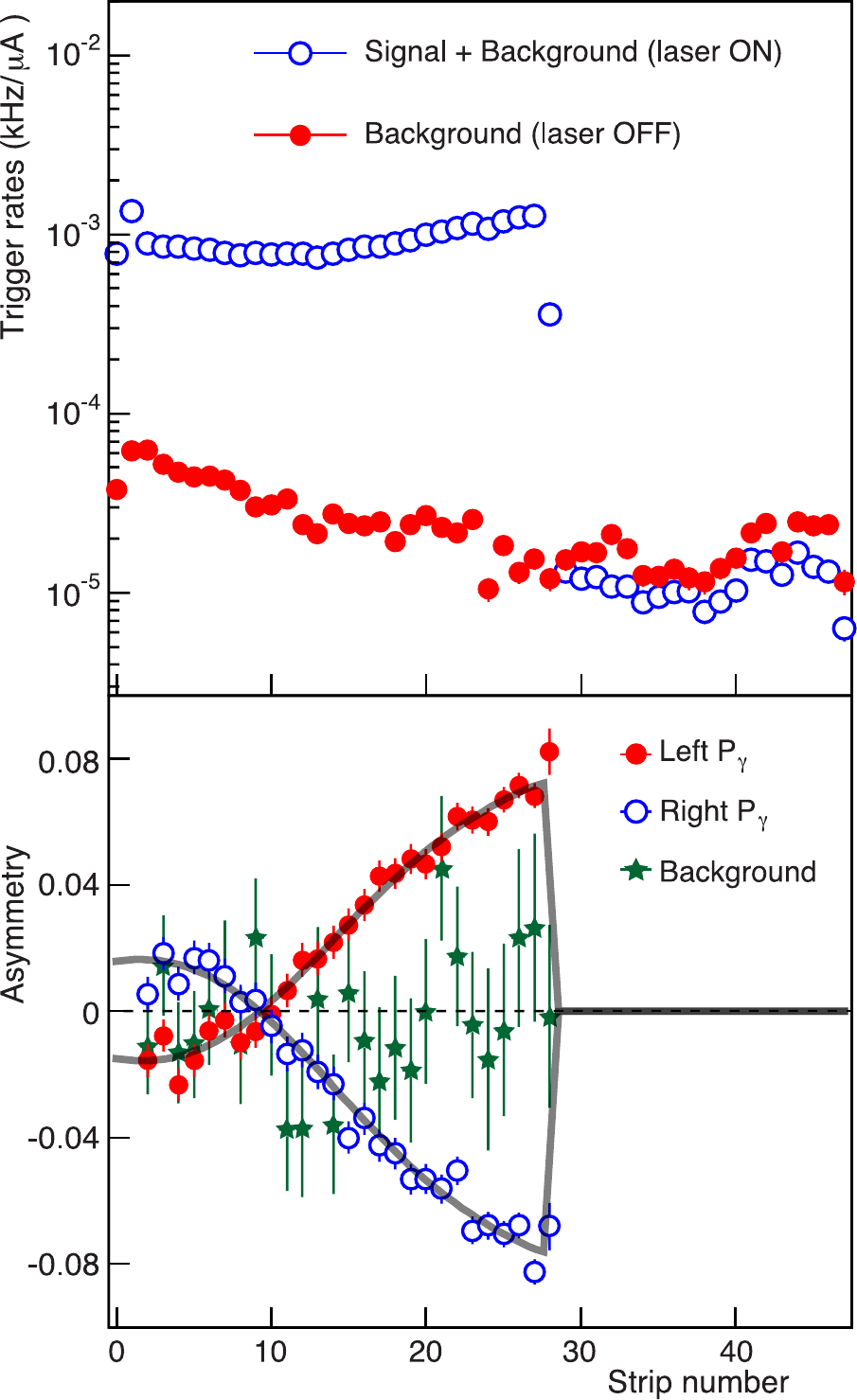}
\caption{(Color online) Signal and background rates (normalized to the electron beam current) in one of the planes of the electron
  detector as a function of strip number (top), and asymmetry measured for
  each of the laser polarization states and the background as a function of strip number (bottom).}
\label{fig:compsample}
\end{figure}

Beam polarization results can be
readily obtained from the electron detector. The electron detector consists of 4 planes
of 48 silicon micro-strips, standing 4.6\,mm above the beam axis during the
DVCS experiment. Figure~\ref{fig:compsample} (top) shows the electron
counting rate versus strip number in one of the detector planes for a
typical Compton run of 3\,h duration. The detector is
located behind the third dipole of the Compton chicane and the strip number
gives the position of the scattered electron along the dispersive axis with
a resolution of 200\,$\mu$m. Hence the horizontal axis of the plot is proportional to
the energy lost by the electron (and given to the photon). The Compton
energy spectrum shows up as a flat rate on the first strips. The background
spectrum has a $1/E$ shape, like bremsstrahlung. The differential asymmetry as a function of the electron energy (strip
number) is shown in the bottom plot of Fig.~\ref{fig:compsample} for each of the laser polarization states.

The systematic error in the polarization measurement due to the
uncertainty of the laser polarization is 0.7\%. The maximum deviation 
of scattered electrons for a beam energy of $\sim$5.75\,GeV is 21.5\,mm at the
electron detector plane, which makes a calibration error of
200\,$\mu$m/21.5\,mm=0.93\%, which propagates to 1.9\% to the polarization
measurement. The total systematic error associated to the beam polarization measurement is 2\%.

\begin{figure}
\centering\includegraphics[bb=14 3 512 410, clip, width=1\linewidth]{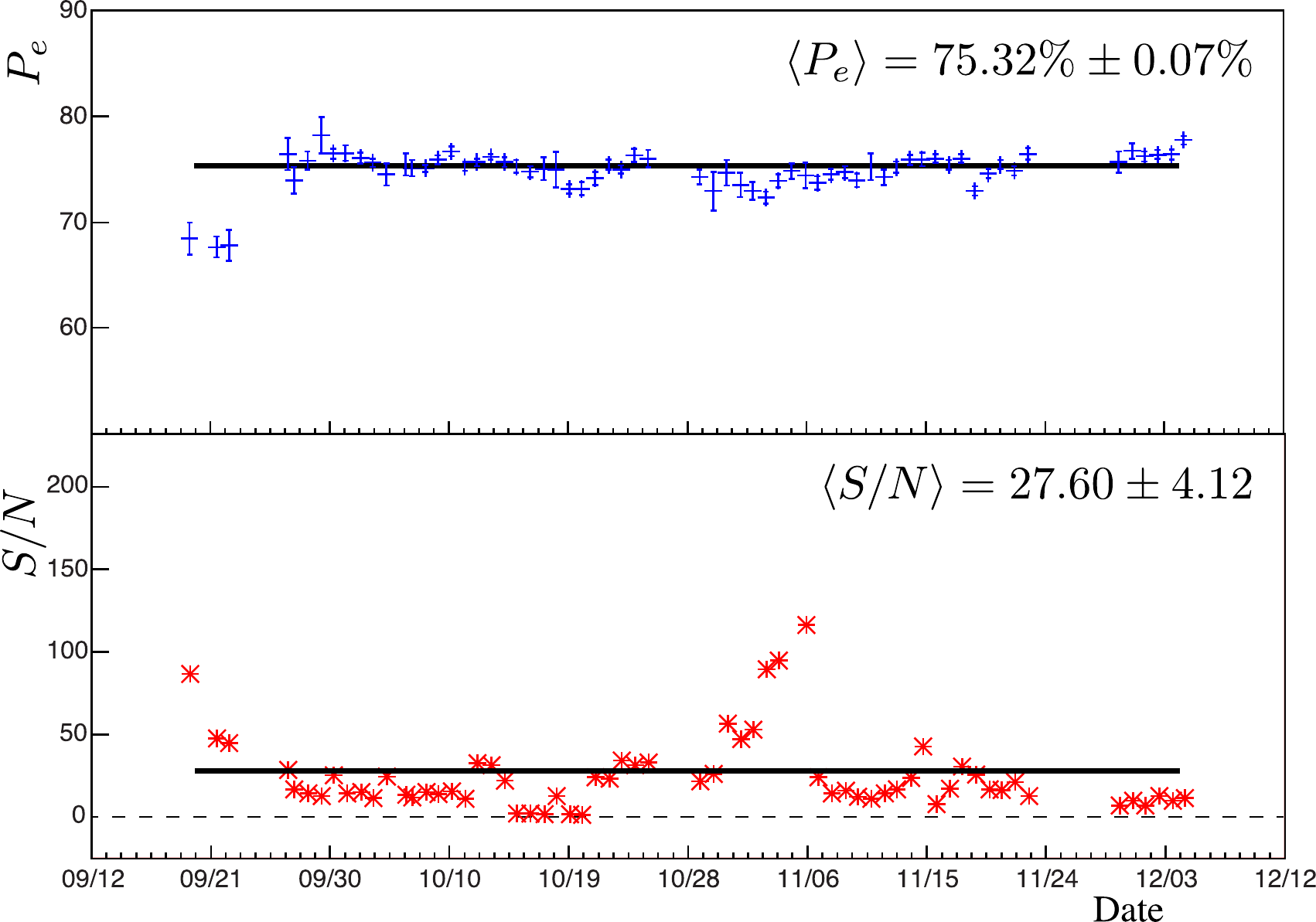}
\caption{(Color online) Compton polarimeter results, using only the electron
  detector. Beam polarization is shown in the upper plot. Lower plot shows
  the signal to background ratio. The
  first three points in the beginning of the experiment correspond to a non-optimal
  Wien angle setting.}
\label{fig:comptonres}
\end{figure}

Figure~\ref{fig:comptonres} shows the Compton polarimeter
results for the full experiment duration, where only the electron
detector was used in the analysis. The beam polarization was
$75.3\pm 0.1_\text{stat}\pm 2.0_\text{syst}\%$ in average during the experiment.
The low polarization values at the beginning of the experiment (first 3
points in Fig.~\ref{fig:comptonres}) correspond to the period when the polarization was not yet optimized for Hall~A.

\subsection{Liquid Hydrogen Target}
\label{ssec:target}
The standard Hall A cryogenic target system~\cite{Suleiman:1998} was mounted inside a scattering chamber custom built for the E00-110 experiment. Its walls were
thinner than those of the usual Hall~A scattering chamber. Also, a larger
exit beam pipe was constructed. The scattering chamber was made of a 1\,cm spherical shell of aluminum, allowing for low energy protons to go through (minimum 
momentum of 305\,MeV/$c$, corresponding to a cut on the kinematic variable $t$ of -0.091\,GeV$^2$). Moreover, the new scattering chamber 
accommodates the spherical symmetry of the reaction and makes energy losses
nearly independent of the scattering angle, except for extended target effects. The larger exit beam pipe reduces the background.

The cryogenic target has three target loops, two of which
were used for the DVCS experiment: a liquid hydrogen (LH$_2$) loop and a
liquid deuterium (LD$_2$) loop~\footnote{The deuterium loop was only used during
  experiment E03-106, running just after the one
  described herein.}. Each of the two liquid loops had an aluminum
cylindrical target cell, 15 cm long. An additional solid target ladder was attached to the system for calibration purposes.
The targets are arranged in a vertical stack, which can be moved from one position to
another by remote control. The solid target ladder contained the following
targets:
\begin{itemize}
\item {Optics:} Seven 1-mm-thick carbon foils used for optics
  calibration of the HRS.
\item {Two dummy targets:} $\pm$2\,cm and $\pm$7.5\,cm Al foils to
  study target walls effects.
\item {Cross hair:} Aluminum foil with a milled cross, used to measure
  beam position with respect to the target.
\item {BeO:} It allows to see the beam spot at the target through a
  camera installed in the scattering chamber.
\item {C: } 1\,mm thick carbon, serving as a point-like target.
\item {Empty:} Position used to reduce radiation on detectors while
  beam was used for other purposes (beam size measurements
  using wire scanners and other beam tunings).
\end{itemize}

\subsection{Hall A Spectrometer}
\label{ssec:hrs}

	The High Resolution Spectrometers (HRS) in Hall~A consist of 4 superconducting magnets in the configuration QQDQ. In the E00-110 experiment, the left HRS was used to detect the scattered electron and therefore define the virtual photon kinematics in an accurate way. The main components of the detector stack are as follows: a set of two scintillator planes called S1 and S2m giving very fast and good timing signals; two vertical drift chambers for track reconstruction; a gas \v Cerenkov counter for $\pi$/$e$ discrimination and a Pion Rejector composed of  two layers of Lead Glass blocks which is used in addition to the \v Cerenkov detector to select a clean sample of electrons. A fast signal from S2m in coincidence with the 
	\v Cerenkov detector was used as a level 1 trigger for the rest of the electronics. It is useful to recall the angular acceptance of the left HRS for electrons : $\pm 30$~mrad horizontal, $\pm 60$~mrad vertical, and $\pm 4.5\%$ in momentum.

\subsection{Calorimeter}
\label{ssec:calo}

\begin{figure}
\centering\includegraphics[width=0.8\linewidth]{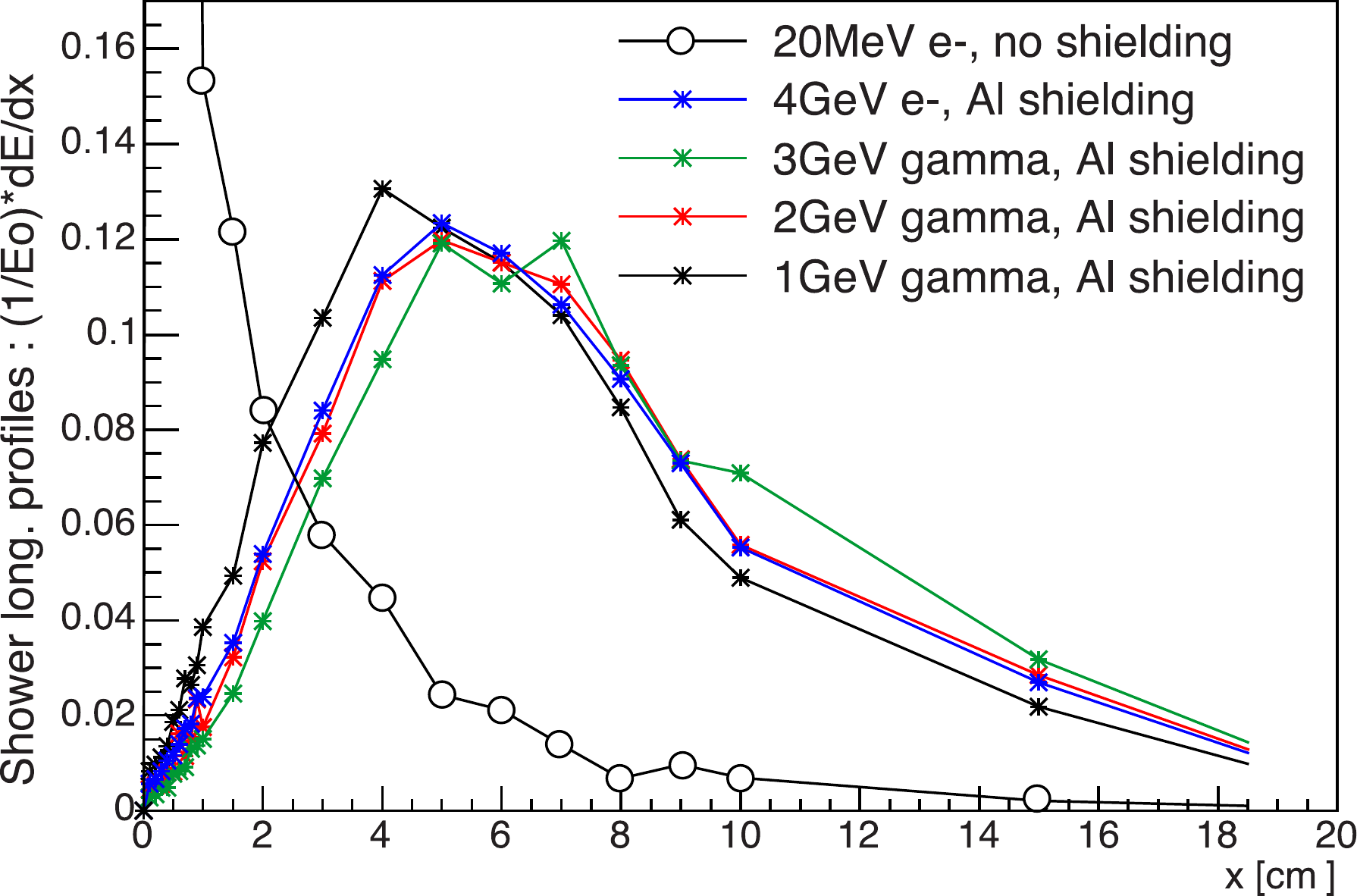}
\caption{(Color online) Shower longitudinal shower profile for different incoming particles into the DVCS calorimeter, obtained from the Monte-Carlo simulation. The aluminum shielding corresponds to the total thickness of material between the target and the calorimeter crystals, and includes both the scattering chamber and some additional Al shielding in front of the calorimeter front face.}
\label{fig:profile}
\end{figure}

	One of the key elements of this experiment was a dedicated  electromagnetic calorimeter, consisting of a 11$\times$12 array of lead fluoride 
	crystals, each 3$\times$3$\times$18.6\,cm$^3$. The crystals were purchased from SICCAS (Shanghai). PbF$_2$ was selected as a pure \v Cerenkov medium, to minimize hadronic backgrounds and to obtain the shortest possible signal without exponential tails. The size of the blocks is adapted to the radiation length and Moli\`ere radius of PbF$_2$ so that a shower is almost completely contained in a cluster of 9 blocks, both longitudinally and transversally. Each block was equipped with a Hamamatsu R7700 fine-mesh photomultiplier tube (PMT). During the experiment, the relative gains of the PMTs were periodically monitored using a cluster of LEDs that  could be moved across the calorimeter face on an X-Y stage.  However, the large luminosity of the experiment  induced radiation damage  near the front face of the blocks. Since the LED light was injected in the front face, the LED method proved  unreliable to measure the true signal  variation of high energy photon or electron showers. Indeed, the \v Cerenkov light from a multi-GeV $\gamma$-ray is mostly produced deeper in the crystal, avoiding most of the damaged area. Figure~\ref{fig:profile} shows simulated shower profiles for 4~GeV electrons, at a typical energy for the electromagnetic background and for photons of various energies expected from DVCS photons. 

\begin{figure}
\centering\includegraphics[width=1\linewidth]{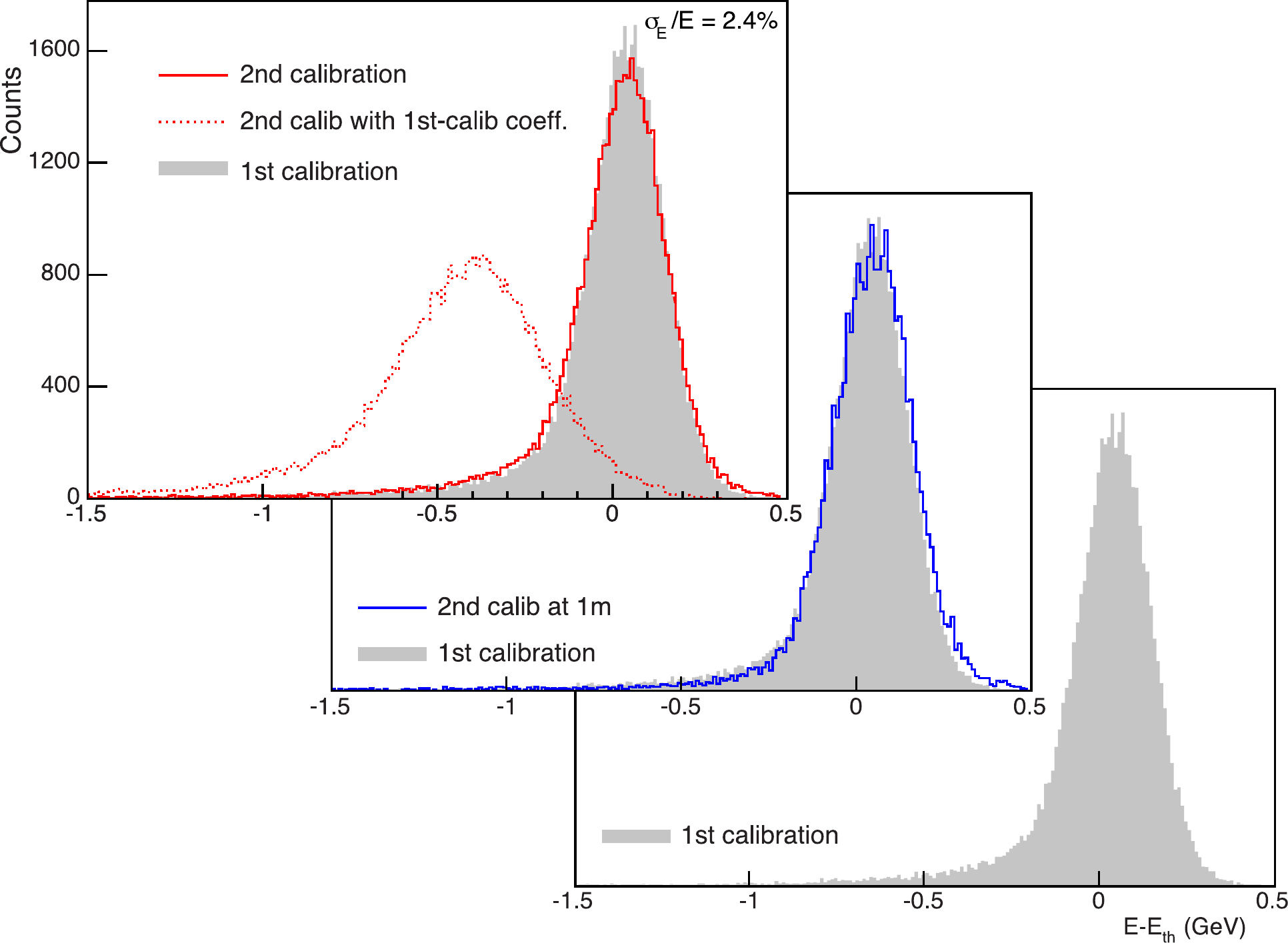}
\caption{(Color online) Energy measured in the calorimeter minus energy expected from elastic kinematics during elastic calibrations runs. In both elastic calibration periods, we obtained 2.4\% energy resolution at an elastic energy of 4.2\,GeV. The results of the
  second calibration when first calibration coefficients are used are also
  plotted to show the necessity of a careful monitoring of the coefficients
  in between these two calibration points.}
\label{fig:deltaE}
\end{figure}

The crystal-by-crystal calibration coefficients were obtained from kinematically over-constrained elastic
	scattering:  H$(e,e'_\text{Calo} p_\text{RHS})$ in which the electron is detected in the calorimeter and the proton is detected in the HRS. In order to illuminate the full acceptance of the calorimeter with elastic electrons, it was necessary to move the calorimeter back to a distance of 5.5~m from the target center during these runs. Data at 1~m, covering only the center part of the calorimeter were taken additionally as a consistency check. Figure~\ref{fig:deltaE} shows the energy resolution of the calorimeter as measured during the elastic calibration runs. Two elastic
calibrations were made, one a few weeks after the start of the experiment and
another one a few weeks before it finished. The calibration
coefficients changed by a considerable amount for some blocks, but the energy resolution did not degrade during the almost 3 months of data taking.
Since calibration coefficients changed with time, in order to keep a good energy resolution all
along the experiment, we interpolated these coefficients between
the two calibrations runs, together with an extrapolation
before and after them. This was done based on the radiation dose
accumulated by each block. This dose is proportional to the beam current
and depends on the block polar angle with respect to the beam line and also
on the target type (LH$_2$ or LD$_2$). The relative dose
accumulation for each block was estimated from its PMT anode current monitoring~\cite{caloextra}. 
In addition, the calibrations were  monitored {\em in situ} using the missing mass peak of the reaction D$(e,e_\text{Calo}^\prime \pi^-_\text{HRS})pp$ and both the missing mass and $(\gamma\gamma)$ invariant mass peaks of the reaction H$(e,e'\pi^0)p$.  Overall, the calibration coefficients were known for any given time at the 1\%-level.

\subsection{Proton Array}
\label{ssec:pa}

	In order to detect the full exclusive final state, a recoil detector was built  to tag the DVCS proton. The recoil proton direction for an exclusive event can be inferred from the information of the HRS and the calorimeter, therefore one can check in the Proton Array (PA) if the proton was actually at the right position. The main difficulties of such a detector is that it needs to detect low-momentum protons in a large acceptance, close to the beam line, with as high an efficiency as possible.  The PA subtended an acceptance (relative to the nominal direction of the virtual photon) of $18^\circ < \theta_{\gamma^* p} < 38^\circ$ and $45^\circ < \phi_{\gamma^* p} < 315^\circ$, arranged in 5 rings of 20 detectors as shown in Fig.~\ref{fig:setuphallA}. The scintillator blocks were fabricated by Eljen Technology as 5 distinct tapered trapezoids, each 30~cm-long, in order to form
	a hermetic ring pointing at the target center.  Each scintillator is equipped with a Photonis XP2972 PMT and a custom voltage divider/pre-amplifier circuit.  This allowed us to operate the PMTs at low gain, to accommodate  the high backgrounds in this open geometry. The 90$^\circ$ cut-off in $\phi_{\gamma^* p}$ corresponds to the exit-beam pipe in the kinematic setting where the detector stack is the closest to the beamline.

\subsection{Sampling Electronics}
\label{ssec:elec}

	The E00-110 experiment was designed with open detectors at low angles (the blocks of the calorimeter closest to the beam line were at 6.5$^\circ$) with limited shielding running at high luminosity. High singles rates up to 10~MHz were expected and also measured in a test run during the design phase. In this environment, regular ADCs even with a reduced gate are strongly affected by pile-up. We therefore chose to use digitizing electronics for all the electronic channels of the dedicated detectors (PbF$_2$ calorimeter and Proton Array), namely a custom 6U 16 VME (A24/D32) module sampling system based on the Analog Ring Sampler (ARS) CMOS ASIC developed at CEA-Saclay 
	\cite{Feinstein:2003vi,Druillole:2001dm}.
	
\begin{figure}[ht]
	\centerline{\includegraphics[width=\linewidth]{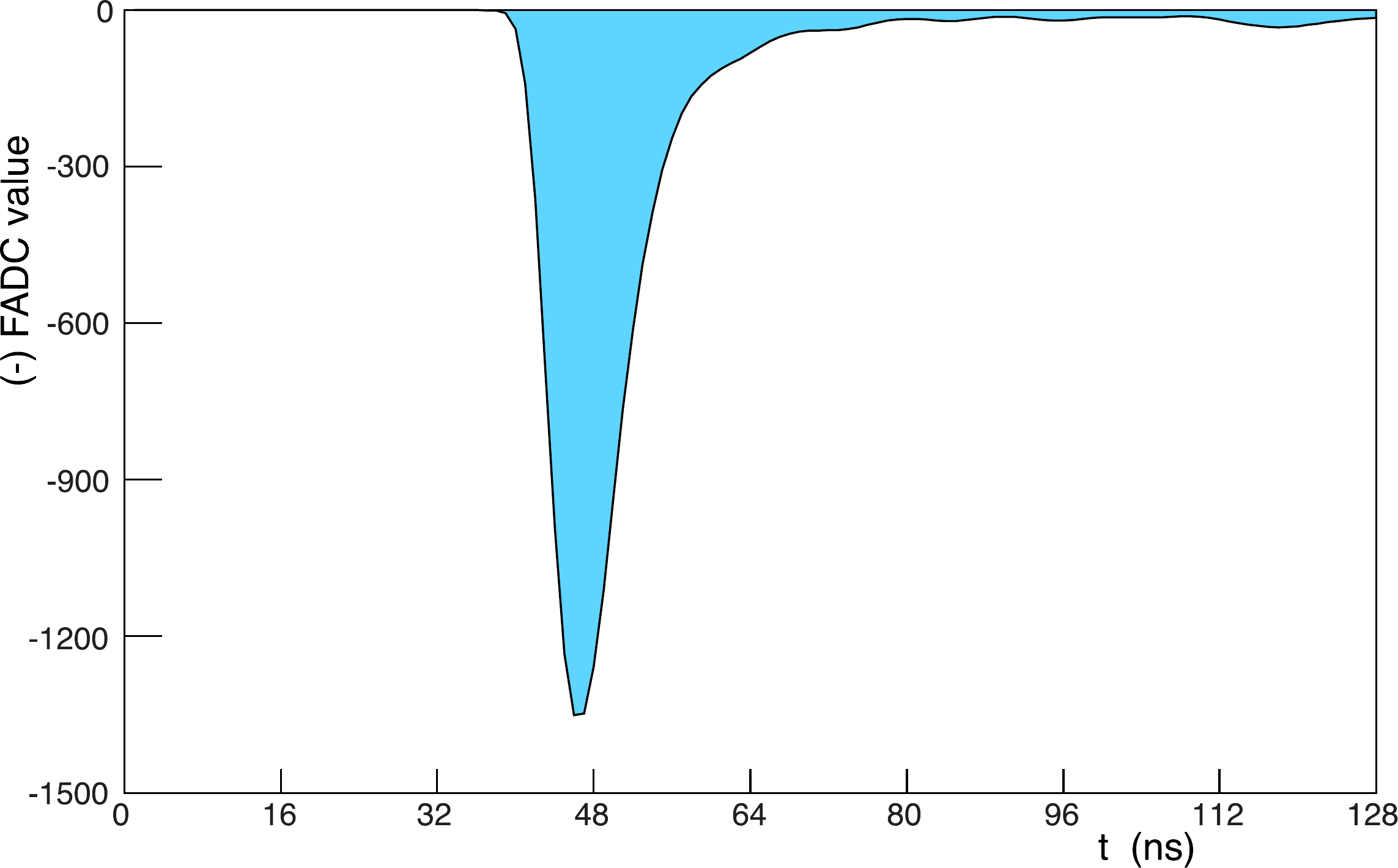}}
	\caption{Flash ADC value as a function of time recorded by the ARS system for a typical calorimeter pulse.} \label{fig:ars}
	\end{figure}

	The ARS uses the concept of analog memories to sample data at a clock rate of 1~GHz: each channel contains a circular array of 128 capacitors: every 1~ns, the ARS points the signal to the next capacitor, eventually overwriting itself after 128~ns. When a trigger is issued, the capacitor array is isolated and the previous 128 samples are stored.  During the next 500~ns, a separate trigger module (described below) decides whether or not to digitize the event.  Following a validation from the trigger, each capacitor array is digitized in parallel using a 12-bit Flash ADC
 at a rate of $1\,\mu$s per sample, for a total of $128\,\mu$s per channel.  During this long digitization period, we observe an exponential RC decay of the samples. This is compensated by a stable baseline included in the pulse waveform analysis.   
  Each ARS ASIC contains four channels, and four ARS ASICs were implemented onto each VME board for a total of 16 channels per board. 
  Figure~\ref{fig:ars} shows a typical calorimeter signal as a function of time, read out by the ARS system.

\subsection{Trigger}
\label{ssec:trigger}

	The Data Acquisition (DAQ) trigger for this experiment was a two-level system.  A standard HRS  electron  trigger was formed from the coincidence of the \v Cerenkov and S2m signals.  This Level-1 signal generated the ``Stop'' to freeze the analog data in the ARS. The ``Validation'', or Level-2 signal is generated by a dedicated DVCS Trigger module.    The DVCS Trigger includes a large backplane, containing FPGA logic.
	Each PbF$_2$ signal is first sent to a Trigger Daughter card, where it is split with one branch going to an ARS input and the
	second branch passing to a Fast ADC chip on the Daughter card.  Each Daughter card has 4 channels.  The 132 trigger ADCs are gated
	by the Level-1 signal, with a programmable width generally set to 60~ns.  Following digitization, the FPGA logic forms local $2\times 2$ 
	overlapping cluster sums.  If a cluster sum is found above a programmable threshold (generally set to 1 GeV equivalent), then the Level-2
	Validation signal is set to true.  This is completed within $\sim500$~ns.  In the absence of a validation signal at the end the 500 ns window, 
	a fast clear is issued to the DVCS Trigger and ARS.  In this way, for random Level-1 triggers, the deadtime is only $\sim 500$ ns, and the
	full readout is incurred only for a genuine H$(e,e'\gamma)X$ coincidence (including accidentals). Notice that only calorimeter channels belonging to $2\times 2$ clusters above threshold were digitized and recorded for each event.

The PA was not in the trigger and was read at every HRS--calorimeter coincidence. As the virtual photon has an almost fixed direction, the approximate region of the proton detector
that a DVCS proton would hit, can be inferred from only the calorimeter
region hit by the photon. The list of proton detector channels to be read out can then be made on the fly out of the information provided by the calorimeter trigger module. 
The accuracy of this prediction is dominated by a convolution of the HRS
acceptance and the
calorimeter energy resolution. Processes such as multiple scattering in the 
target and the scattering chamber can also affect the accuracy of
the prediction. The look-up table of proton detector channels to read as a
function of calorimeter ones was computed using the Monte-Carlo simulation, which included a realistic description of all the
elements of the experimental setup~(see section~\ref{ssec:simu}). The detector
inefficiency for DVCS protons due to this online readout choice is smaller than 0.1\%.
Depending on the kinematic setting, only between 15\% and 30\% of the proton
detector blocks needed to be read out in average. This reduced the amount of
data to record and therefore the acquisition deadtime. Note that no threshold was
set in any of the proton detector channels, so that even very low-energy
protons could be detected. The communication between the calorimeter and proton array crates necessary for this block selection in the proton detector was made possible by a custom multiplexer module (MUX) which allowed the calorimeter trigger module to send its data to the proton
array crate.

	The HRS DAQ functions in the standard way for a Hall A coincidence experiment, with all HRS analog PMT signals sent through delay cables corresponding to 880~ns. Scintillator signals go in common start into a LeCroy 1875 high resolution TDC. The delayed HRS signals arrive at their respective ADC and TDC inputs after the Level-2 decision is made.
	
	Even though the ARS represent a considerable advantage for this type of experiment it has an obvious drawback: the amount of data to transfer is about a hundred times higher than a regular HRS event. If the 232 ARS channels of the E00-110 experiment were to be recorded at every event, the event size would be 232$\times$128$\times$16/8$\sim$60kB. Typically, only $\sim 40$ ARS channels were recorded at every event.

\section{Data analysis}
\label{sec:Xsec}

As mentioned before, the selection of the $e p \to e p \gamma$ final state is based on a missing-mass analysis of the $e p \to e \gamma X$ event sample. This is made possible by the excellent momentum resolution of the Hall~A HRS and the fair energy and position resolutions of our dedicated electromagnetic calorimeter. The following subsections describe the selection of electron candidates from the HRS, and the analysis of the calorimeter in order to select the final-state photon. We will then focus on the final steps to ensure that our $e p \to e p \gamma$ selection is efficient and its purity close to perfect. Finally, we will describe the normalization procedure, the Monte-Carlo simulation and the method used to extract cross sections from our data and Monte-Carlo events.

\subsection{HRS Analysis}
\label{sssec:hrsana}

\begin{figure}
\centering\includegraphics[width=0.95\linewidth]{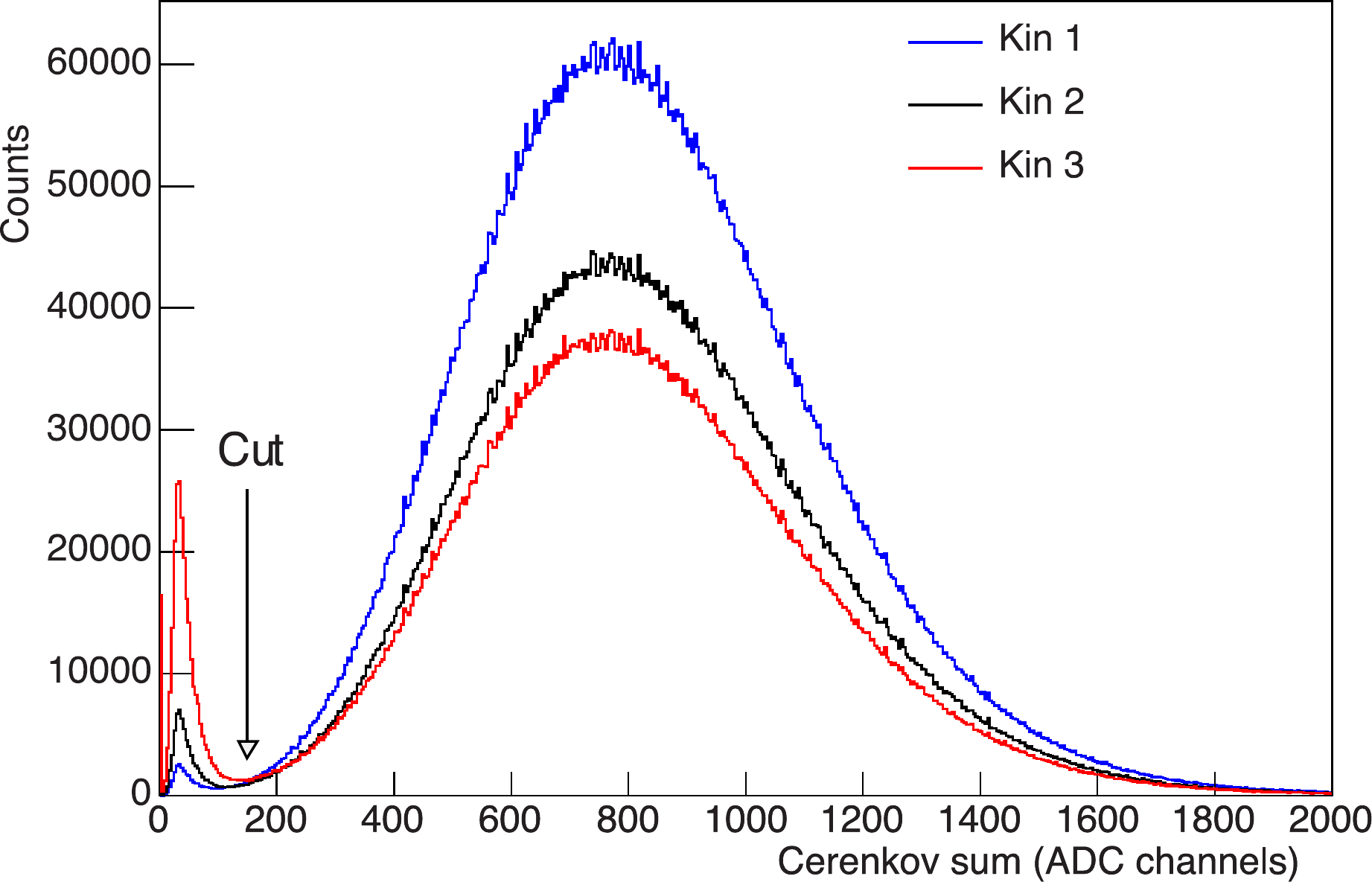}
\caption{(Color online) Distribution of the sum of all 10 \v Cerenkov PMT ADC values, for each
  kinematic setting. The cut applied to remove the 1-photoelectron signal from data is also shown.}
\label{fig:cer}
\end{figure}

The HRS \v Cerenkov detector was used for the electron identification. 
The number of photoelectrons detected is 7 on average so that the distribution is
Poissonian. Figure~\ref{fig:cer} shows the distribution of
the sum of all 10 PMTs (in ADC channels). The first 'peak' in the spectrum
is the tail of the electronic noise in the pedestal.
 We remove 1-photoelectron
events (either thermal emission in the PMT or $\delta$-rays from
pions) by applying a cut at 150 ADC channels.  The 1-photoelectron peak is
only visible if the \v Cerenkov signal is removed from the trigger, and a
cut is made on the Pion Rejector to select minimum ionizing
particles ({\it i.e.} pions).

\begin{figure}
\centering\includegraphics[width=0.95\linewidth]{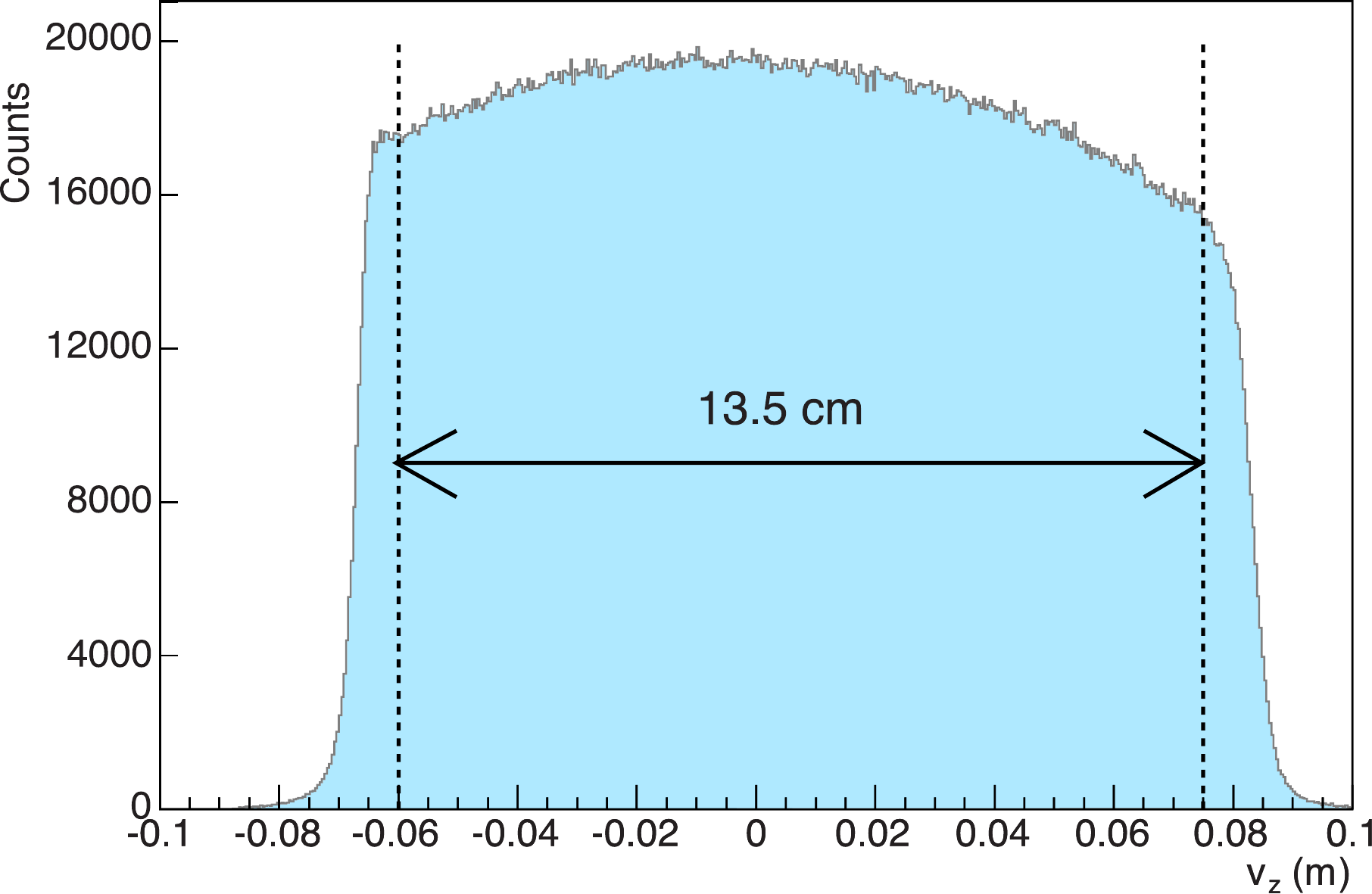}
\caption{(Color online) Reaction point along the beam reconstructed by the HRS. The cut
  on the target length applied is shown by the vertical lines. The 
  7.8\,mm downstream shift of the target observed during the experiment is also evident.}
\label{fig:vertex}
\end{figure}

Figure~\ref{fig:vertex} shows the distribution of the reaction point along
the beam $v_z$ reconstructed by the HRS. The target center relative to
the Hall center was determined to be 7.8\,mm downstream. A cut in order to avoid the contribution from the
target cell wall was applied to the data: $-6.00\,\text{cm}<v_z<7.50\,\text{cm}$.

\begin{figure}
\centering\includegraphics[width=0.95\linewidth]{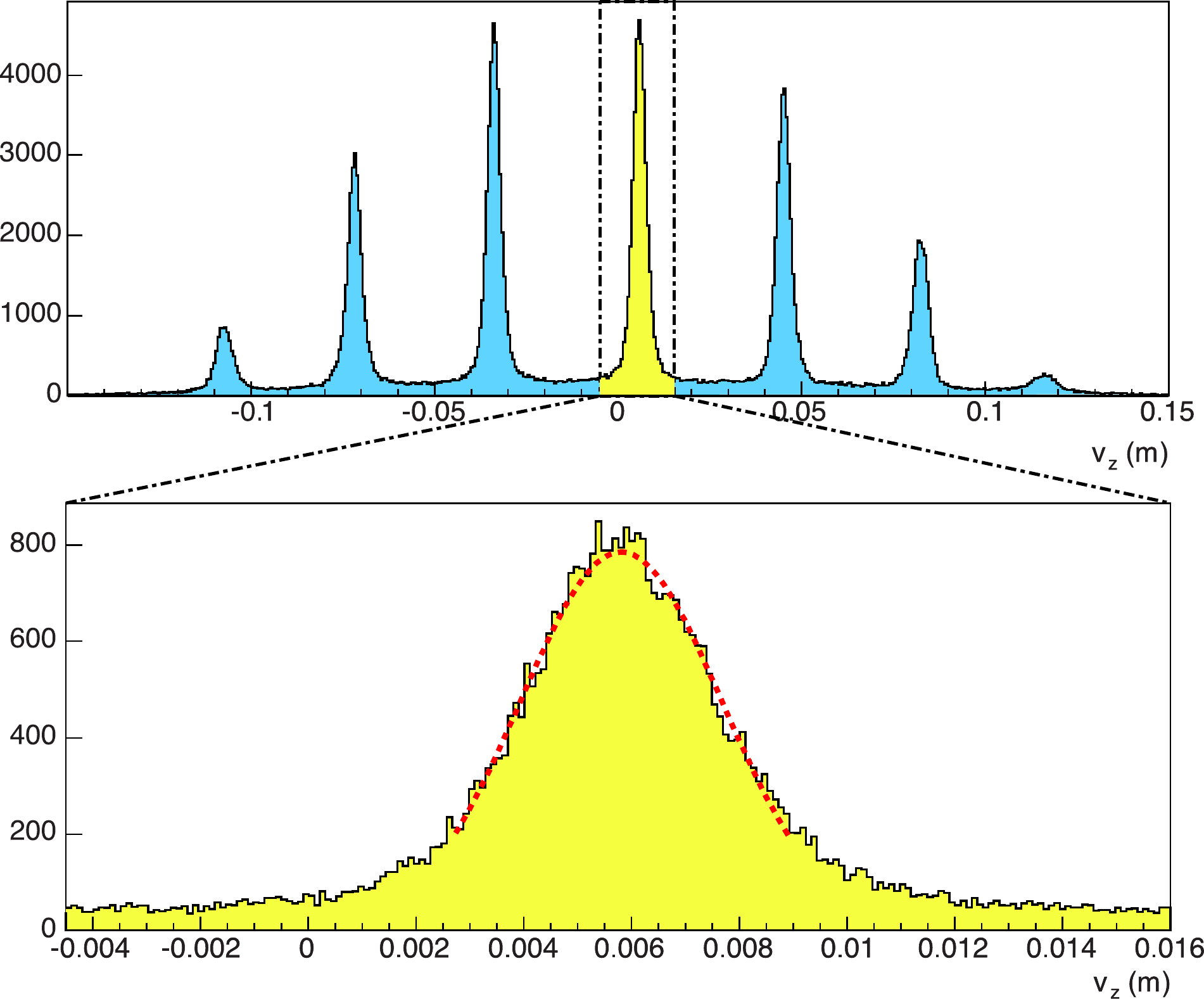}
\caption{(Color online) Top: Resolution of the vertex reconstruction from a multi-foil target. Bottom: Close-up of the central foil fit, which results in a $\sigma=1.9$\,mm resolution. The foil
  thickness is 1\,mm and the HRS was at 37.69$^\circ$ during this run.}
\label{fig:vreso}
\end{figure}

Figure~\ref{fig:vreso} shows the resolution on the vertex reconstruction as
measured with a carbon multi-foil target. The thickness of each foil is
1\,mm, and the HRS was at 37.69$^\circ$ during this run. The HRS vertex
resolution varies as:
\begin{equation}
\sigma=\frac{\sigma_{90^\circ}}{\sin{\theta_\text{HRS}}}\,.
\end{equation}
The $\sigma$ measured at 37.69$^\circ$ is 1.87\,mm, which means 1.2\,mm at
90$^\circ$ (the $\sigma$ introduced by the foil thickness is
$(1/\sqrt{12})$\,mm and can be neglected).

The HRS acceptance is a hypervolume depending on 5 correlated variables: $x_{tg}$ and $\theta_{tg}$ (the position of the particle and the tangent of the angle made by its trajectory along the dispersive direction), $y_{tg}$ and $\phi_{tg}$ (the position and the tangent of the angle perpendicular to the dispersive direction), and $\delta_{tg}$ (the fractional deviation of the particle momentum with respect to the central momentum of the HRS). Trajectories of higher-momentum particles have lower curvature in the dipole, and in order for them to fit into the spectrometer they need to have lower $\theta_{tg}$. The dipole magnet has a trapezoidal cross section and higher-momentum particles tending to fly closer to its shorter base (high
magnetic field) side, which makes the accepted range of $\phi_{tg}$
smaller for higher $\delta_{tg}$. Finally, increasing $y_{tg}$ requires
decreasing $\phi_{tg}$ in order for the particle to get into the spectrometer
entrance window. Making cuts independently in each of the variables to limit
events to flat acceptance regions in each of them is thus very inefficient. Instead, we used an acceptance function~\cite{Rvachev:2001}, which allows to place a 4-dimensional cut ($x_{tg}=0$ is assumed). This procedure is almost
twice more efficient than the traditional sequential acceptance cuts. This function takes
the arguments $y_{tg}$, $\theta_{tg}$, $\phi_{tg}$ and $\delta_{tg}$ and returns a so-called {R-value} which is the minimum distance (in radians) to the $(\theta_{tg},\phi_{tg})$ solid angle acceptance region appropriate for a given value of $y_{tg}$ and $\delta_{tg}$. A value of 5~mrad was used in order to constrain a well-defined region of the HRS acceptance. The cross-section results varied by no more than 1\% when increasing the R-function cut. This value was used as an estimate of the HRS acceptance systematic uncertainty.
 
\subsection{Calorimeter Analysis}

The calorimeter analysis is done in two steps: first, the recorded ARS waveforms are analyzed in order to extract the time and energy information. Then, an algorithm is used to aggregate the block information into photon clusters with a measured position, time and total energy.

\subsubsection{ARS Waveform Analysis}
All the detector channels of the electromagnetic calorimeter were equipped with ARS electronics, which allowed to save the full waveform of blocks that were recorded during a trigger, in a manner similar to a digital oscilloscope. In order to extract time and amplitude information from the ARS, a waveform analysis is needed which is performed offline.

Each pulse as a function of time is described by a reference pulse multiplied by an amplitude. For an ideal event without noise, the amplitude of the pulse and its arrival time are free parameters. For any given arrival time $t$, the amplitude $a(t)$ which best fits the signal $\{x_i\}$
is simply given by the one which minimizes:
\begin{equation}
\chi^2(t)=\sum_{i_{min}}^{i_{max}} (x_i-a(t)h_{i-t}-b(t))^2\,,
\label{eq:wfchi2}
\end{equation}
where $\{h_i\}$ is the reference shape. Notice that we also fit a flat baseline $b(t)$. Reference shapes for each individual PMT are determined experimentally from data, using elastic calibration runs, where the probability of pile-up is very small. In order to reduce the impact of accidental events, only $i_{max}-i_{min}$=80 ARS samples were used in the calorimeter analysis, centered around the expected arrival time of DVCS events, which because of cable lengths, varies slightly from one channel to another. The partial derivatives of $\chi^2(t)$ with respect to  $a(t)$ and $b(t)$ yield a linear set of equations in order to obtain the best amplitude for any given arrival time $t$. If the minimum value of $\chi^2(t)$ found for all the possible $t$ is above a given analysis threshold $\chi^2_1$, the algorithm will fit a second pulse to the waveform by minimizing:
\begin{multline}
\chi^2(t_1,t_2)=\sum_{i_{min}}^{i_{max}} (x_i-a_1(t_1,t_2)h_{i-t_1}-\\a_2(t_1,t_2)h_{i-t_2}-b(t_1,t_2))^2\,,
\end{multline}
for every combination of $t_1$ and $t_2$. For every pair of $t_1$ and $t_2$ and the corresponding fitted amplitudes and baseline, a reduced $\chi^2$ is also computed in a time window of $\pm 20$\,ns around the minimum of the pulse. The minimum reduced $\chi^2$ found determines the amplitudes and arrival times of the pulses. 
Pulses were searched in a $[-20,25]$\,ns interval around the expectation arrival time of events, in steps of 1\,ns. An improved time resolution is obtained by interpolating around the time that minimizes the $\chi^2$ for any time $t=t_1,t_2$:
\begin{equation}
t=t(\chi^2_{min})+\frac{\chi_{t-1}^2-\chi_{t+1}^2}{2(\chi_{t+1}^2+\chi_{t-1}^2-2\chi^2_{min})} \,.
\end{equation}
The threshold value $\chi^2_1$ used for the analysis corresponded to an effective missed pulse of $\sim$280\,MeV for each particular calorimeter block (which translates to slightly different ARS channel thresholds due to the different calibration of each block). Also, if the $\chi^2$ of a fit by a flat-line $b$ was below an equivalent energy of $\chi^2_0 \sim 40$~MeV, no pulse was fitted and the signal was discarded. Finally, if 2 pulses were found with a relative arrival time smaller than 4\,ns, the algorithm returned the best single pulse fit since 2-pulse results proved to be unstable in those cases.

The waveform analysis of the proton array ARS data used the same algorithm, but with slightly different parameters. Energy thresholds were set to $\chi^2_0\sim$~2\,MeV and $\chi^2_1\sim$~15\,MeV in order to best fit the much smaller recoil proton energies in the detector. Due to the high counting rate in the detector, only 30 ARS samples were used for the fit. Also, time windows to search for pulses were set to $-20\leq t_1,t_2\leq 20$\,ns around the expected event signal.

Overall, the waveform analysis of ARS signals increases the energy
resolution in the DVCS calorimeter by a factor of 2--3 (depending on
the background level) with respect to results obtained integrating the
signal in a 60~ns window. We found about 8\% of events in the
calorimeter with some pile-up from accidentals.

\subsubsection{Clustering Algorithm}
The algorithm used to separate clusters in the electromagnetic calorimeter is based on a cellular
automata, as described in~\cite{cell}, and uses only pulses arriving within a [-3,3]~ns interval. This coincidence time window is more than 6 times the time resolution of the detector ($\sim$0.8\,ns). For each cluster found, the  total photon energy $E$ is taken to be the sum over the deposited energy
$E_i$ in each of the cluster blocks:
\begin{equation}
E=\sum_i E_i\qquad\qquad E_i=C_iA_i\,,
\end{equation}
where $A_i$ is the signal amplitude collected in block $i$ and $C_i$ its
calibration coefficient. The impact position $x_{clus}$ is calculated as the sum of blocks positions $x_i$
weighted logarithmically by the relative energy deposition in each of them \cite{Awes:1992yp}:
\begin{equation}
x_{clus}=\frac{\sum_i w_i\ x_i}{\sum_i w_i}\quad w_i=\max{\{0,
 {W_0}+\ln (E_i/E) \}}\,.
\label{eq:pos}
\end{equation}
The parameter $W_0$ allows a further tuning of the relative weight between
blocks: as $W_0\to\infty$ the weighting becomes uniform regardless of the
energy deposited in each block, whereas small values of $W_0$ give a larger
relative weight to blocks with large energy deposition. The value of $W_0$ fixes
the energy threshold for blocks to be taken into account in the position
determination: blocks with a relative energy deposition less than
$e^{-W_0}$ are neglected in the calculation.

The calorimeter was placed at 110\,cm from the 15-cm-long
target. The incidence angle of particles on the front face of the calorimeter could therefore vary by significant amounts: corrections due to the vertex position in the target needed to
be applied. Furthermore, the electromagnetic shower does not begin at the
surface of the calorimeter, but at a certain depth as shown in Fig.~\ref{fig:profile}. This depth is, to first
approximation, independent of the incident particle energy. Taking
these two effects into account, the position $x_{clus}$ given by
equation~(\ref{eq:pos}) is corrected by:
\begin{equation}
x_{corr}= x_{clus}\bigg (1-\frac{a}{\sqrt{L_{vc}^2+x^2}}\bigg )
\end{equation}
where $L_{vc}$ is the distance from the vertex to the calorimeter and $a$
is the distance of the electromagnetic shower centroid to the calorimeter
front face, taken along the direction of its propagation.
The algorithm depends on two parameters $W_0$ and $a$, which have been
optimized to $W_0=4.3$ and $a=7$~cm by Monte-Carlo simulation and real data from the elastic runs, where a 2\,mm position resolution ($\sigma$) 
at 1.1\,m and 4.2\,GeV was measured, compatible with the one
obtained from Monte-Carlo simulations. Position resolution when two partially overlapping clusters are present, is slightly worse than in the case of a single cluster: simulated data show in this case a 4\,mm spatial resolution.

\subsection{Event Selection}
\label{sssec:DVCSana}

\begin{figure}
\centering
\epsfig{file=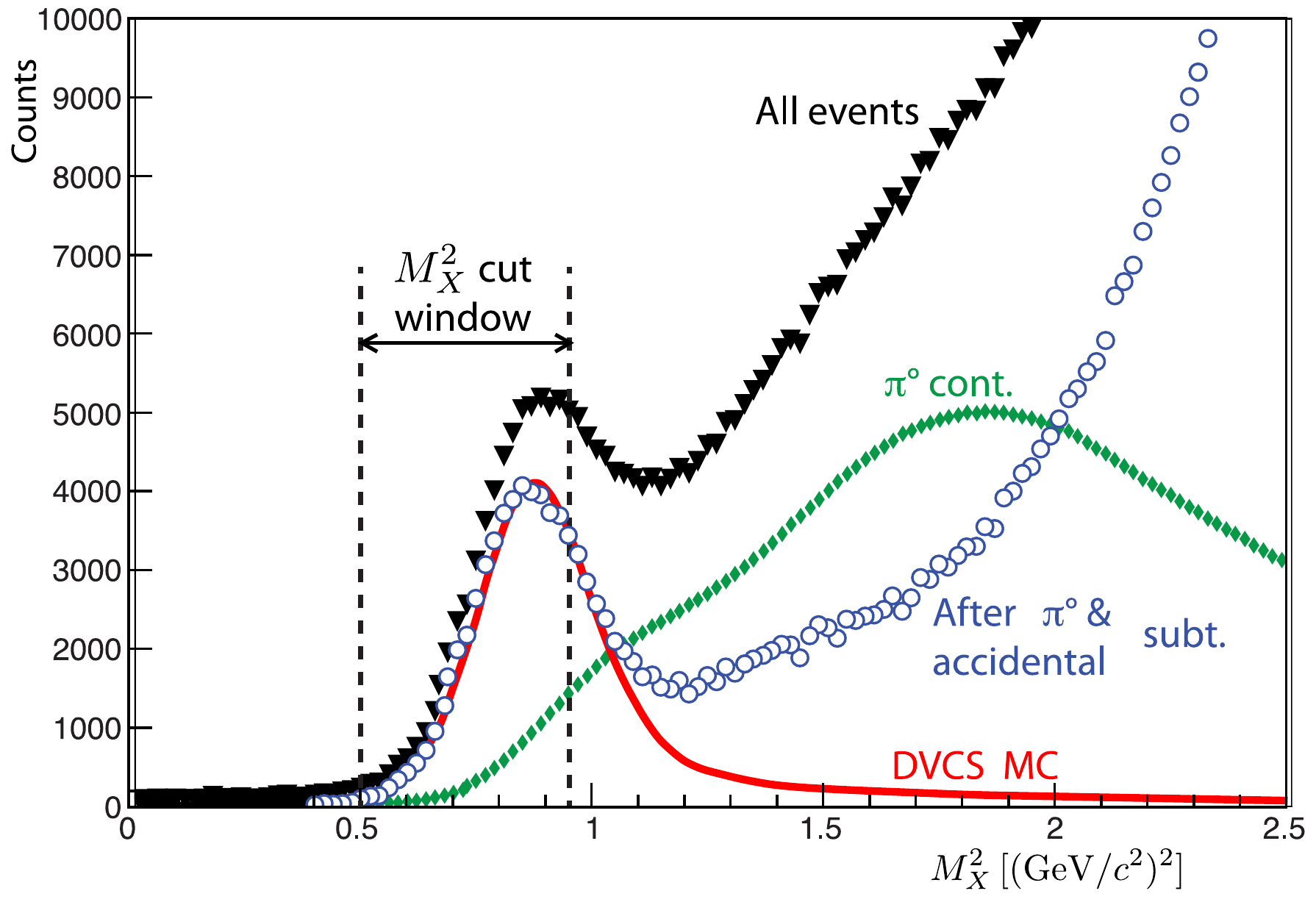,width=0.95\linewidth,clip=} 
\caption{\label{mm_kin} (Color online) Squared missing mass associated with the reaction $ep\rightarrow e\gamma X$ for Kin2. Total events for Kin2 are represented as inverted black triangles, the estimated $\pi^0$ contamination is represented as green diamonds, the distribution after the subtraction of accidentals and $\pi^0$'s is shown as blue open circles. Finally, it is compared with the DVCS Monte-Carlo shown as a red solid line. In order to remove unnecessary uncertainties due to low-missing-mass-squared accidental events, we apply a cut requiring a missing mass squared higher than 0.5~GeV$^2$ for all kinematics.
}
\end{figure}

\begin{figure}
\centering\includegraphics[width=0.95\linewidth]{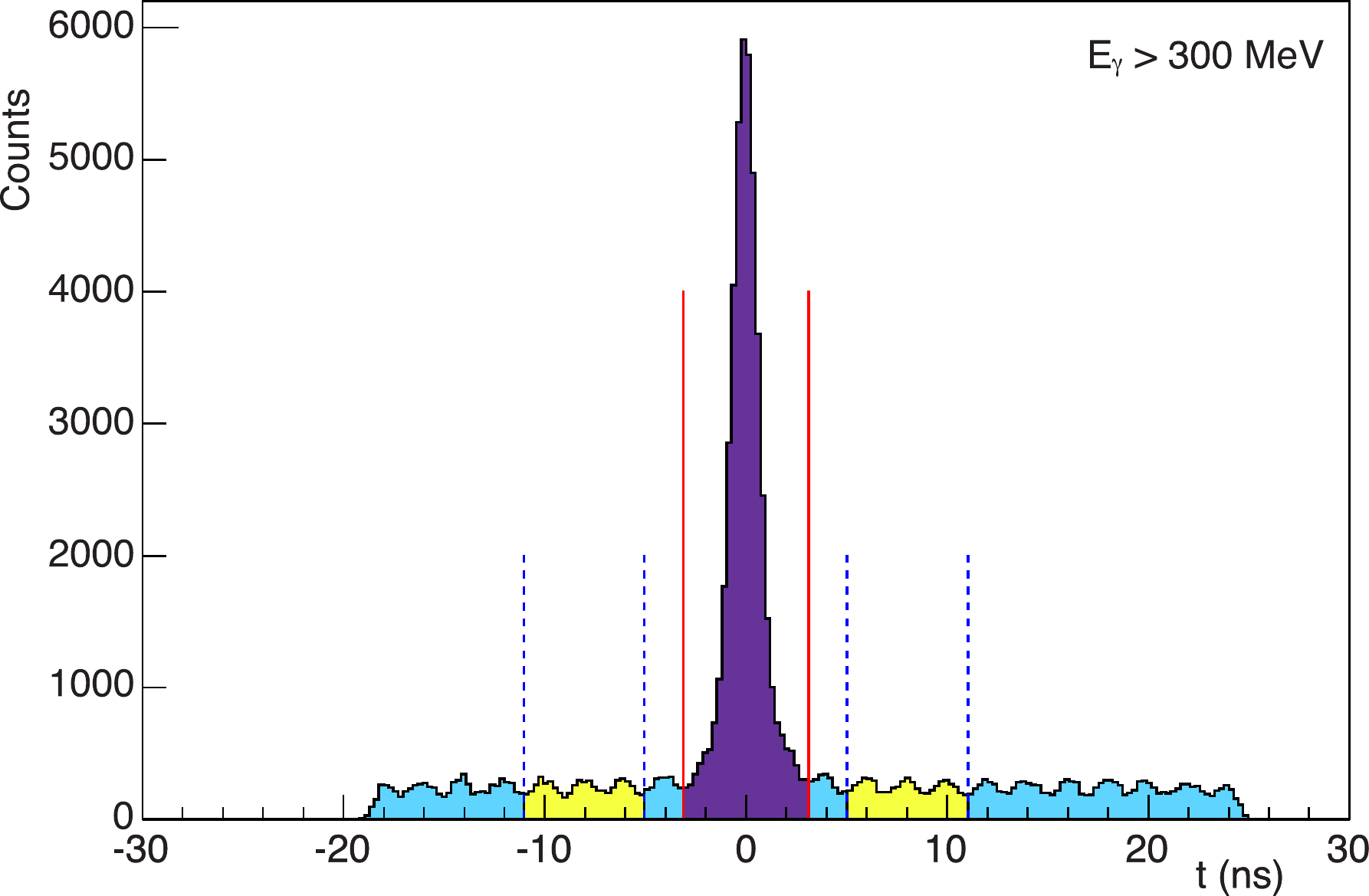}
\caption{(Color online) Time spectrum of blocks with $E>300$\, MeV in the Kin3 setting. It shows the 45\,ns
  time window of the waveform analysis. The 2\,ns CEBAF beam structure is clearly visible. The coincidence
  [-3,3]\,ns window
  used for clustering is shown by the solid line. Dashed lines show windows
  used for the HRS--calorimeter accidental subtraction.}
\label{fig:time_calo}
\end{figure}

The $e p \to e p \gamma$ events are selected among the calorimeter 1-cluster events. A software energy threshold of 1.1~GeV was applied to calorimeter clusters, slightly above the hardware threshold of $\sim 1$~GeV. Fiducial cuts were used to discard events hitting blocks at the edges of the calorimeter. Figure~\ref{mm_kin} shows the $ep\rightarrow e\gamma X$ missing-mass-squared distribution of the data. Accidental coincidences were estimated by analyzing events in [-11,-5] and [5,11]~ns time windows, the same width as the coincidence clustering window but shifted in time (see Fig.~\ref{fig:time_calo}). The use of two intervals to estimate the accidental sample reduces its statistical uncertainty.

Neutral pion decays with only one photon reaching the calorimeter form an important source of background to the DVCS sample. This background is subtracted using $\pi^0$ events where the two photons are detected in the calorimeter. For each detected $\pi^0$, its isotropic decay in its center-of-mass frame is simulated $n_{dec}=5000$ times, and the decay photons are projected onto the calorimeter acceptance. This simulation allows us to make a statistical subtraction of the $\pi^0$ background to the DVCS signal, including both exclusive and inclusive $\pi^0$ events. The subtraction is obtained from the simulated decays in which one of the photons is emitted close to the pion momentum direction. Note that this background subtraction scheme could not be applied in Kin1 as the energy of $\pi^0$ decay photons is too close to the calorimeter threshold to ensure an efficient background subtraction.

\begin{figure}[!htp]
\centering
\epsfig{file=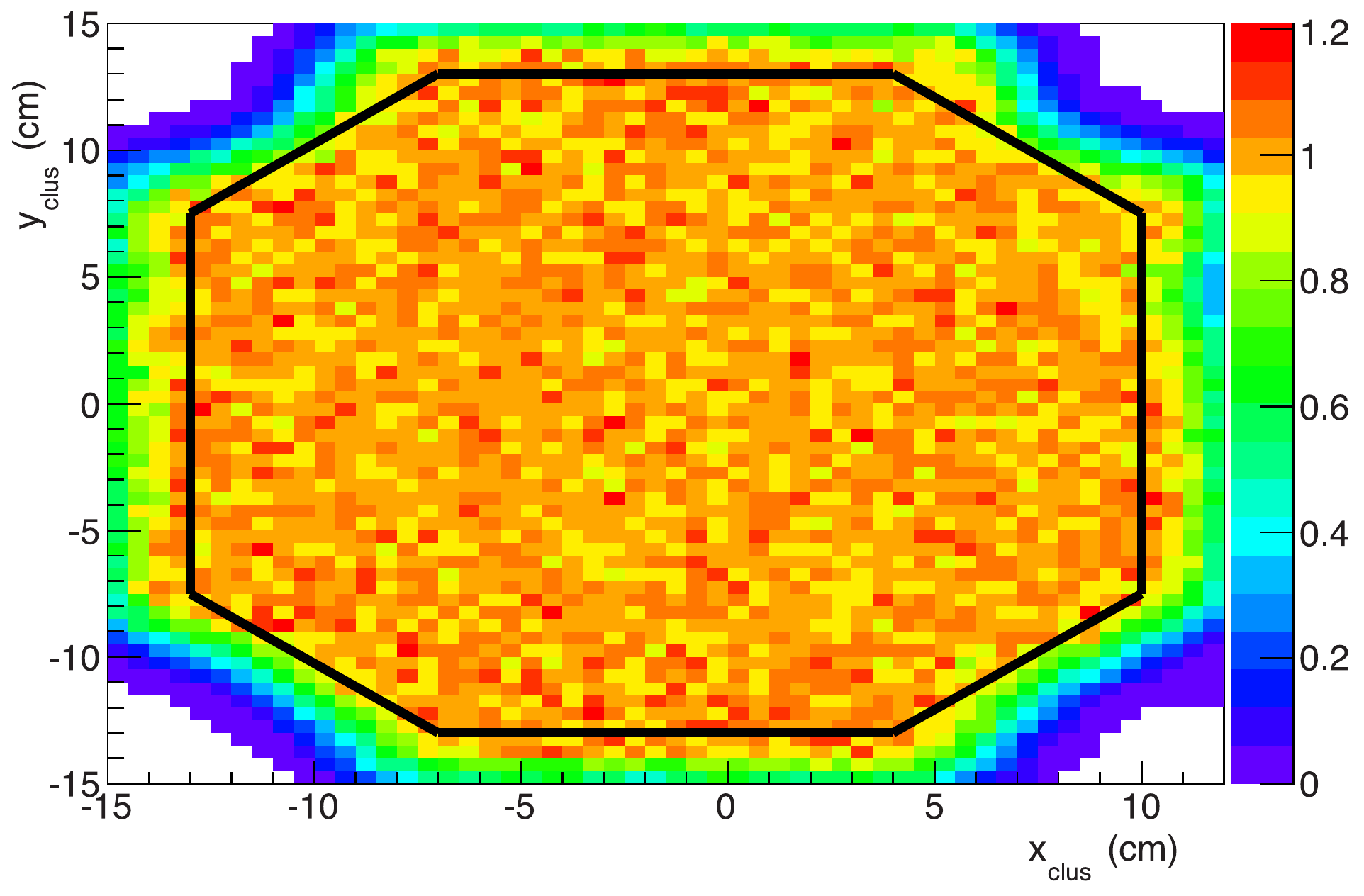,width=0.95\linewidth,clip=}
\caption{\label{pi0sub} 
(Color online) The octagonal cut on the front face of the calorimeter applied to all events in order to ascertain a high efficiency in the $\pi^0$ subtraction. The general shape and size of the cut can be understood from the size of the front face and the width of the shower profile (Fig.~\ref{fig:profile}).}  
\end{figure}

A self-consistency check of the $\pi^0$ subtraction method was performed using a Monte-Carlo. $\pi^0$'s were generated over the acceptance and classified into two categories: the one-photon-detected and the two-photon-detected events. After applying the $\pi^0$-subtraction method described above to the two-photon category, we obtained a number of one-photon events and compared it to the one-photon-detected category. The result is presented on Fig~\ref{pi0sub}. This efficiency ratio is close to 1 except in the corners or close to the edges. Therefore we applied a geometrical cut on the cluster in the data and the Monte-Carlo simulation, also shown on the figure.

\begin{figure}[!htp]
\centering
\epsfig{file=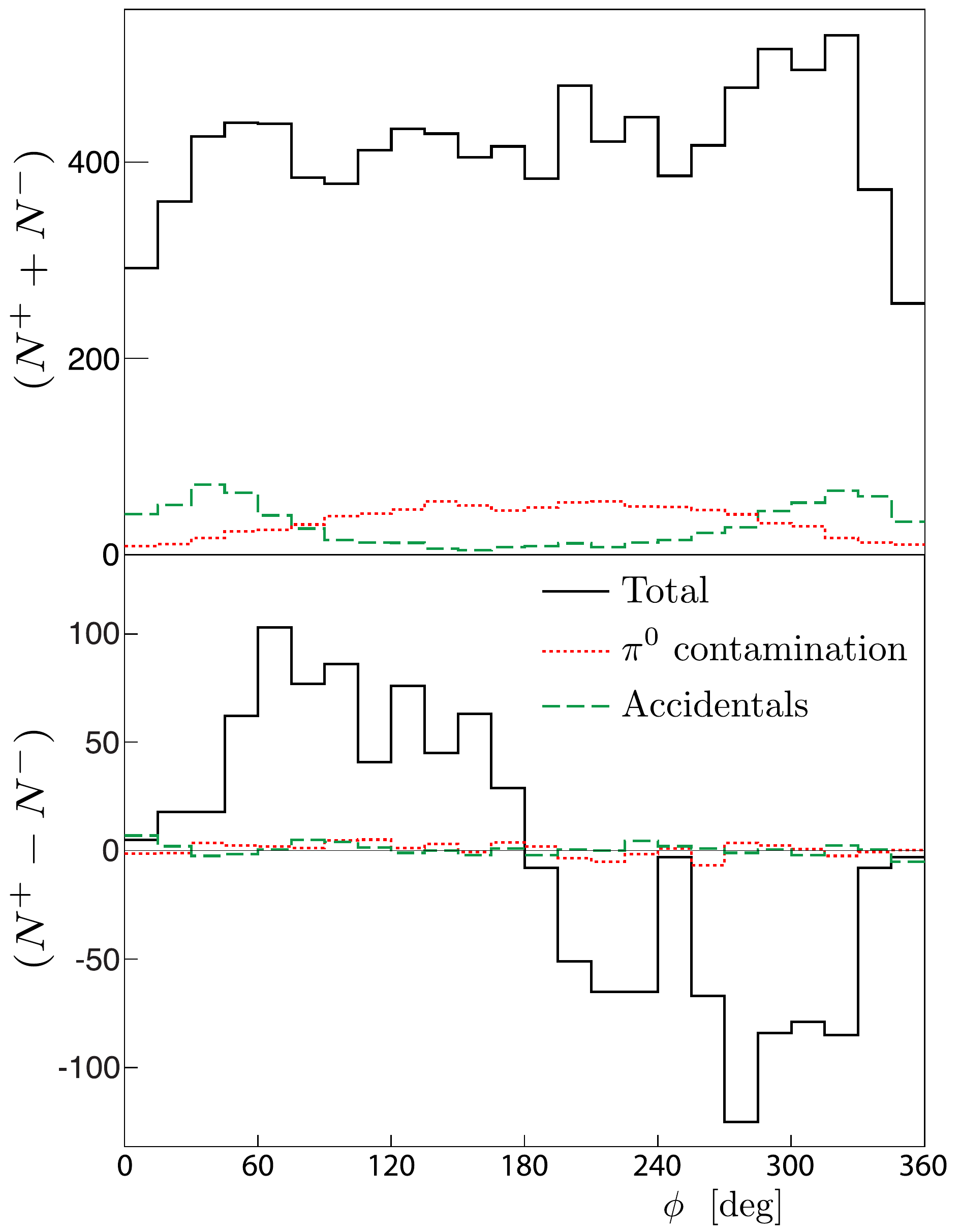,width=0.95\linewidth,clip=}
\caption{\label{fig:counts} 
(Color online) Total counts (top) and difference of counts for opposite helicities as a function of $\phi$ (bottom) for the Kin3 bin $x_B=0.37$, $Q^2=2.36$~GeV$^2$ and $-t=0.32$~GeV$^2$. The solid curve histogram in black corresponds to the distribution of events after all analysis cuts have been performed. The estimated contribution corresponding to accidental events is shown as a dashed green histogram. The estimated $\pi^0$ contribution is represented as a red dotted histogram.}  
\end{figure}

Figure~\ref{fig:counts} (top) shows the total distribution of events in a Kin3 bin along with the accidentals and $\pi^0$ contributions. Accidental events reside close to $\phi=0^\circ$ which corresponds to the beamline side of the calorimeter, where higher single rates are observed. The contribution of $\pi^0$ events, however, is larger around $\phi=180^\circ$. This feature remains true for most experimental bins. The bottom plot of Fig.~\ref{fig:counts} shows the helicity-dependent distribution of events for the same bin. The contribution of accidental events cancels in this difference of counts, as they are essentially helicity-independent. The fact that the same feature is observed for $\pi^0$'s is not trivial. As it turns out, exclusive $\pi^0$ events are known to have a small beam-spin asymmetry at Jefferson Lab kinematics~\cite{DeMasi:2007id}, and the $\pi^0$ events we subtract may include semi-inclusive $\pi^0$'s that have an even smaller asymmetry~\cite{Aghasyan:2011ha}. We have checked that in all our experimental bins the contributions of both accidental and $\pi^0$ events to the difference of counts for opposite helicities are compatible with 0 within statistical error bars. We have therefore decided to not subtract these contributions in the computation of the helicity-dependent cross sections. In this way, even though we were unable to evaluate the unpolarized cross section for Kin1, we did succeed in evaluating the helicity-dependent cross section.

After the $\pi^0$ subtraction, the only remaining channels (other than exclusive DVCS) are kinematically constrained to $M_x^2>(M+m_\pi)^2$. However, resolution effects may cause these channels to contribute
below the $M_x^2$ cut. This contamination was strongly suppressed by the tight missing-mass-squared cut and will be evaluated in section~\ref{sssec:systexclu}.

All selection cuts in this analysis are summarized in Tab.~\ref{tab:selec}.
 \begin{table}
 \begin{tabular}{lc}
 \hline \hline
 {\bf HRS} & \\ \hline \hline
Track mult. & $N_{tr} = 1$ \\
 R-function & R-value$>0.005$ \\
 \u{C}erenkov & ADC$_{sum}>150$ \\
Vertex & $-6.0$~cm$<v_z<7.5$~cm \\
 \hline \hline
  {\bf Calorimeter} & \\
   \hline \hline 
Cluster mult. & $N_{clus}= 1$ \\
Energy & $E_{clus}>1.1$~GeV \\
Position &  $-13<x_{clus}<10$\\
(in cm) & $|y_{clus}|<13$ \\ 
& $y_{clus}<0.92 \times \left(x_{clus} +13\right)+7.5$ \\
& $y_{clus}<-0.92 \times \left(x_{clus}-4\right)+13$ \\
 & $y_{clus}>-0.92 \times \left(x_{clus} +13\right)-7.5$ \\
 & $y_{clus}>0.92 \times \left(x_{clus} -4\right)-13$ \\ \hline \hline
 {\bf Exclusivity} & \\
  \hline \hline 
All settings & $M^2_{X}>$ 0.5~GeV$^2$/c$^4$  \\ 
Kin1-2 & $M^2_{X}<$ 0.95~GeV$^2$/c$^4$  \\
Kin3 & $M^2_{X}<$ 1.09~GeV$^2$/c$^4$  \\
 \hline \hline
 \end{tabular}
 \caption{\label{tab:selec} Summary of the $e p \to e p \gamma$ selection cuts.}
 \end{table}

\subsection{Efficiencies and Normalization}
\label{sssec:eff_cor}

The efficiency of the scintillators that were used for the electron trigger was monitored during dedicated runs along the experiment. An efficiency of 99.95\% was measured over the duration of the experiment. The efficiency of the \u{C}erenkov counter used to discriminate electrons from negative pions was measured to be 99\%. The purity of the electron sample was estimated at 98.8\%, further enhanced by the missing-mass-squared cut on H$(e,e'\gamma) X $. We estimated that a maximum of 0.5\% of electrons may still be misidentified and consider this value as the systematic uncertainty on the electron identification.

The dead time associated with the data acquisition is determined by comparing the number of pulses from two clocks running both at 62.5 MHz: one is always running, the other one is veto-ed when the DAQ is busy. The integrated luminosity is corrected for the dead time on a run-by-run basis, with an associated systematic error estimated to be 1\% for Hall~A \cite{Alcorn:2004sb}. The average dead time varied between 14\% and 40\% depending on the kinematic setting.

When multiple tracks were detected in the HRS, events were discarded due to the unreliability of the reconstruction. These events represent between 7\% and 10\% of the total statistics, depending on the kinematic setting. However, most of these multi-track events show a very low energy in the pion rejector, indicating that most of them contain secondary tracks from showers generated in the exit region of the Q3 magnet or pions that trigger the DAQ with $\delta-$rays. The number of multi-track events corresponding to good electrons was estimated by requiring a $\sim$1.7~GeV minimum energy deposited in the pion rejector. The number of good electron events with 2 or more tracks in the VDCs amounts to only $\sim$2\% of the total number of events for all kinematics. The 0.5\% associated systematic error has been evaluated by changing the energy threshold of the pion rejector.

Similarly, multi-cluster events in the DVCS calorimeter are discarded from the analysis. They represent from 1\% to 5\% of the statistics, depending on the kinematic setting. In order to apply a correction for this, 2-cluster events were thoroughly studied. All selection cuts were applied to each of the two photons, and a correction was computed, based on the number of events that remain after the cuts are applied. Two-cluster events with an invariant mass between 100 and 170~MeV/c$^2$ were not  included in the sample used to calculate this correction as they are mostly decay photons from neutral pions. In rare cases where both photons fulfilled all selection cuts, they contribute to the correction with a relative weight based on the accidental rate measured in their respective kinematical bin. We attribute a systematic uncertainty to the multi-cluster correction based on the number of events with more than 2 clusters in the calorimeter, which were not considered in our analysis. This number represents 7\% of the 2-cluster events, and therefore an associated systematic uncertainty of less than 0.4\% overall.

Table~\ref{tab:corlist} gives a summary of efficiency factors applied to experimental yields.

\begin{table}[htp]
\centering
\begin{tabular}{lc}
Source &  Correction to yield \\
\hline\hline
\v Cerenkov & 1.01 \\
Multi-track & 1.02 \\
Multi-cluster & 1.02 \\
\hline\hline
Total & 1.05 \\
\end{tabular}
\caption{\label{tab:corlist} Summary of efficiency factors to be applied multiplicatively to experimental yields. The multi-cluster correction depends on the kinematic setting and the experimental bin, only the average value is listed in this table.}
\end{table}

\subsection{Monte-Carlo Simulation}
\label{ssec:simu}

The experimental setup was implemented in a {\sc Geant4} Monte-Carlo simulation. The HRS geometrical acceptance was modeled by a collimator window placed at the entrance of the spectrometer.  Its acceptance was simulated by applying the same R-value cut that is used for the experimental data (R-value$\, >\, $5~mrad).
The PbF$_2$ DVCS electromagnetic calorimeter geometry was implemented in detail, including all active and passive materials of the experimental setup. Only the energy deposit of particles in the calorimeter is digitized in our simulation, as the generation and tracking of \v Cerenkov photons requires unrealistic simulation times and proves to be unreliable due to the difficulty to define optical surfaces accurately. Detector offsets were adjusted following geometrical surveys of the experimental equipment.

Events were generated following a flat distribution in $Q^2$, $x_B$, $t$, $\phi$ and $\phi_e$. In addition, the $z$-position of the vertex was randomized within the full length of the target cell. The ranges of $Q^2$ and $x_B$ are defined by the angular and momentum acceptance of the HRS. The hadronic part of the reaction ($\gamma^*p\rightarrow\gamma p$) is computed in its center-of-mass and final-state particles are then boosted to the laboratory frame. The generation range in $t$ is kinematically constrained event-by-event by the values of $Q^2$ and $x_B$. The angle $\phi$ is then generated uniformly inside 2$\pi$. Finally, all particles in the final state are rotated around the beam axis by $\phi_e$, chosen large enough to cover the full vertical acceptance of the HRS for all positions along the length of the target. Each event is then weighted by a phase-space factor $\Delta \Gamma=\Delta x_B \Delta Q^2 \Delta \phi \Delta t(x_B,Q^2)\Delta\phi_e/N_{gen}$, where $N_{gen}$ is the total number of generated events.

\begin{figure}[htp]
\centering
\epsfig{file=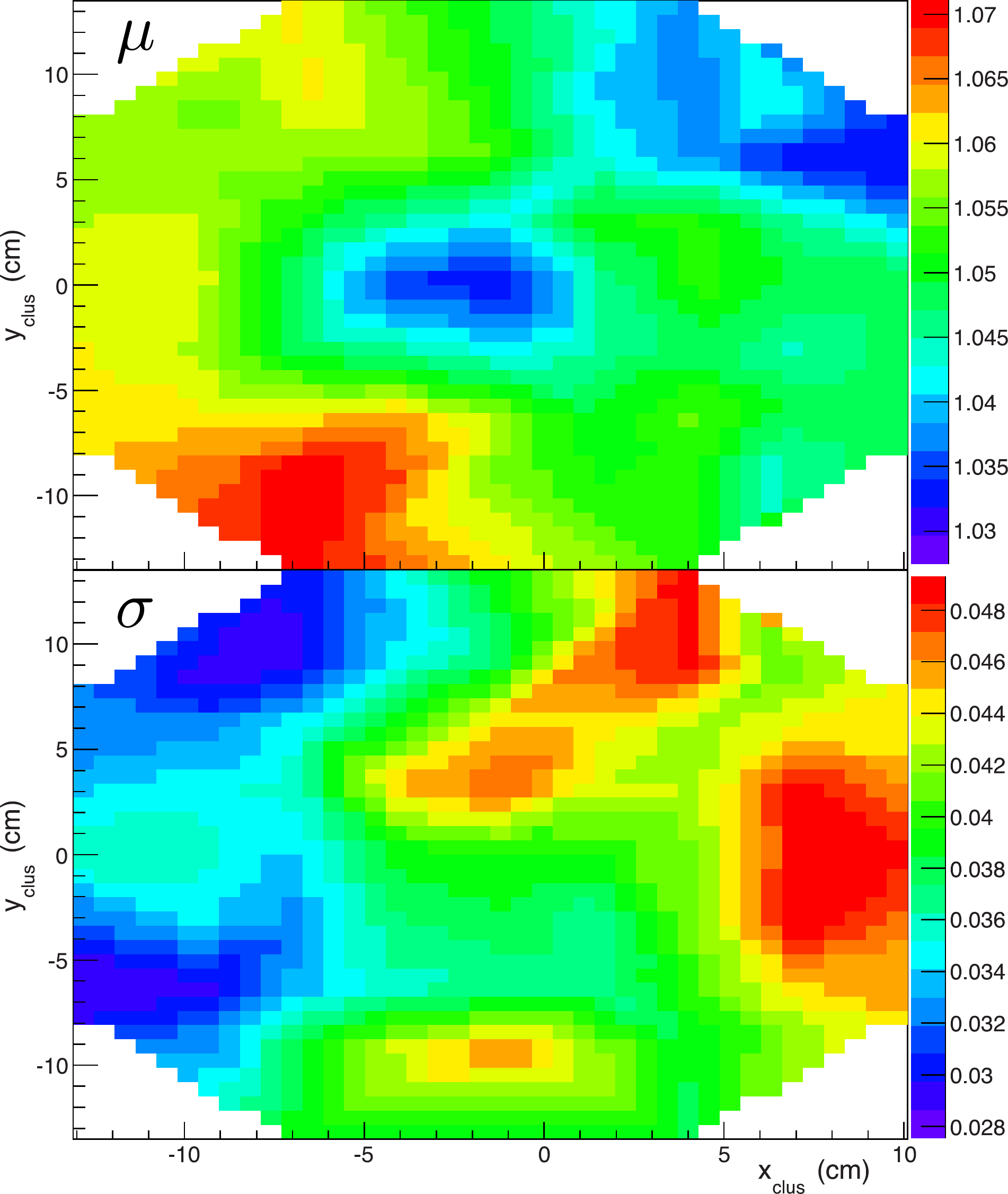,width=0.95\linewidth}
\caption{\label{mapmeansig} (Color online) Mean $\mu$ (top) and standard deviation $\sigma$ (bottom) of the gaussian distribution used to smear the simulated photon data of Kin3 viewed on the calorimeter surface $y_{clus}$ vs. $x_{clus}$ (beam on the right side). A worse energy resolution is observed at small angles with respect to the electron beam (positive $x_{clus}$). We also notice areas of fluctuating resolution corresponding to varying quality of the PbF$_2$ crystals.}
\label{fig:smearing}
\end{figure}

Because of Bremsstrahlung energy losses and resolution effects, the missing-mass-squared cut removes a significant fraction of exclusive events. This is corrected through the Monte-Carlo simulation by applying the same cut in the simulated data. However, the experimental resolution of the calorimeter and the imperfections of the calibration procedure have to be reproduced by the Monte-Carlo simulation. 
In order to achieve this, the detector is divided into 49 partially overlapping areas. From the photon four-momentum in the Monte-Carlo simulation the following smearing transformation is applied:
\begin{center}
\begin{equation}
\begin{pmatrix} q_x\\ q_y \\ q_z \\E\end{pmatrix} \longmapsto gaus(\mu,\sigma) \times \begin{pmatrix} q_x\\ q_y \\ q_z \\E\end{pmatrix}\,.
\end{equation}
\end{center}
In each area, the parameters $\mu$ and $\sigma$ are fitted in order to best match the $M^2_X$ spectra of the simulated and the experimental data in the exclusive region. The final values of $\mu$ and $\sigma$ used to smear the simulated events are interpolated event-by-event according to the impact point of the photon in the calorimeter. Figure~\ref{fig:smearing} shows the resulting values of $\mu$ and $\sigma$ for Kin3, interpolated across the calorimeter surface, and within the fiducial region defined by the octogonal cut shown in Fig.~\ref{pi0sub}. The parameter $\mu$ corrects  imperfections in the estimation of the energy in the Monte-Carlo simulation compared to the data. The parameter $\sigma$ accounts for different resolutions on the different areas of the calorimeter. The latter can be due to either different levels of background or different quality of the crystals.

The missing-mass-squared cut to ensure exclusivity is chosen as the value where the Monte-Carlo and the data spectra start to differ due to contamination by non-exclusive events. This leads to two different values of missing-mass-squared cut: 0.95~GeV$^2$ for Kin1 and Kin2, 1.09~GeV$^2$ for Kin3. A study of the systematic uncertainty on the exclusivity is presented in section~\ref{sssec:systexclu}.

\subsection{Cross Section}
\label{ssec:fit_proc}

To derive differential cross sections from the measured data, the solid
angle (or acceptance) $\Omega$ of the detection apparatus has to be accurately
known. In the expression of the photon electroproduction cross section, Compton Form Factor (CFF) combinations $\mathcal{F}(Q^2,x_B,t)$ appear multiplied by different kinematical factors $\Gamma(Q^2,x_B,t,\phi)$, which also vary within the bin width. In addition, Bethe-Heitler itself is a rapidly varying cross section, especially as a function of $\phi$, $x_B$ and $t$. Since all the kinematic dependences besides the intrinsic CFF ones are known, we decided to use a method which directly extracts the CFF from data by disentangling all effects in a combined data-Monte-Carlo fit. This method has the additional advantage of automatically handling bin migration effects that may occur. The extraction method is formally described in the following.

Let
\begin{equation}
{\bf x}_v= \left(
\begin{array}{c}
E_b \\
x_B  \\
Q^2  \\
t \\
\phi \\
\phi_e
\end{array}
\right)_v
\end{equation}
represent the kinematic variable vector at the vertex in the simulation. The
incident electron energy $E_b$ is included in order to treat the radiative
tail. Let
\begin{equation}
{\bf x}_e= \left(
\begin{array}{c}
E_b \\
x_B  \\
Q^2  \\
t \\
\phi \\
\phi_e
\end{array}
\right)_e
\end{equation}
represent the reconstructed event variables. In the Monte-Carlo
simulation, we define the mapping
\begin{equation}
K({\bf x}_e|{\bf x}_v)
\end{equation}
as the conditional probability distribution to observe an event at the
kinematic point ${\bf x}_e$ starting from vertex point ${\bf x}_v$. The
experimental acceptances, intrinsic detector efficiencies and resolutions, and real-radiative effects
are included in $K({\bf x}_e|{\bf x}_v)$. This conditional probability,
which we compute using the Monte-Carlo simulation, takes into account the potential
bin migration due to detector resolutions and radiative effects.
The binning vector
\begin{equation}
{\bf i}_e = \left(
\begin{array}{c}
i_{x_B} \\
i_{Q^2}  \\
i_t \\
i_\phi
\end{array}
\right)_e\end{equation}
labels a set of bins in the corresponding event kinematics, after
integration over $\phi_e$ since unpolarized target observables depend on only one azimuthal angle $\phi$. The
binning vector
\begin{equation}
{\bf j}_v= \left(
\begin{array}{c}
j_{x_B} \\
j_{Q^2}  \\
j_t 
\end{array}
\right)_v
\label{eq:jv}
\end{equation}
labels a similar set of bins in the vertex variables. The helicity-dependent and helicity-independent cross sections can
be written as a sum of several harmonic
contributions as described by Eq.~(\ref{eq:BHPhi}--\ref{eq:DVCSPhi}):
\begin{eqnarray}
\sigma({\bf x}_v) &=&\sum_{\Lambda} \Gamma^\Lambda({\bf x}_v)
X^\Lambda_{{\bf j}_v}\label{eq:kinfac}\,,
\end{eqnarray}
where $\Gamma^\Lambda({\bf x}_v)$ represent some
kinematical factors and $X^\Lambda_{{\bf j}_v}$ are some combinations
of CFFs that are unknown and that parametrize the DVCS cross section.
Notice that, as shown by Eq.~(\ref{eq:jv}), the variable $\phi$ is not
binned at the vertex. This is because the full $\phi$--dependence
of the cross section is known and contained in the kinematical factors
$\Gamma^\Lambda({\bf x}_v)$. Thus, the unknowns $X^\Lambda_{{\bf j}_v}$
are independent of $\phi$. The fact that the total number of bins in
the reconstructed event variables is significantly higher than the
number of bins in the vertex variables is precisely what makes the fit
described below possible.

The number of counts per bin at the vertex is
\begin{multline}
N({\bf j}_v)=\mathcal{L}\int_{{\bf
    x}_v\in\text{Bin}({\bf j}_v)} \sum_{\Lambda} \Gamma^\Lambda({\bf
    x}_v) X^\Lambda_{{\bf j}_v} d{\bf
    x}_v=\\\mathcal{L}\sum_{\Lambda} X^\Lambda_{{\bf j}_v} \int_{{\bf
    x}_v\in\text{Bin}({\bf j}_v)} \Gamma^\Lambda({\bf
    x}_v) d{\bf
    x}_v\,,
\end{multline}
where $\mathcal{L}$ is the integrated luminosity. In the experimental bin
${\bf i}_e$, the yield is
\begin{widetext}
\begin{equation}
 N({\bf i}_e)=\int\displaylimits_{{\bf
     x}_e\in\text{Bin}({\bf i}_e)} \!\!\!\! d{\bf
     x}_e \sum_{{\bf j}_v} N({\bf i}_v) K({\bf x}_e|{\bf x}_v)
  = \mathcal{L} \sum_{{\bf j}_v} \sum_{\Lambda}  X^\Lambda_{{\bf j}_v}\int\displaylimits_{{\bf
      x}_e\in\text{Bin}({\bf i}_e)} \!\!\!\! d{\bf x}_e \int\displaylimits_{{\bf
      x}_v\in\text{Bin}({\bf j}_v)} \!\!\!\! d{\bf x}_v \Gamma^\Lambda({\bf
      x}_v)K({\bf x}_e|{\bf x}_v)
\end{equation}
\end{widetext}

We define a bin mapping function:
\begin{equation}
K_{{\bf i}_e,{\bf j}_v}^\Lambda=\!\!\!\!\!\!\!\!\!\!\int\displaylimits_{{\bf
    x}_e\in\text{Bin}({\bf i}_e)}\int\displaylimits_{{\bf x}_v\in\text{Bin}({\bf j}_v)}
    \!\!\!\!\!\! d{\bf x}_e\,d{\bf x}_v \ K({\bf x}_e|{\bf x}_v)\Gamma^\Lambda ({\bf x}_v)\,.
\label{eq:binmap}
\end{equation}
This function is basically the solid angle weighted by the kinematic
factors $\Gamma^\Lambda ({\bf x}_v)$, where the effects of bin migration are taken into
account through the function $K({\bf x}_e|{\bf x}_v)$.
Generally, the number of
counts per bin can thus be written as:
\begin{equation}
N^\text{MC} ({\bf i}_e)=\mathcal{L}\sum_{{\bf
    j}_v,\Lambda} K^\Lambda_{{\bf i}_e,{\bf j}_v} X^\Lambda_{{\bf j}_v}\,.
\label{eq:ymc}
\end{equation}
Note the summation over all ${\bf j}_v$. All bins at the vertex might
contribute to a given experimental bin ${\bf i}_e$, with a certain
probability or weight given by the function $K^\Lambda_{{\bf i}_e,{\bf
    j}_v}$, computed in the simulation.
We construct a $\chi^2$ which we minimize to extract the
${\bf X}_{{\bf j}_v}$:
\begin{equation}
\chi^2=\sum_{{\bf i}_e}\frac{[N^\text{Exp}({\bf i}_e)-N^\text{MC}({\bf
      i}_e)]^2}{[\delta^\text{Exp}({\bf i}_e)]^2}\,,
\label{eq:chi2charles}
\end{equation}
where $\delta^\text{Exp}({\bf i}_e)$ are the experimental statistical uncertainties in
each bin.

The coefficients ${\overline{{\bf X}}}_{{\bf j}_v}$ are defined as the
values of ${\bf X}_{{\bf j}_v}$ that minimize $\chi^2$:
\begin{eqnarray}
0 & = & -\left.\frac{1}{2}\frac{\partial\chi^2}{\partial X_{{\bf
      j}_v}^\Lambda} \right|_{\overline{\bf X}_{{\bf j}_v}}\nonumber\\ 
       \phantom{toto}\nonumber\\ 
0 & = & \sum_{{\bf
      j}'_v,\,\Lambda '} \alpha_{{\bf j}_v,\,{\bf j}'_v}^{\Lambda
      ,\,\Lambda '}\  {\overline{X}}^{\Lambda '}_{{\bf
      j}'_v} -\beta_{{\bf j}_v}^\Lambda\quad \forall \ {\bf j}_v,\,\Lambda\,.
\label{eq:linsys}
\end{eqnarray}
The linear system is defined by:
\begin{eqnarray}
\alpha_{{\bf j}_v,\,{\bf j}'_v}^{\Lambda
      ,\,\Lambda '} & = & \sum_{{\bf i}_e} \mathcal{L}^2\frac{K_{{\bf
      i}_e,\,{\bf j}_v}^\Lambda\,  K_{{\bf
      i}_e,\,{\bf j}'_v}^{\Lambda '}}
{[\delta^\text{Exp}({\bf i}_e)]^2}\,,\\
\beta_{{\bf j}_v}^\Lambda & = & \sum_{{\bf i}_e} 
 \mathcal{L}\frac{ N^\text{Exp}({\bf i}_e) \, K_{{\bf
      i}_e,\,{\bf j}_v}^\Lambda}
{[\delta^\text{Exp}({\bf i}_e)]^2}\,.
\end{eqnarray}
The fit parameters are:
\begin{equation}
{\overline{ X}}^{\Lambda }_{{\bf j}_v}= \sum_{{\bf
      j}'_v,\,\Lambda '} [\alpha^{-1}]^{\Lambda ,\,\Lambda '}_{{\bf j}_v,\,{\bf
      j}'_v}\,\beta_{{\bf j}'_v}^{\Lambda '}\,.
\end{equation}
The covariance matrix of the fitted parameters is:
\begin{equation}
V_{{\bf j}_v,\,{\bf
      j}'_v}^{\Lambda ,\,\Lambda '} = [\alpha^{-1}]^{\Lambda ,\,\Lambda '}_{{\bf j}_v,\,{\bf
      j}'_v}\,.
\end{equation}

Finally, the cross-section values (and associated error bars) at the
point $ \overline{x}_{{\bf i}_e}$ are obtained as:
\begin{equation}
\frac{d^4\sigma(\overline{x}_{{\bf i}_e})}{dx_BdQ^2dtd\phi}=
\frac{d^4\sigma^\text{Fit}_{{\bf j}_v}(\overline{x}_{{\bf
i}_e})}{dx_BdQ^2dtd\phi}\cdot\frac{N^\text{Exp}({\bf
i}_e,{\bf j}_v)}{N^\text{MC}({\bf i}_e,{\bf j}_v)}\,,
\end{equation}
where
\begin{equation}
\frac{d^4\sigma^\text{Fit}_{{\bf j}_v}(\overline{x}_{{\bf
i}_e})}{dx_BdQ^2dtd\phi}
=\sum_\Lambda {\Gamma}^{\Lambda} (\overline{x}_{{\bf i}_e})
\overline{X}^{\Lambda}_{{\bf j}_v}\,,
\end{equation}
is defined by the fit parameters of the bin ${\bf j}_v$ which has the
same bin limits in $x_B$, $Q^2$ and $t$ as the experimental bin ${\bf
i}_e$. The number of counts $N^\text{MC}({\bf i}_e,{\bf j}_v)$ and
$N^\text{Exp}({\bf i}_e,{\bf j}_v)$ corrected from bin migration are
given by:
\begin{eqnarray}
& & N^\text{MC} ({\bf i}_e,{\bf j}_v)=\mathcal{L}\sum_{\Lambda}
K^\Lambda_{{\bf i}_e,{\bf j}_v} X^\Lambda_{{\bf j}_v} \, , \\
& & N^\text{Exp} ({\bf i}_e,{\bf j}_v)= N^\text{Exp} ({\bf i}_e) \nonumber \\
&& \phantom{N^\text{Exp} ({\bf i}_e,{\bf j}_v)=} -\mathcal{L}\sum_{\Lambda}\sum_{{\bf j}'_{v}\neq{\bf j}_{v}}
K^\Lambda_{{\bf i}_e,{\bf j}'_{v}} X^\Lambda_{{\bf j}'_{v}}\,.
\label{eq:yexp}
\end{eqnarray}

Note that we study the harmonic coefficients of the cross section which are the sum of $(c,s)_n^{\mathcal I}$ and $(c,s)_n^{DVCS}$, themselves involving several combinations of twist-2 and twist-3 CFFs. The $\phi$-dependence is therefore not enough to separate all linear/bilinear combinations of CFFs. The $\phi$-dependence of the cross section can thus be properly described by different choices of free parameters. 

In this analysis, we chose to parametrize the DVCS helicity-independent cross section by the three following combinations of effective CFFs: $\mathcal C^{DVCS}(\mathcal F,\mathcal F^*)$ (Eq.~\ref{eq:cdvcs}), $\Real[\mathcal C^{\mathcal I}(\mathcal F)]$ (Eq.~\ref{eq:ci}) and $\Real[\mathcal C^{\mathcal I}(\mathcal F_{eff})]$ (Eq.~\ref{eq:ceff}). The helicity-dependent cross section is fitted using the $\Imag[\mathcal C^{\mathcal I}(\mathcal F)]$ and $\Imag[\mathcal C^{\mathcal I}(\mathcal F_{eff})]$. Three reasons have led to this choice:
\begin{itemize}
\item The contributions to the cross section associated to each of these parameters have a distinct $\phi$-dependence, minimizing the correlations among them,
\item We keep the dominant twist DVCS$^2$ contribution,
\item Higher twist contributions are kinematically suppressed.
\end{itemize}
While this is the most physical choice of parameters, any other choice that provides a good fit ($\chi^2/dof\sim 1$) to the $\phi$-dependence of the number of counts is an equally valid choice as far as the cross-section extraction is concerned. The fitted parameters, though, would have a less straightforward physics interpretation in that case. We have tested the stability of our cross-section results against a different choice of free parameters and results are discussed in section~\ref{sssec:systexclu}.

\subsection{Global Normalization}
\label{sssec:beamana}

In the previous section, we defined $\mathcal L$ as the integrated luminosity. It is computed from the average total charge $Q$ recorded by the BCMs as:
\begin{equation}
\int\ \frac{d \mathcal{L}}{dt}\,dt=\frac{Q}{e}\frac{{N_A\rho}l}{A_H}\,,
\end{equation}
where $e=1.602\cdot 10^{-19}$\,C is the electron charge, $A_H=1.0079$\,g/mol is
the atomic mass of H, and $N_A=6.022\cdot 10^{23}$\,mol$^{-1}$ is Avogadro's
number. The LH$_2$ target length was $l=15$~cm and was operated at 19\,K and a pressure of 25\,psi, which gives a density of $\rho=0.07229$\,g/cm$^3$. Table~\ref{tab:lumi} shows the integrated luminosity (corrected by the acquisition deadtime) recorded for each of the kinematic settings.

\begin{table}[htp]
\centering
\begin{tabular}{|c|c|c|c|c||c|}
\hline
Kin & $Q_{+}$(C) & $Q_{-}$(C) & $Q$(C) & $Q_{asy}$ ($10^{-3}$) & $\mathcal L$ (fb$^{-1}$)\\
\hline\hline
1 & 0.3732 & 0.3733 & 0.7464 &  -0.1 & 3059 \\
\hline
2 & 0.4057 & 0.4064 & 0.8121 &  -0.7 & 3328 \\
\hline
3 & 0.6913 & 0.6937 & 1.385 &  -2.4 & 5676 \\
\hline
\end{tabular}
\caption{Helicity-correlated charge ($Q_\pm)$, total charge ($Q=Q_+ + Q_-$) and charge asymmetry 
$Q_{asy} = (Q_+ - Q_-)/Q$ for the three kinematics settings. The last column shows the integrated luminosity including events for which the helicity bit is undefined. In all cases, the charge and the luminosity have been corrected for the deadtime.}
\label{tab:lumi}
\end{table}

\section{Radiative Corrections}
\label{ssec:rc}

The diagrams of Fig.~\ref{fig:epepg} only include the lowest order QED amplitude for the DVCS process.
The experimental cross section necessarily includes higher-order QED processes.
Radiative corrections to the $ep\rightarrow ep\gamma$ reaction have been studied in several papers
\cite{Vanderhaeghen:2000ws, Bytev:2003qf, Afanasev:2005pb,Akushevich:2012tw}.  In this analysis
we follow the approach of \cite{Vanderhaeghen:2000ws}.

In this analysis we fit a model cross section to the experimental yield, bin by bin. We separate the radiative corrections into
terms that are dependent on the missing-mass-squared $M_X^2$ cut we impose on the H$(e,e'\gamma)X$ spectra,
and terms that are independent of this cut.  
  The external radiative effects on the incident electron, and internal real radiative effects at the vertex
  are treated in the equivalent radiator approximation
  \cite{Mo:1968cg,Tsai:1973py}.
  Pre-scattering radiation is modeled by generating an event-by-event energy loss $\Delta E_{\rm in}$ of the incident electron ($E_b$) following a distribution ($b \simeq 4/3$):
  \begin{equation}
    I_{\rm in}(E_b, \Delta E_{\rm in}, t_{\rm in}) = \frac{bt_{\rm in} + \delta_S/2}{\Delta E_{\rm in}} \left[ \frac{\Delta E_{\rm in}}{E_b} \right]^{bt_{\rm in} + \delta_S/2}
  \end{equation}
  with
  \begin{equation}
    \delta_S = \frac{2\alpha}{\pi} \left[ \ln\frac{Q^2}{m_e^2} - 1 \right] \, ,
  \end{equation}
  where $t_{\rm in}$ is the event-by-event target thickness (in radiation lengths) traversed by the electron
  before the scattering vertex. 
  The Schwinger term $\delta_S$ models the internal pre-scattering radiation.
  The scattered energy at the vertex is $E_v^{\prime} = E_b-\Delta E_{\rm in}-Q^2/(2 M_p x_B)$.
  Internal post-scattering radiation is modeled by a similar distribution in the post-scattering radiated energy $\Delta E_{\rm out}$:
  \begin{equation}
    I_{\rm out} = \frac{ \delta_S/2}{\Delta E_{\rm out}} \left[ \frac{\Delta E_{\rm out}}{E_v^{\prime}} \right]^{\delta_S/2} \, .
  \end{equation}
  These radiative effects are treated within the peaking approximation.
  External post-scattering radiation by the scattered electron is evaluated with the Monte-Carlo simulation by transporting the electron to the entrance of the spectrometer.
  Kinematic shifts ({\em e.g.} in either the norm or the direction of $\vec{q}$) from external and internal radiations are fully included in the simulation and thereby unfolded from the extracted cross sections.
  
  In addition to these radiative effects incorporated into our Monte-Carlo, we correct the data for 
  internal virtual radiation  as well as the cut-off independent
  effect of unresolvable soft real radiation, given by Eqs.~58-62 of \cite{Vanderhaeghen:2000ws}.
  The virtual corrections to the VCS amplitude are model independent, in the sense that they do not depend on the dynamics
  of the $\gamma^\ast p \to \gamma p$ process.  These corrections (vacuum polarization and vertex renormalization) are essentially equivalent to the corrections to elastic
  $ep$ scattering, with suitable adjustment to the kinematics.  On the other hand, the vacuum polarization and vertex 
  corrections to the BH amplitude differ
  by several percent relative to the VCS corrections, and the BH amplitude also has self-energy corrections to the virtual 
  electron propagators.  We calculate separately the  radiative corrections to the helicity-independent and helicity-dependent cross sections based on a code
  derived from   \cite{Vanderhaeghen:2000ws} which includes the leading-twist DVCS amplitude with a fully factorized 
  GPD ansatz \cite{Guichon2007,Guichon:1998xv}.
  The correction factors vary by less than $0.5\%$ over $\phi$ and by $\approx 1\%$ over the $[x_B,Q^2]$
  acceptance of each kinematic setting.  
  We assign a $2\%$
  systematic error to the combined real- and virtual-radiative corrections.  This is based on the variation of the correction 
  over the acceptance, ambiguities over whether or not to exponentiate the correction, and the model-dependence of
  the relative contributions of the $\left|\mathcal T^{BH} \right|^2$, interference and $\left| \mathcal T^{DVCS}\right|^2$ terms in the unpolarized cross sections.  Over our five kinematic settings, the average corrections varied by less than $0.5\%$.
  Since this is less than the uncertainty of the correction, we apply the following global corrections to all cross-section bins:
  \begin{align}
d^4\sigma^\text{Born} &= (0.948\pm 0.02)\  d^4\sigma^\text{Exp},
\nonumber \\
 \Delta^4\sigma^\text{Born} &= (0.973\pm 0.02)\  \Delta^4\sigma^\text{Exp}.
\end{align}

\section{Systematic Uncertainties}
\label{ssec:syst}

Systematic uncertainties are divided into uncorrelated (or point-to-point) and correlated (or normalization) uncertainties. The largest source of uncorrelated error in this experiment was associated to the missing-mass-squared cut. The correlated uncertainties have been described before and a summary table is shown in this section.

\subsection{Missing-mass-squared Cut}
\label{sssec:systexclu}

\begin{figure}[htp]
\epsfig{file=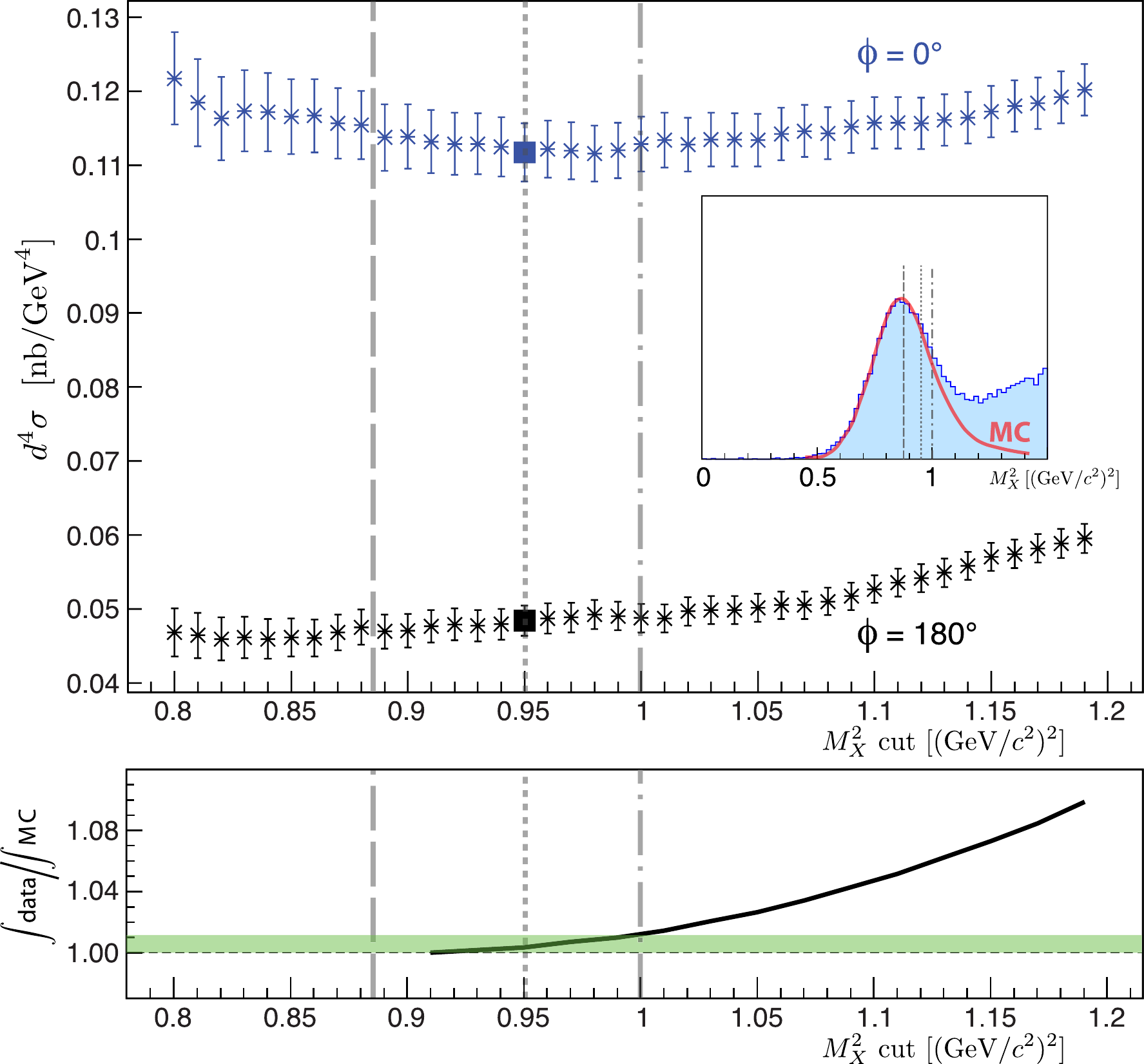,width=\linewidth}
\caption{\label{mm_kin_cut} (Color online) Top: Variation of the $ep\rightarrow ep\gamma$ cross section for Kin2, $-t=0.17$~GeV$^2$, as a function of the missing-mass-squared cut, for $\phi=0^\circ$ (upper blue points) and $\phi=180^\circ$ (lower black points). The dotted vertical line corresponds to the nominal cut. The systematic errors are evaluated bin by bin in $\phi$ and $t$ for each kinematic setting by studying the variation of the cross section between the nominal and the lower missing-mass-squared cut (dashed line). The insert represents the same cuts on the missing-mass plot. Bottom: Ratio of the integrals of the experimental and  Monte-Carlo missing-mass spectra, as a function of the missing-mass-squared cut. By varying the cut up to 1~GeV$^2$/c$^4$ , represented by the dotted-dash line, the observed contamination remains smaller than 1\% (green band).}
\end{figure}

Two systematic effects are associated with the missing-mass-squared cut. The first comes from semi-inclusive events contaminating our sample. These events have larger missing-mass-squared values induced by extra missing particles. Indeed, even if the cut is supposed to keep this contamination minimal, a small fraction of such events may remain below the missing-mass-squared cut. In order to evaluate an upper value for this systematic error, we examined the ratio of the integrals of the experimental and  Monte-Carlo missing-mass-squared spectra. As seen in Fig.~\ref{mm_kin_cut}, this ratio increases significantly with the missing-mass-squared cut, which is expected since the Monte-Carlo only contains exclusive events. By varying the cut from the nominal value 0.95~GeV$^{2}$/c$^4$ up to 1~GeV$^2$/c$^4$, the observed contamination remains smaller than 1\%, which we took as the systematic uncertainty on the cross section.

The second effect induced by the missing-mass-squared cut arises from a mismatch on the position and shape of the missing-mass-squared peaks between data and Monte-Carlo. This is due to our limited ability to reproduce perfectly the response of our calorimeter. This mismatch increases as the missing-mass-squared cut decreases and is maximal around the maximum of the distribution. We estimate the corresponding error by looking at the variation of the cross section between the nominal cut and a lower cut value. This lower bound is chosen such that the loss of statistics is 15\%, ensuring that the observed variations are not statistical in nature. The systematic error is evaluated for each $(t,\phi)$ bins of each kinematic setting and may reach up to a few percent. These point -to-point uncertainties are included in the data tables~\ref{tab:Kin2}--\ref{tab:diffKinX3}.

\subsection{Cross-section Parametrization}
\label{sssec:systpar}

\begin{figure}[htp]
\epsfig{file=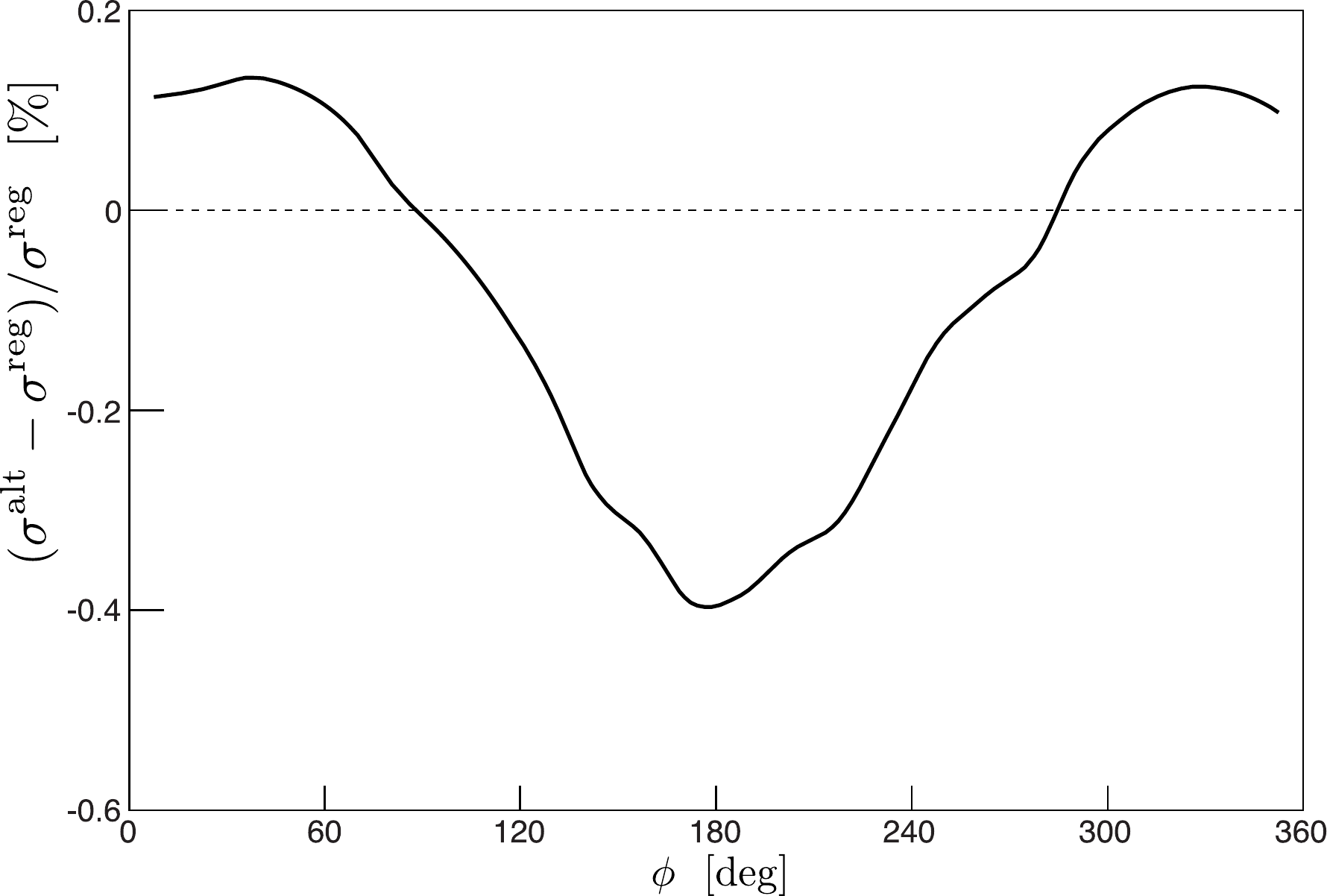,width=\linewidth}
\caption{\label{err_fit} Difference in \% between the cross section extracted with the squared DVCS amplitude term and with the $\Real\mbox{ } \mathcal{C}^{\mathcal{I},V}$ term for $x_B=0.37$, $Q^2=2.36$~GeV$^2$ and $-t=0.33$~GeV$^2$. The $\phi$-profile of the difference is a consequence of the small $\cos\phi$ and $\cos 2\phi$ dependences of the $\Real\mbox{ } \mathcal{C}^{\mathcal{I},V}$ kinematic coefficient. Both extractions give almost the same reduced $\chi^2/dof$=0.94 (nominal) and 0.93 (alternate) for the entire Kin2 setting.}
\end{figure}

As mentioned in section~\ref{ssec:fit_proc}, the cross-section results should be independent of the choice of parametrization in the extraction method. To evaluate the impact of this choice, we used a different parameter set by replacing the squared DVCS amplitude term by the interference term $\Real\mbox{ } \mathcal{C}^{\mathcal{I},V}$, which yields an equally good fit to the data. A difference in the cross-section value of up to 1\% appears locally depending on the kinematic bin, as shown in Fig~\ref{err_fit}. As a consequence we estimated the systematic error from the parameter choice to be 1\%. 

\subsection{Correlated Uncertainties}
\label{sssec:norm}
Table~\ref{tab:sys} presents the systematic uncertainties on the cross section stemming from normalization effects, which are considered 100\% correlated bin-by-bin. Note that the helicity-dependent cross sections have an additional uncertainty coming from the beam polarization measurement. The determination of these uncertainties are discussed in the associated section listed in the table.
\begin{table}
\centering
\begin{tabular}{lrr}
{\bf Systematic uncertainty} & {\bf Value} & {\bf Section}\\
\hline
\hline
HRS acceptance cut & 1\% & \ref{sssec:hrsana} \\
Electron ID & 0.5\% &  \ref{sssec:eff_cor} \\
HRS multitrack & 0.5\% & \ref{sssec:eff_cor}\\
Multi-cluster & 0.4\% & \ref{sssec:eff_cor}\\
Corrected luminosity & 1\% & \ref{sssec:eff_cor}\\
Fit parameters & 1\% & \ref{sssec:systpar}\\
Radiative corrections & 2\% & \ref{ssec:rc} \\
Beam polarization  & 2\% & \ref{sssec:beampol}\\ 
\hline
\hline
Total (helicity-independent) & 2.8\%\\
\hline
Total (helicity-dependent) & 3.4\%\\
\hline
\end{tabular}
\caption{Normalization systematic uncertainties in the extracted photon electroproduction cross sections. The systematic error coming from the fit parameter choice is not a normalization error per se, but we consider that 1\% is an upper limit for this error on all kinematic bins. The helicity-dependent cross sections have an extra uncertainty stemming from the beam polarization measurement. The last column gives the section in which each systematic effect is discussed.}
\label{tab:sys}
\end{table}

\section{Results}
\label{sec:results}

The cross-section extraction procedure described in section~\ref{ssec:fit_proc} was applied to all data sets, for both the unpolarized and the helicity-dependent cases. In addition to the $Q^2$-dependence of the helicity-dependent cross sections, we were able to measure the $Q^2$-dependence of the unpolarized cross section at two values of $Q^2$=1.9 and 2.3~GeV$^2$. The $x_B$-dependence of helicity-dependent and -independent cross sections were studied using the KinX2 and KinX3 settings. Note that an extra bin in $t$ was analyzed compared to our previous publication~\cite{MunozCamacho:2006hx} for all $(x_B,Q^2)$ settings.

\begin{figure}
\centering\includegraphics[width=\linewidth]{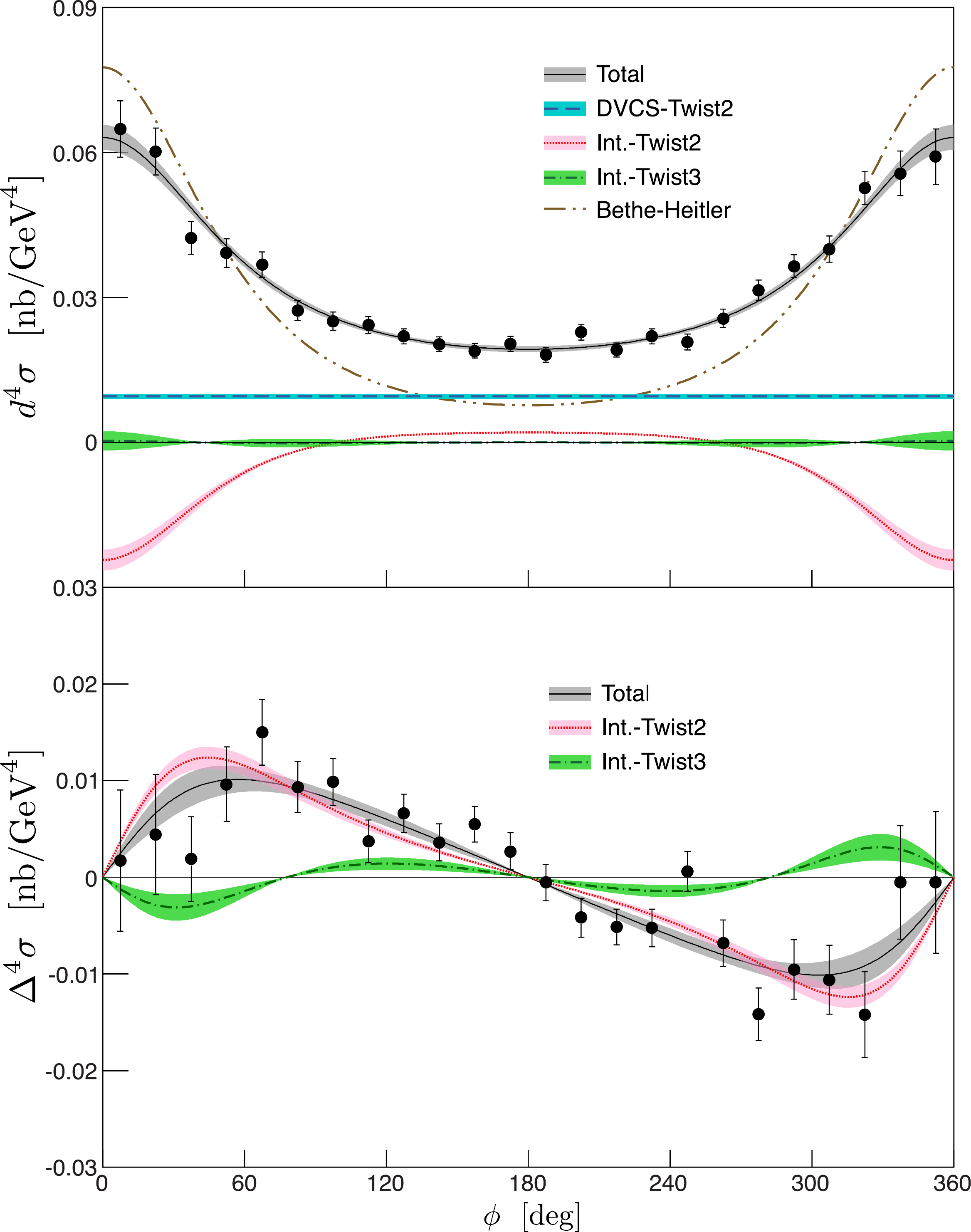}
\caption{(Color online) Unpolarized (top) and helicity-dependent (bottom) cross-section extraction for the Kin3 bin $-t=0.32$~GeV$^2$. The error bars on the data points are statistical only. The shaded areas represent the statistical uncertainty for each contribution.}
\label{fig:fitresult}
\end{figure}

An example of the cross-section extraction is presented in Fig.~\ref{fig:fitresult} for $x_B=0.37$, $Q^2=2.36$~GeV$^2$ and $-t=0.32$~GeV$^2$, along with the different contributions resulting from the fit, which gave an overall $\chi^2/dof$ of 1.1. For the unpolarized cross section, one observes a significant contribution from the term associated with $\left| \mathcal T^{DVCS}\right|^2$, in addition to a large contribution from the interference term. Note that the $\left| \mathcal T^{DVCS}\right|^2$ contribution is $\phi$-independent in contrast to the BH and interference contributions. Indeed, it does not contain the $\mathcal P_1(\phi)\mathcal P_2(\phi)$ electron propagators as shown in Eq.~\ref{eq:DVCSPhi}. The precision of the data is such that other contributions than the Bethe-Heitler are obviously necessary to explain the observed cross section. The helicity-dependent cross section is dominated by the twist-2 interference term, as noticed before in this experiment 
\cite{MunozCamacho:2006hx}  and elsewhere \cite{Airapetian:2012pg,Girod:2007aa}. 
These conclusions extend to all bins in our analysis, whose results are shown in section~\ref{ssec:figures}. Tab.~\ref{tab::chi2} lists the  $\chi^2/dof$ resulting from the extraction method for all kinematics settings.
\begin{table}
\begin{tabular}{lcc}
\hline\hline
Settings & ~$\chi^2_{pol}/dof$~ & ~$\chi^2_{unp}/dof$~\\
\hline\hline
Kin1 & 0.88 & - \\
Kin2 & 1.00 & 1.16 \\
KinX2 & 0.96 & 0.82 \\
Kin3 & 1.15 & 0.99 \\
KinX3 & 1.08 & 1.28 \\
\hline
\end{tabular}
\caption{\label{tab::chi2} $\chi^2/dof$ resulting from the extraction method for all kinematics settings. The subscript "pol" stands for polarized cross sections, "unp" for unpolarized cross sections.}
\end{table}

\subsection{Scan in $Q^2$}
\label{ssec:extract}

\begin{figure*}[!htp]
\centering
\epsfig{file=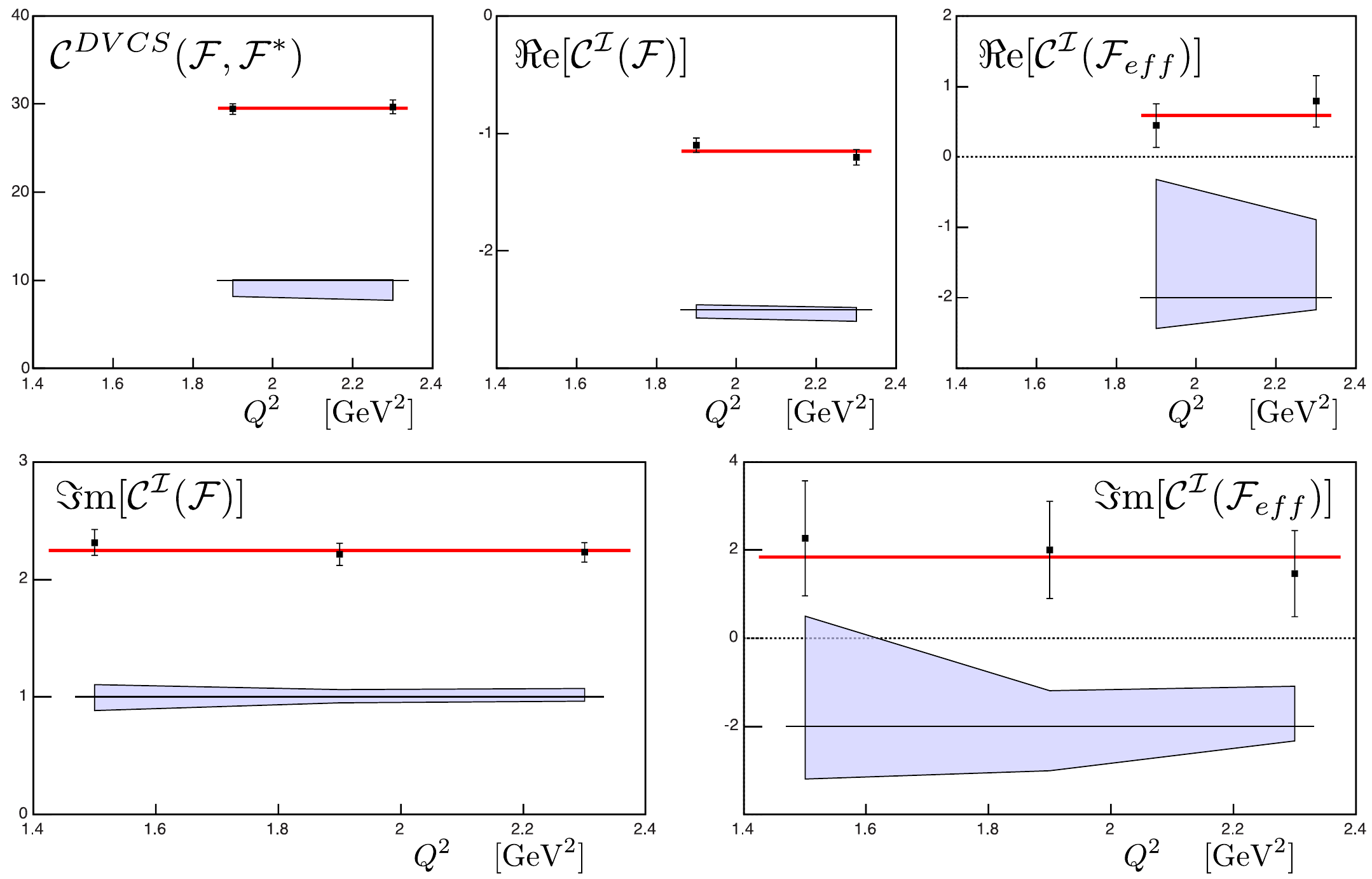,width=0.95\linewidth}
\caption{\label{fig:scaling} Combinations of effective CFFs extracted from the fitting procedure described in section~\ref{ssec:fit_proc} using the formalism developed in \cite{Belitsky:2010jw}, integrated over $t$ and plotted as a function of $Q^2$. The top three plots show the effective CFFs resulting from the unpolarized cross section fit (Kin2 and Kin3), whereas the bottom plots show the effective CFFs resulting from the helicity-dependent cross section fit (Kin1--3).
The shaded areas represent systematic errors.}
\end{figure*}

The combinations of effective CFFs which have been extracted from the fitting procedure for Kin1--3 using the formalism developed in \cite{Belitsky:2010jw} are shown integrated over $t$ in Fig.~\ref{fig:scaling}. With the choice of parameters used to describe the kinematical dependence of the cross sections (as explained in section~\ref{ssec:fit_proc}), the contribution associated with the $\left| \mathcal T^{DVCS}\right|^2$ term is large for the unpolarized case. The twist-2 interference term is significant and the contribution of the twist-3 interference term is often found to be small, with large systematic errors. For the polarized case, the twist-2 interference term is dominant, the twist-3 contribution is small, again with large systematic errors.

Overall, the extracted parameters show no $Q^2$--dependence for either the helicity-dependent or the helicity-independent cases over our $Q^2$--range. Note that the logarithmic $Q^2$--evolution can safely be neglected within this $Q^2$ lever arm at this $x_B$.

The full set of results for settings Kin1--3 are presented in Fig.~\ref{kin2}--\ref{diff_kin3} in section~\ref{ssec:figures}.

\subsection{Scan in $x_B$}
\label{ssec:xB_result}
The results from KinX2 and KinX3 showing the $x_B$-dependence of the cross sections are presented in Fig.~\ref{kinX2}--\ref{kinX3diff} in section~\ref{ssec:figures}. KinX3 has a limited acceptance close to 0$^\circ$, which 
increases the correlation between the different fit parameters describing the azimuthal dependence of the cross section. Indeed, the separation of the real part of the twist-2 interference and $\left| \mathcal T^{DVCS}\right|^2$  contributions in the fit is particularly sensitive to the relative value of the cross section measured around both $\phi=0$ and 180$^\circ$. These difficulties have basically no impact on the determination of the cross sections themselves. The measured $x_B$-dependence  will set interesting constraints on GPD models and parametrizations, especially thanks to the relatively high accuracy of our data.
 
\subsection{Comparison with GPD Models}
\label{ssec:comparGPD}
 
\begin{figure}
\centering\includegraphics[width=\linewidth]{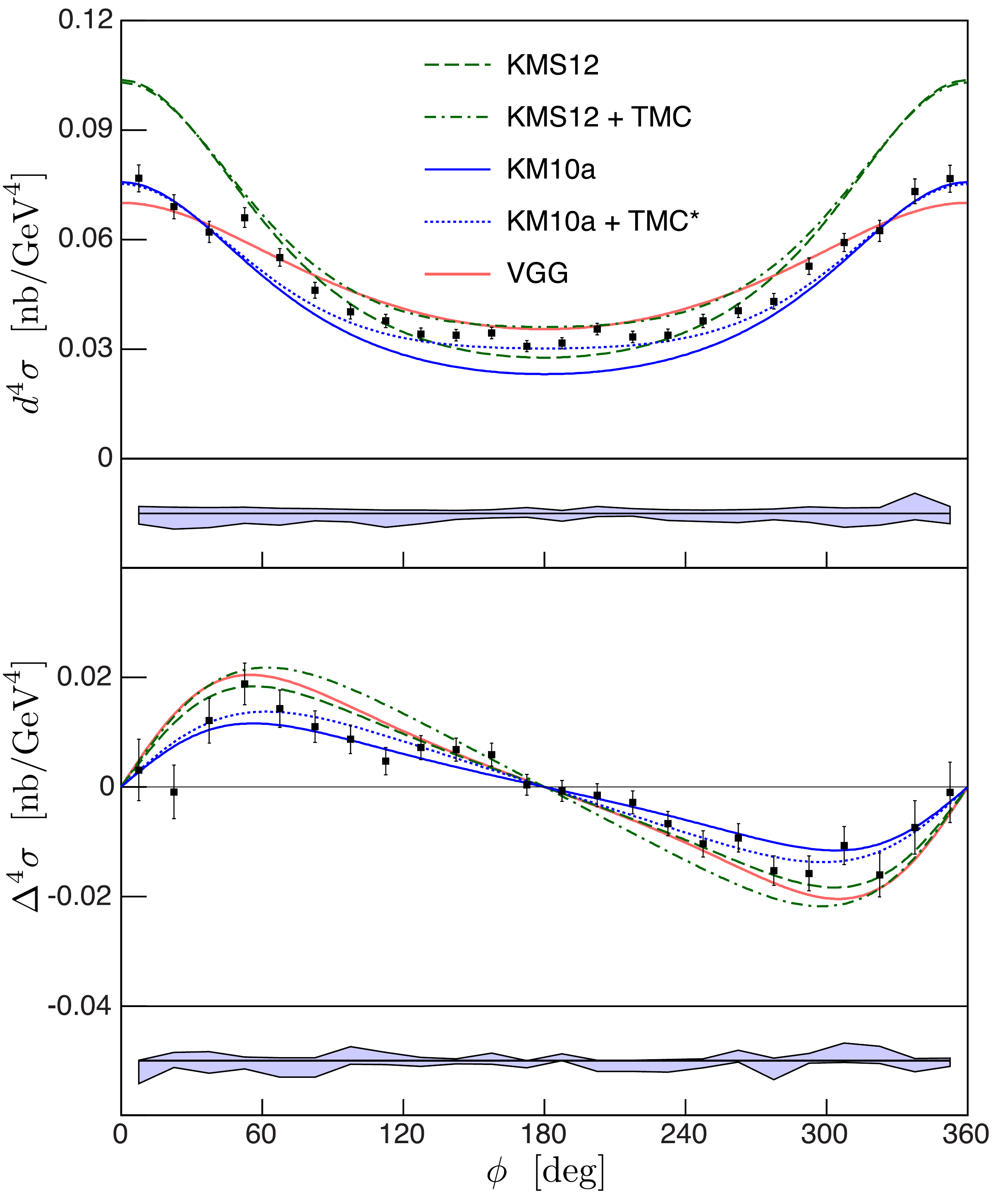}
\caption{(Color online) Unpolarized (top) and helicity-dependent (bottom) cross sections for the  Kin2 bin $-t=0.23$~GeV$^2$. The light blue area represents the point-to-point systematic uncertainties added linearly to the normalization error. The predictions from the distribution-based models KMS12 and VGG are shown  as the dashed green and solid red curves, respectively. The KM10a fit is represented as the solid blue line. The target-mass and finite-$t$ corrections are included in the KMS12 model and shown as the dotted-dash curve. The correction is then applied to the KM10a model shown as the dotted blue line.}
\label{fig:compar_model}
\end{figure}

In Fig.~\ref{fig:compar_model}, we compare our results with various models and previous fits to data. We have chosen to use two different kinds of double-distribution GPD models, namely the VGG~\cite{Goeke:2001tz}  and KMS12~\cite{Kroll:2012sm} models. Note that in contrast to VGG, the KMS12 model was tuned using vector meson data at low to very-low $x_B$, and is not considered adapted yet to the valence quark region. In any case, one observes that both models overshoot the helicity-dependent cross section data in this  Kin2 bin, whereas VGG is more adequate for the unpolarized data.

In addition, we have compared our data with the KM10a model \cite{Kumericki:2009uq}, which fits some of its parameters to all DVCS data available worldwide except for the previously published results from a subset of the present experiment. The consequence is that no absolute DVCS cross-section data in the valence region were used for this fit. The KM10a model is clearly very close to the helicity-dependent data, which is not a surprise considering that the CLAS asymmetry data in the same kinematic region were used to constrain this model. However, this same model significantly underestimates the DVCS unpolarized cross section around $\phi=180^\circ$.

Recently, kinematic twist-4 target-mass and finite-$t$ corrections (TMC) have been calculated for DVCS on the proton and estimated for the KMS12 model \cite{Braun:2012hq,Braun:2014sta} (shown in Fig.~\ref{fig:compar_model}). Since this model is not adapted to the valence quark region, we have extracted the correction factor and applied it to the KM10a parametrization~\footnote{In principle, the full calculation of TMC can only be evaluated knowing the GPDs in the entire region $x>\xi$. KM10a however uses a dispersion relation fit for the valence region by parametrizing the GPD $H$ on the cross-over $x=\xi$ line and a subtraction constant. Moreover, even if the main part of the TMC for unpolarized observables could in principle be evaluated by a change of conventions to CFFs and the $\xi$ variable \cite{Braun:2014paa}, the KM10a parametrization is currently only available as a binary package giving directly the photon electroproduction cross section.}. This allows us to gauge the effect of such corrections in the most realistic model available to us. It is striking that the lack of strength observed at $\phi=180^\circ$ for the KM10a model is largely compensated by the TMC, giving a surprisingly good agreement between this modified KM10a model and our data. 
 
 An update of the KMS12 model, taking into account the DVCS data in the valence region, would allow for a much stronger statement about the necessity of target-mass and finite-$t$ corrections at these moderate $Q^2$.  At any rate, we emphasize that the high accuracy of the present data is  crucial to disentangle the different contributions at play in this critical area around 180$^\circ$. There is no doubt that the addition of our new data set to the KM fit will be most interesting, especially in the light of these new higher-twist calculations.

All the features we have described remain true for most of our data bins, which are shown in section~\ref{ssec:figures}. It is interesting to note that for the highest bins in $t$, especially for Kin2 and KinX2 (Figure~\ref{kin2} and \ref{kinX2}), the TMC to the unpolarized cross section is of the same order as the cross section itself around $\phi =180^\circ$. This corresponds to values of $(-t/Q^2 ) \sim 0.15$ or larger. It is not unreasonable to expect that higher-order corrections in $(-t/Q^2)^2$ start to be important at these values, and may compensate the peculiar behavior of the TMC around $\phi =180^\circ$, which is not visible in data. Efforts to achieve a resummation of the $(-t/Q^2)^k$ series to all orders are currently undertaken \cite{Braun:private}.

\subsection{Results for All Kinematics}
\label{ssec:figures}

\begin{figure*}[!htp]
\centering
\epsfig{file=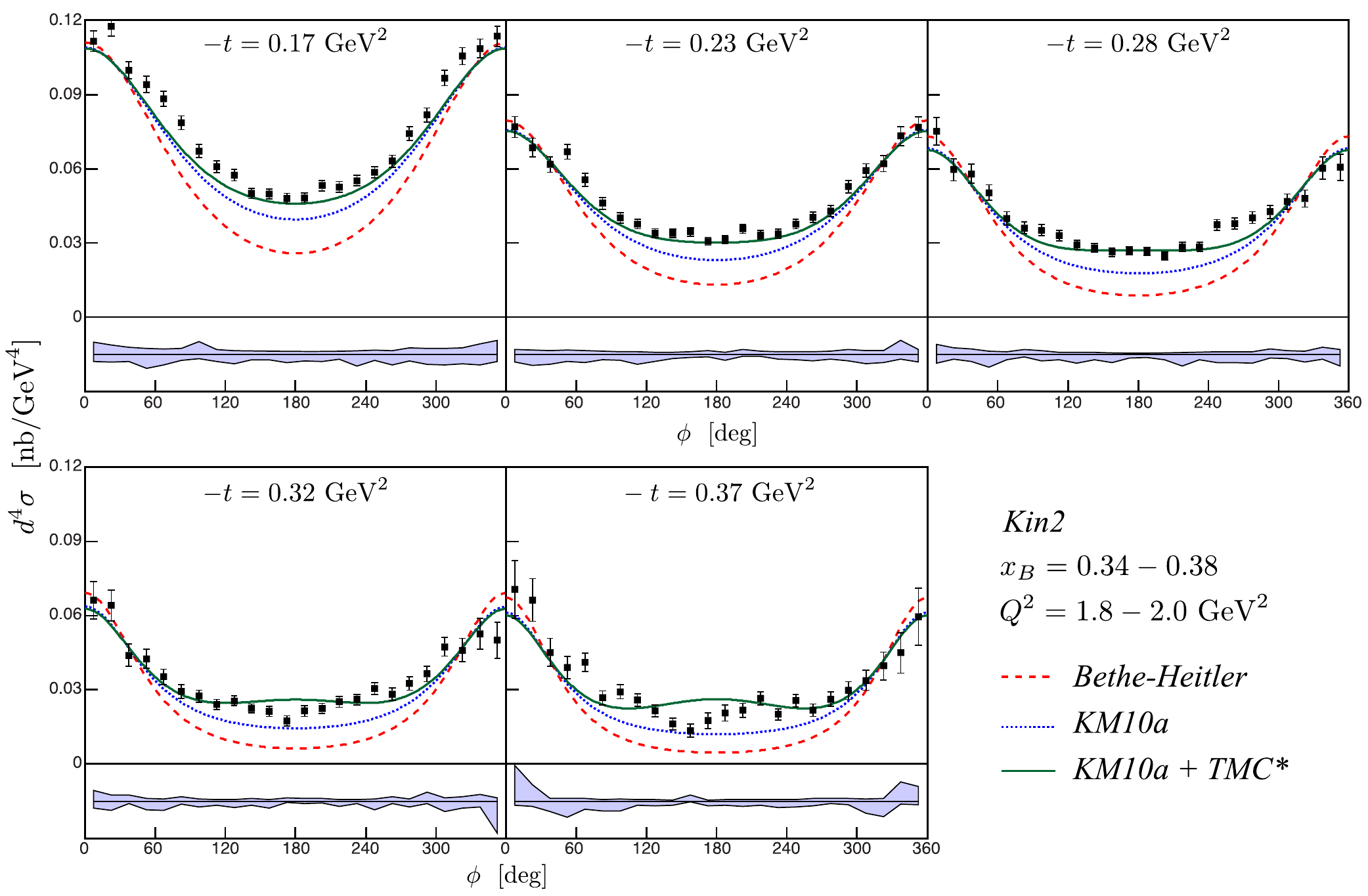,width=\linewidth,clip=}
\caption{\label{kin2}(Color online) Unpolarized cross sections for Kin2. Each $t$-bin corresponds to slightly different average $(x_B,Q^2)$ values; their range is indicated in the legend, their specific values are listed in the data tables. Error bars are statistical only. The light blue area represents the point-to-point systematic uncertainties added linearly to the normalization error. The KM10a model along with its modified version (including the TMC effects) are shown as dotted blue and solid green curves, respectively. The Bethe-Heitler contribution is represented as a dashed red line.}
\end{figure*}

\begin{figure*}[!htp]
\centering
\epsfig{file=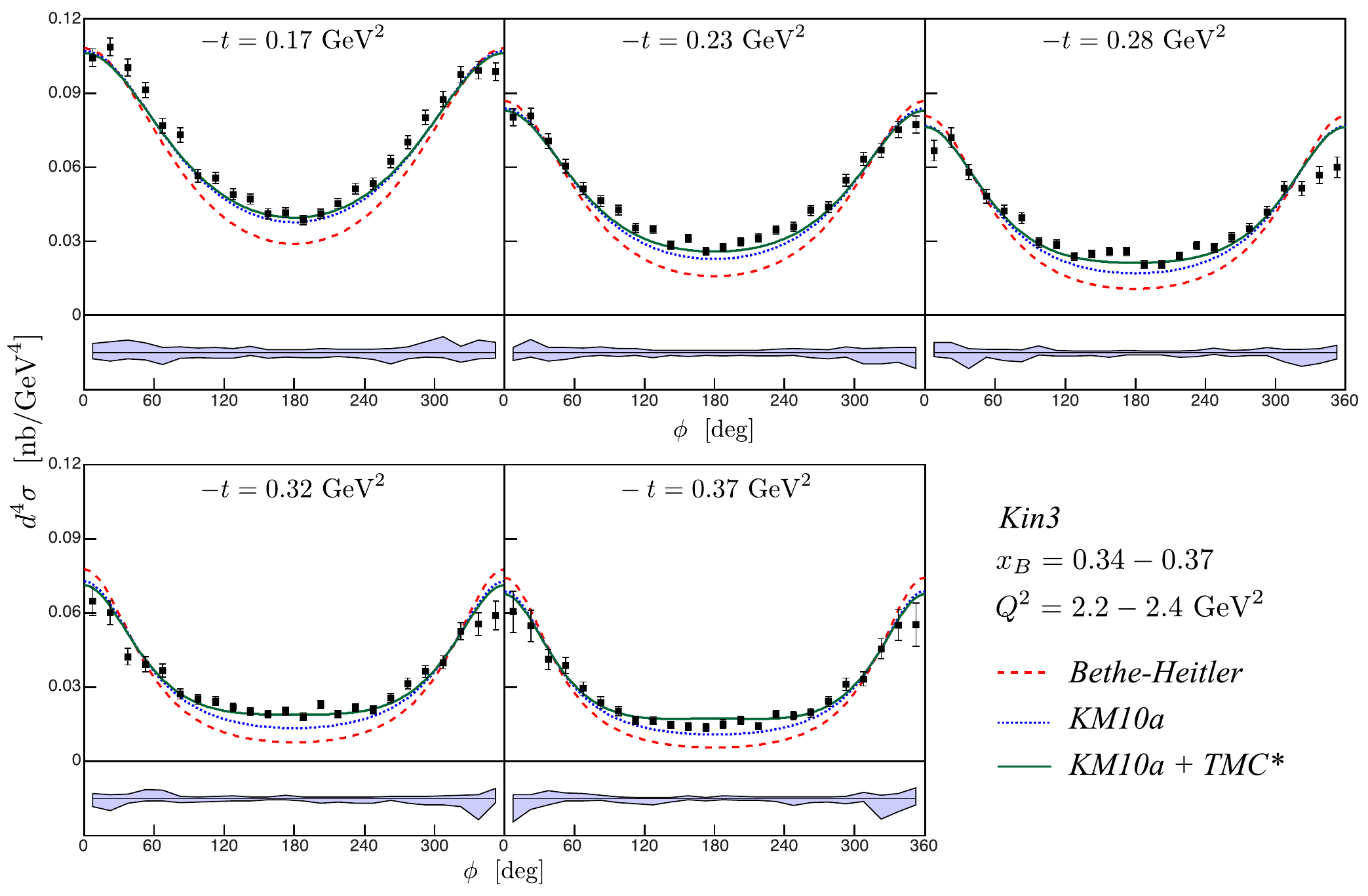,width=\linewidth,clip=}
\caption{\label{kin3}(Color online) Unpolarized cross sections for Kin3. Error bars are statistical only. The light blue area represents the point-to-point systematic uncertainties added linearly to the normalization error. The KM10a model along with its modified version (including the TMC effects) are shown as dotted blue and solid green curves, respectively. The Bethe-Heitler contribution is represented as a dashed red line.}
\end{figure*}

\begin{figure*}[!htp]
\centering
\epsfig{file=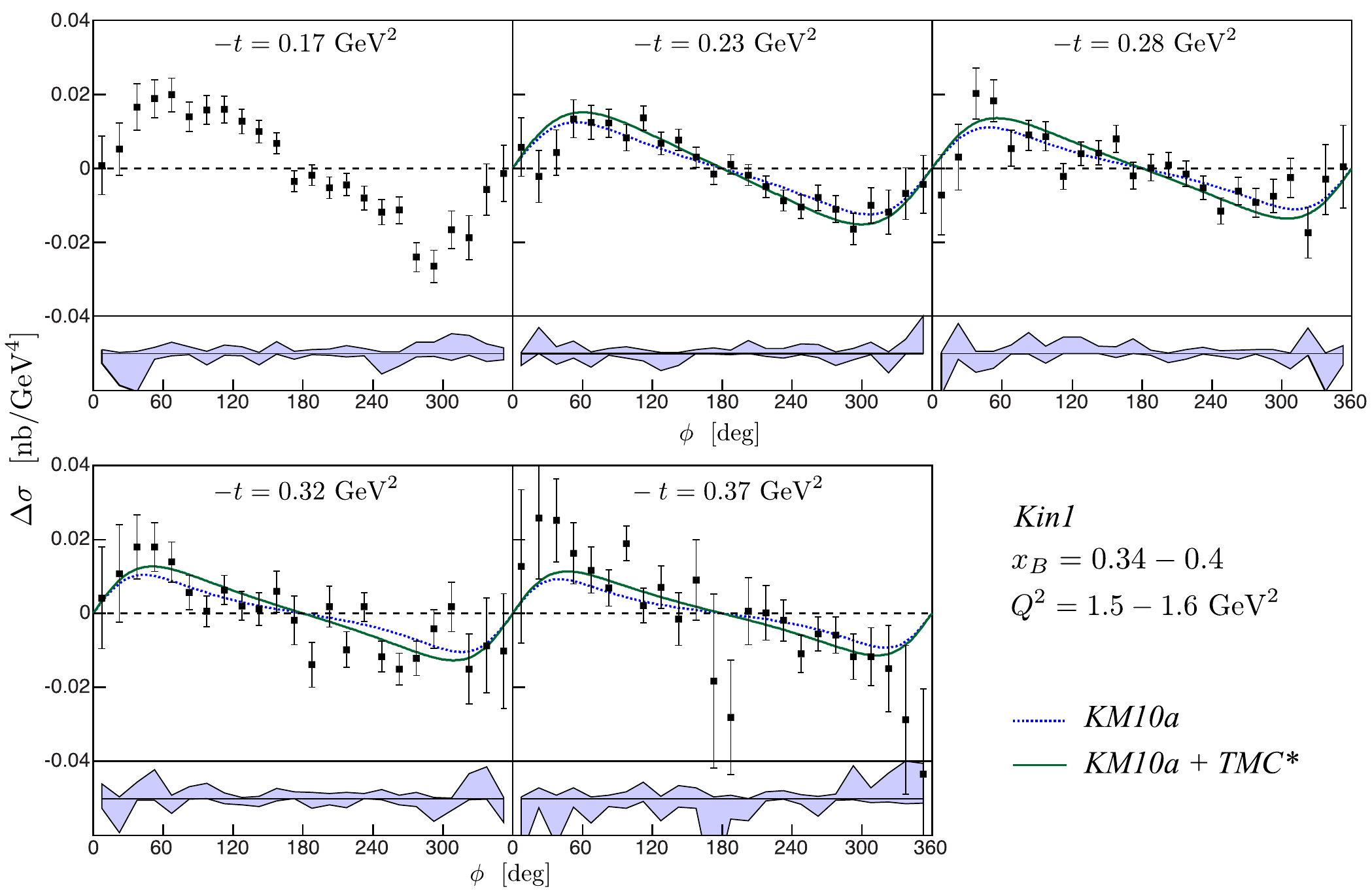,width=\linewidth,clip=}
\caption{\label{diff_kin1}(Color online) Cross-section differences for opposite beam helicities for Kin1.  Error bars are statistical only. The light blue area represents the point-to-point systematic uncertainties added linearly to the normalization error. The KM10a model along with its modified version (including the TMC effects) are shown as dotted blue and solid green curves, respectively, except for the first $t$-bin which is outside the prescribed range of this model.}
\end{figure*}

\begin{figure*}[!htp]
\centering
\epsfig{file=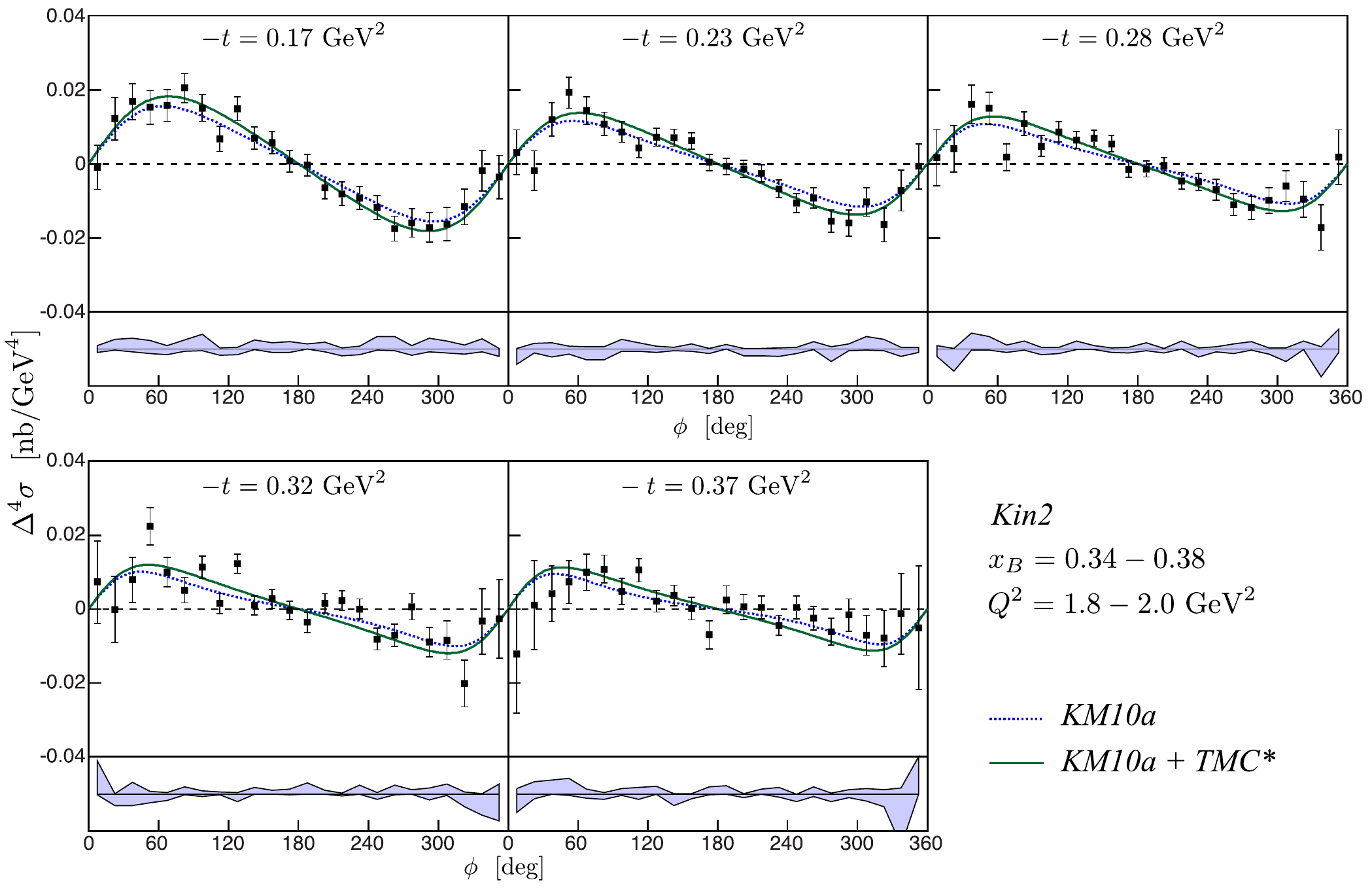,width=\linewidth,clip=}
\caption{\label{diff_kin2}(Color online) Cross-section differences for opposite beam helicities for Kin2.  Error bars are statistical only. The light blue area represents the point-to-point systematic uncertainties added linearly to the normalization error. The KM10a model along with its modified version (including the TMC effects) are shown as dotted blue and solid green curves, respectively.}
\end{figure*}

\begin{figure*}[!htp]
\centering
\epsfig{file=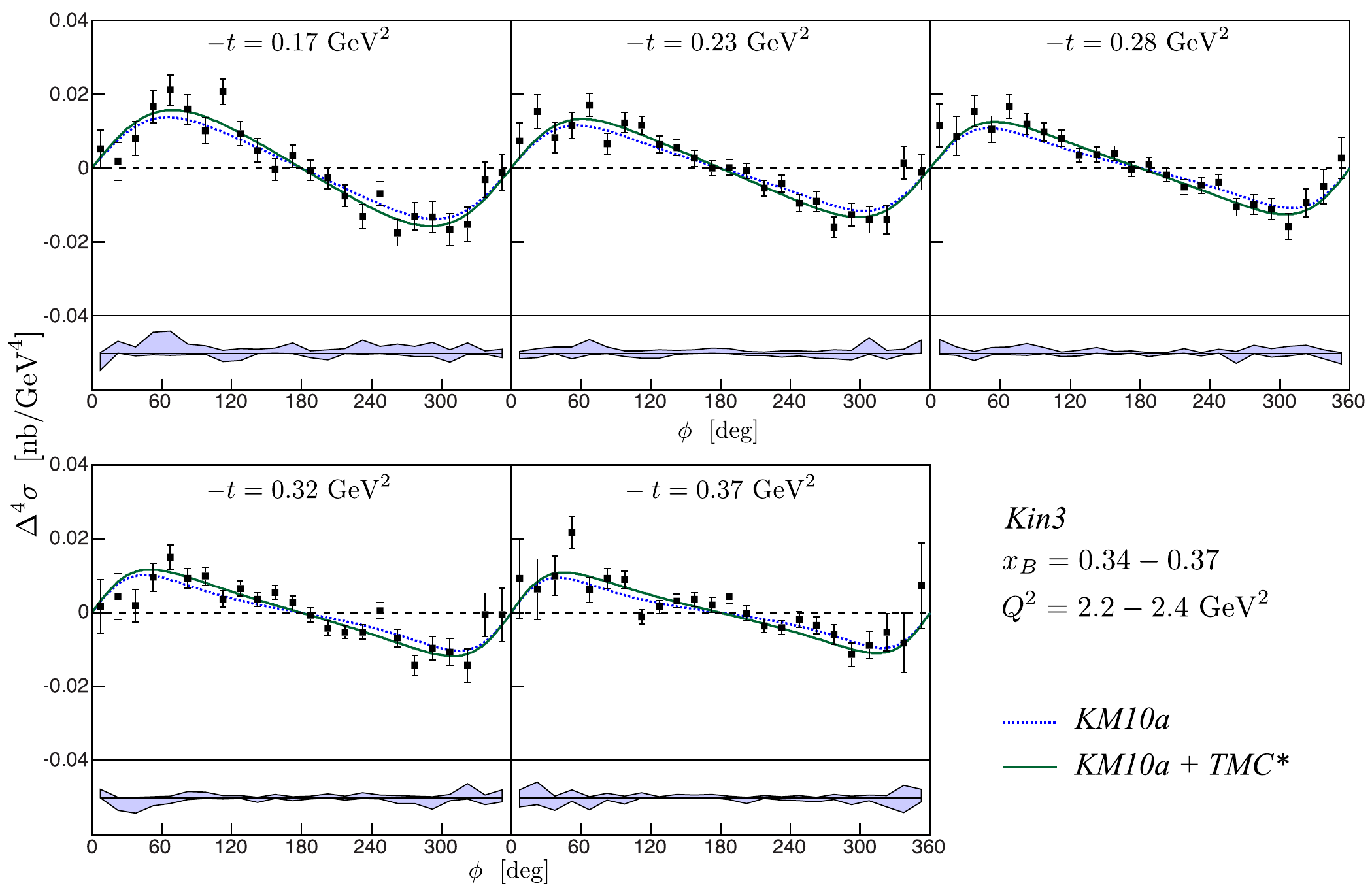,width=\linewidth,clip=}
\caption{\label{diff_kin3}(Color online) Cross-section differences for opposite beam helicities for Kin3.  Error bars are statistical only. The light blue area represents the point-to-point systematic uncertainties added linearly to the normalization error. The KM10a model along with its modified version (including the TMC effects) are shown as dotted blue and solid green curves, respectively.}
\end{figure*}

\begin{figure*}[!htp]
\centering
\epsfig{file=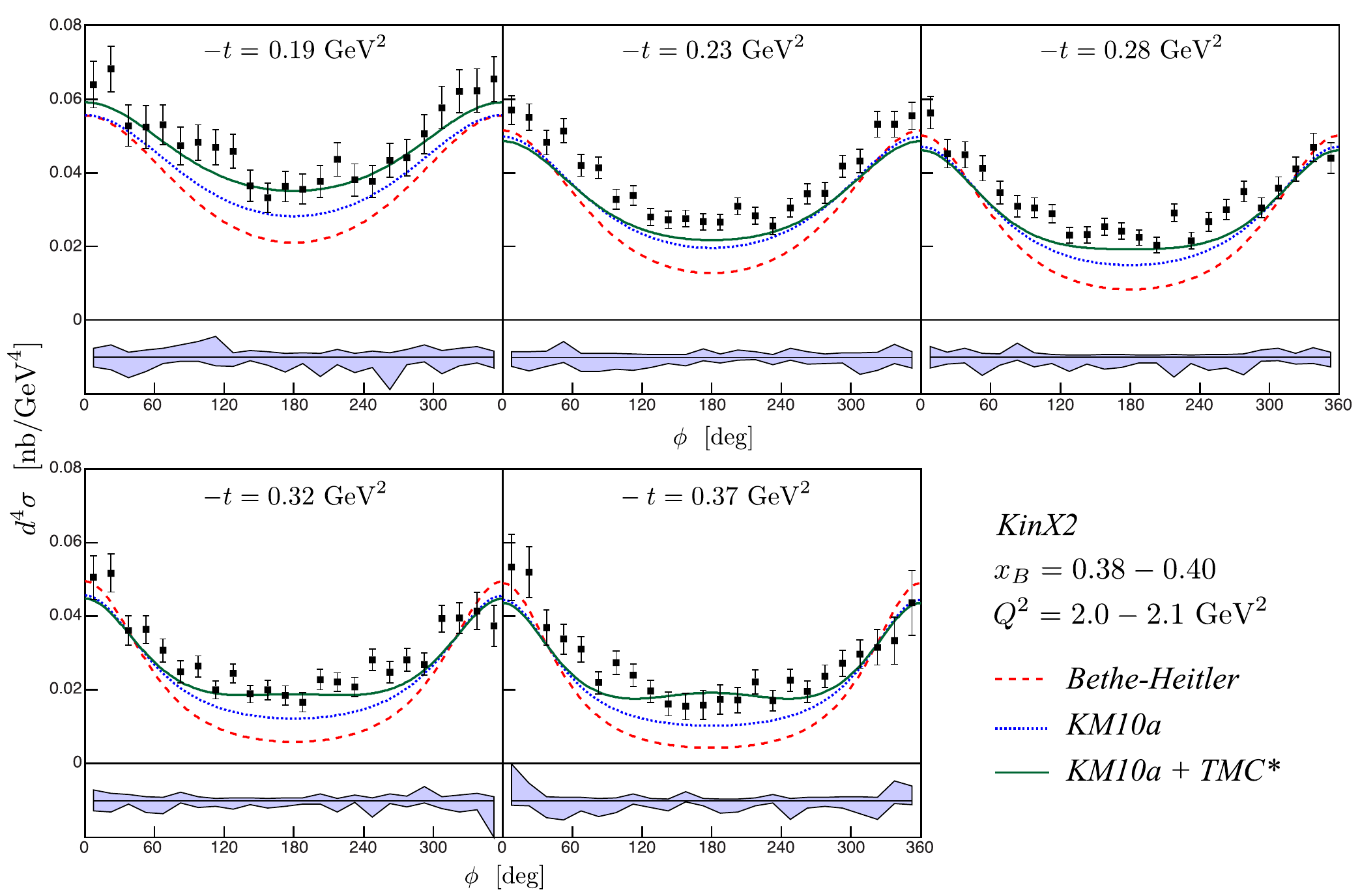,width=\linewidth,clip=}
\caption{\label{kinX2}(Color online) Unpolarized cross sections for KinX2. Error bars are statistical only. The light blue area represents the point-to-point systematic uncertainties added linearly to the normalization error. The KM10a model along with its modified version (including the TMC effects) are shown as dotted blue and solid green curves, respectively. The Bethe-Heitler contribution is represented as a dashed red line.}
\end{figure*}

\begin{figure*}[!ht]
\centering
\epsfig{file=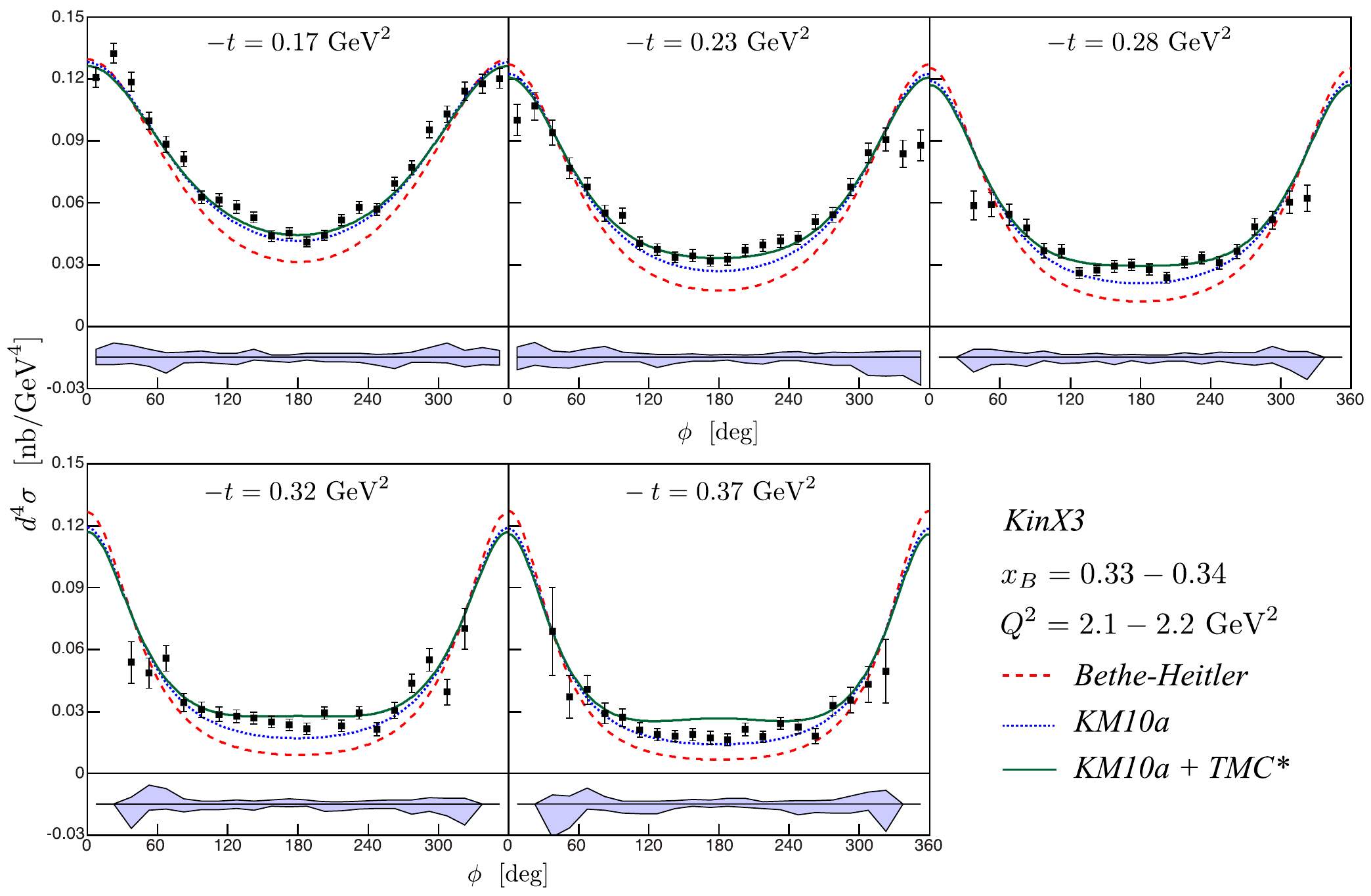,width=\linewidth,clip=}
\caption{\label{kinX3}(Color online) Unpolarized cross sections for KinX3. Error bars are statistical only. The light blue area represents the point-to-point systematic uncertainties added linearly to the normalization error. The KM10a model along with its modified version (including the TMC effects) are shown as dotted blue and solid green curves, respectively. The Bethe-Heitler contribution is represented as a dashed red line.}
\end{figure*}

\begin{figure*}[!ht]
\centering
\epsfig{file=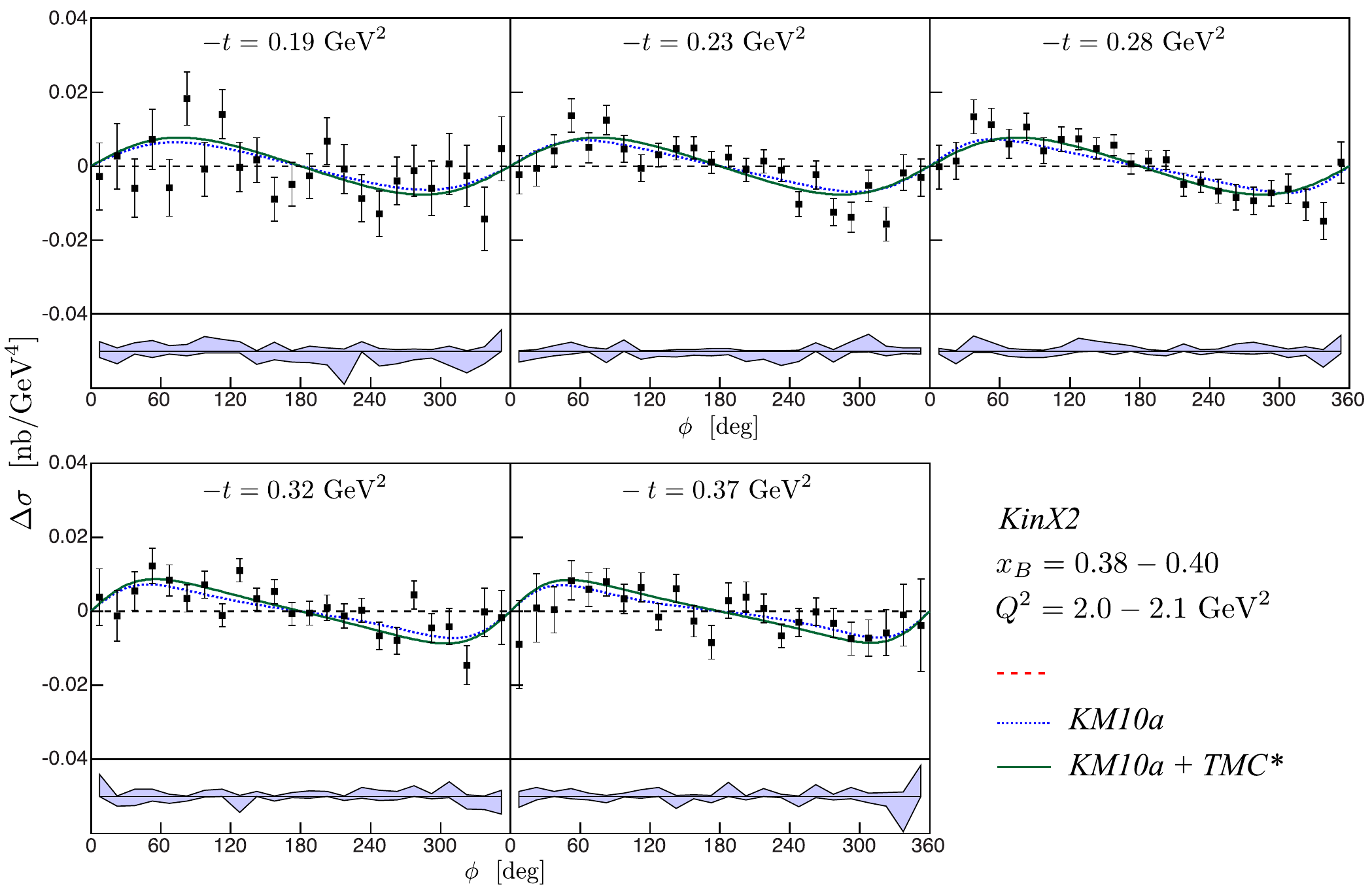,width=\linewidth,clip=}
\caption{\label{kinX2diff}(Color online) Cross-section differences for opposite beam helicities for KinX2. Error bars are statistical only. The light blue area represents the point-to-point systematic uncertainties added linearly to the normalization error. The KM10a model along with its modified version (including the TMC effects) are shown as dotted blue and solid green curves, respectively.}
\end{figure*}

\begin{figure*}[!ht]
\centering
\epsfig{file=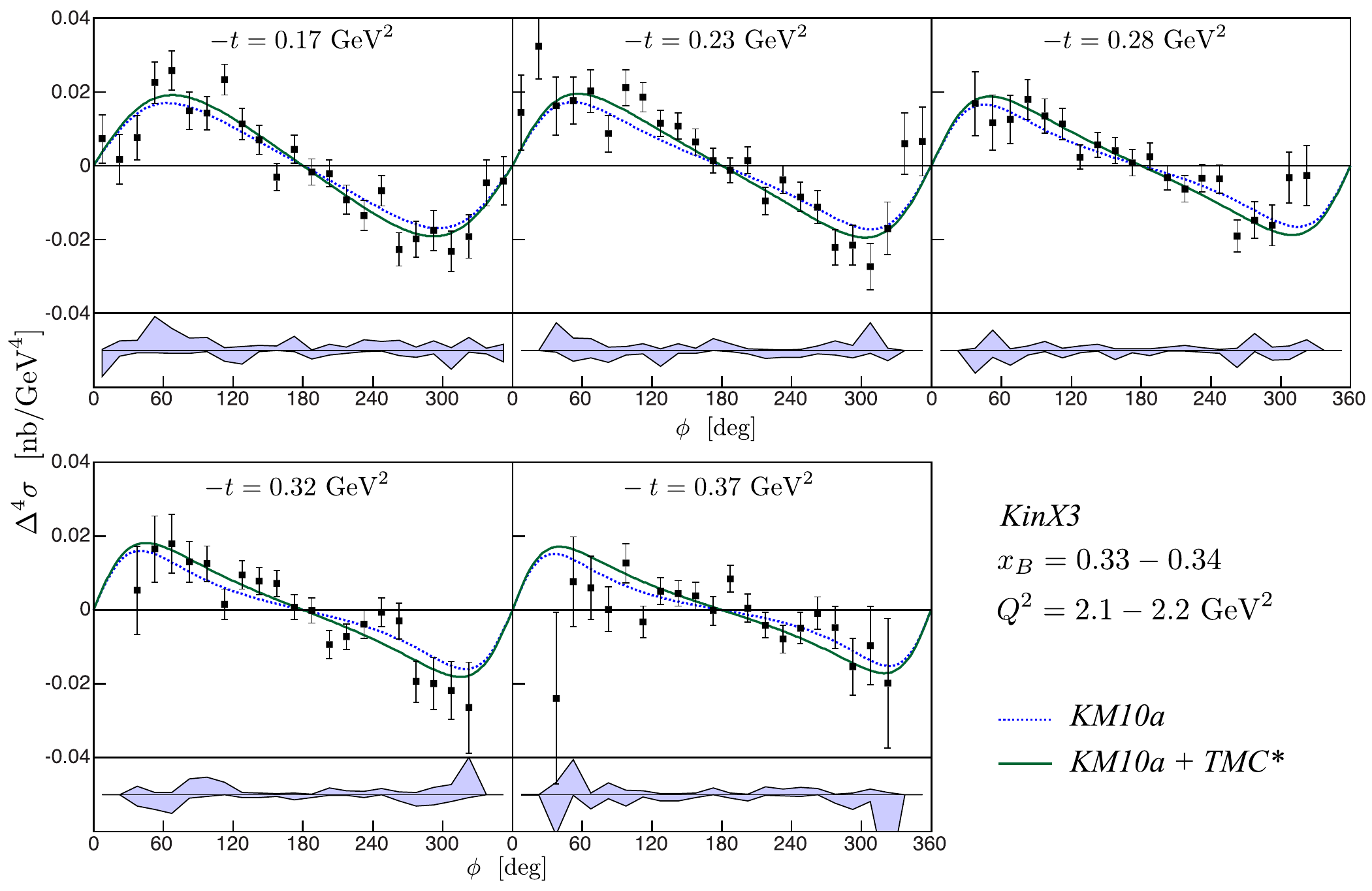,width=\linewidth,clip=}
\caption{\label{kinX3diff}(Color online) Cross-section differences for opposite beam helicities for KinX3. Error bars are statistical only. The light blue area represents the point-to-point systematic uncertainties added linearly to the normalization error. The KM10a model along with its modified version (including the TMC effects) are shown as dotted blue and solid green curves, respectively.}
\end{figure*}

In the following we present the unpolarized cross sections for Kin2, Kin3 as well as KinX2 and KinX3  in Fig.~\ref{kin2}, \ref{kin3}, \ref{kinX2} and \ref{kinX3}, respectively, for a total of 468 experimental bins in $(x_B,Q^2,-t,\phi )$. The cross-section differences for opposite beam helicities are presented for Kin1--3, KinX2 and KinX3 in Fig.~\ref{diff_kin1}--\ref{diff_kin3} and \ref{kinX2diff}, \ref{kinX3diff}, for a total of 588 experimental bins in $(x_B,Q^2,-t,\phi )$. All results are compared to only two models for clarity: the KM10a model and its modified version, including the TMC effects as described in section~\ref{ssec:comparGPD}. All the cross-section data are also listed in Tables~\ref{tab:Kin2}--\ref{tab:diffKinX3} along with their statistical and point-to-point systematic uncertainties. The correlated systematic errors are summarized in Tab.~\ref{tab:sys}.

\section{Summary and conclusion}
\label{sec:conclusion}

We report final results for E00-110, the first dedicated Deeply Virtual Compton Scattering experiment, which ran in  Hall~A of Jefferson Lab. Using new developments in the parametrization of the DVCS reaction, we extracted cross sections including for the first time an evaluation of the DVCS squared amplitude. We showed results for the unpolarized DVCS cross section at two different $Q^2$-values and two different $x_B$-values, thanks to a new analysis which allowed for a reliable evaluation of the $\pi^0$ background in all these kinematics. The effective Compton Form Factors used to described the kinematical dependence of the helicity-dependent and helicity-independent cross section show no $Q^2$--dependence, compatible with the dominance of the leading-twist diagram in this region of moderate $Q^2$ and high $x_B$. Our results were compared with various models based on the Generalized Parton Distributions framework. A relative good agreement was found with the KM10a parametrization. However, this model does not fully match the behavior of the unpolarized cross section for $\phi\sim180^\circ$. We showed that adding an empirical estimate of the target-mass and finite-$t$ corrections to the KM10a model improved the agreement with our data significantly, which may hint at the necessity to include such effects in the analysis of moderate-$Q^2$ data, highly relevant for current and future Jefferson Lab experiments. At any rate, the accuracy of the unpolarized cross-section data around $\phi=180^\circ$ seems absolutely critical to disentangle all contributions of the cross section. 

The significant deviation of the DVCS cross section observed in this experiment with respect to the Bethe-Heitler contribution motivated the subsequent experiment E07-007~\cite{E07-007}, currently under analysis. Its goal is to investigate the nature of this deviation by using the beam-energy dependence of the different terms of the cross section. Indeed, the BH-DVCS interference contribution has a $\sim E_b^3$ dependence whereas the DVCS$^2$ varies as $\sim E_b^2$. In a way similar to a Rosenbluth separation, by measuring the DVCS cross section at exactly the same kinematics but different beam energies, one will be able to tell if this deviation is mainly due to the DVCS$^2$, the BH-DVCS interference terms or higher twist terms. This high accuracy measurement of the cross section at two beam energies will set stringent constraints on GPD models.

\section*{Acknowledgments}
We acknowledge essential work of the JLab accelerator staff and the Hall A technical staff. We also thank V.~Braun for many informative discussions and B.~M.~Pirnay for the access to his TMC code. This work was supported by the Department of Energy (DOE), the National Science Foundation, the French {\em Centre National de la Recherche Scientifique}, the {\em Agence Nationale de la Recherche}, the {\em Commissariat \`a l'\'energie atomique et aux \'energies alternatives} and P2IO Laboratory of Excellence. Jefferson Science Associates, LLC, operates Jefferson Lab for the U.S. DOE under U.S. DOE contract DE-AC05-060R23177.

\begin{table*}
\footnotesize
\centering
\begin{tabular}{|c||rr||rr||rr||rr||rr|}
\hline
\multirow{3}{*}{$\phi$ (deg)} & \multicolumn{2}{|c||}{$x_{B}$ = 0.343} & \multicolumn{2}{|c||}{$x_{B}$ = 0.368} &\multicolumn{2}{|c||}{ $x_{B}$ = 0.375} &\multicolumn{2}{|c||}{$x_{B}$ = 0.379} &\multicolumn{2}{|c|}{ $x_{B}$ = 0.381} \\
 &\multicolumn{2}{|c||}{ $Q^{2}$ = 1.820 GeV$^{2}$} &\multicolumn{2}{|c||}{ $Q^{2}$ = 1.933 GeV$^{2}$} &\multicolumn{2}{|c||}{ $Q^{2}$ = 1.964 GeV$^{2}$ } &\multicolumn{2}{|c||}{ $Q^{2}$ = 1.986 GeV$^{2}$} &\multicolumn{2}{|c|}{ $Q^{2}$ = 1.999 GeV$^{2}$}\\
 & \multicolumn{2}{|c||}{$t$ = -0.172 GeV$^{2}$} & \multicolumn{2}{|c||}{$t$ = -0.232 GeV$^{2}$} & \multicolumn{2}{|c||}{$t$ = -0.278 GeV$^{2}$} &\multicolumn{2}{|c||}{ $t$ = -0.323 GeV$^{2}$} &\multicolumn{2}{|c|}{ $t$ = -0.371 GeV$^{2}$} \\
\hline
\multirow{2}{*}{   7.5}&\multirow{2}{*}{ 111.6$\; \pm \;$4.1}&+ 2.3&\multirow{2}{*}{  76.9$\; \pm \;$4.2}&+ 0.0&\multirow{2}{*}{  75.2$\; \pm \;$5.6}&+ 2.2&\multirow{2}{*}{  66.2$\; \pm \;$7.7}&+ 3.2&\multirow{2}{*}{  70.6$\; \pm \;$11.8}&+ 13.1\\
&&-- 0.0&&-- 1.1&&-- 1.7&&-- 0.9&&-- 0.0\\
\hline
\multirow{2}{*}{  22.5}&\multirow{2}{*}{ 117.6$\; \pm \;$3.9}&+ 0.7&\multirow{2}{*}{  68.6$\; \pm \;$3.7}&+ 0.0&\multirow{2}{*}{  59.7$\; \pm \;$4.5}&+ 1.4&\multirow{2}{*}{  64.1$\; \pm \;$6.3}&+ 0.9&\multirow{2}{*}{  66.3$\; \pm \;$8.7}&+ 5.2\\
&&-- 0.3&&-- 2.8&&-- 0.6&&-- 2.0&&-- 0.4\\
\hline
\multirow{2}{*}{  37.5}&\multirow{2}{*}{  99.8$\; \pm \;$3.5}&+ 0.3&\multirow{2}{*}{  61.7$\; \pm \;$3.2}&+ 0.0&\multirow{2}{*}{  57.9$\; \pm \;$3.8}&+ 0.9&\multirow{2}{*}{  43.9$\; \pm \;$4.5}&+ 1.8&\multirow{2}{*}{  45.1$\; \pm \;$5.7}&+ 0.0\\
\hline
\end{tabular}
\caption{\label{tab:Kin2} Unpolarized cross sections in pb$\cdot$GeV$^{-4}$ with their statistical and asymmetric point-to-point systematic uncertainties for the Kin2 setting, for each bin in $\phi$ (vertical) and $-t$ (horizontal).}
\end{table*}

\begin{table*}
\footnotesize
\centering
\begin{tabular}{|c||rr||rr||rr||rr||rr|}
\hline
\multirow{3}{*}{$\phi$ (deg)} & \multicolumn{2}{|c||}{$x_{B}$ = 0.345} & \multicolumn{2}{|c||}{$x_{B}$ = 0.363} &\multicolumn{2}{|c||}{ $x_{B}$ = 0.368} &\multicolumn{2}{|c||}{$x_{B}$ = 0.371} &\multicolumn{2}{|c|}{ $x_{B}$ = 0.373} \\
 &\multicolumn{2}{|c||}{ $Q^{2}$ = 2.218 GeV$^{2}$} &\multicolumn{2}{|c||}{ $Q^{2}$ = 2.318 GeV$^{2}$} &\multicolumn{2}{|c||}{ $Q^{2}$ = 2.348 GeV$^{2}$ } &\multicolumn{2}{|c||}{ $Q^{2}$ = 2.360 GeV$^{2}$} &\multicolumn{2}{|c|}{ $Q^{2}$ = 2.375 GeV$^{2}$}\\
 & \multicolumn{2}{|c||}{$t$ = -0.176 GeV$^{2}$} & \multicolumn{2}{|c||}{$t$ = -0.232 GeV$^{2}$} & \multicolumn{2}{|c||}{$t$ = -0.279 GeV$^{2}$} &\multicolumn{2}{|c||}{ $t$ = -0.325 GeV$^{2}$} &\multicolumn{2}{|c|}{ $t$ = -0.372 GeV$^{2}$} \\
\hline
\multirow{2}{*}{   7.5}&\multirow{2}{*}{ 104.3$\; \pm \;$3.6}&+ 1.1&\multirow{2}{*}{  80.3$\; \pm \;$3.5}&+ 0.0&\multirow{2}{*}{  66.6$\; \pm \;$4.2}&+ 3.1&\multirow{2}{*}{  64.8$\; \pm \;$5.8}&+ 0.5&\multirow{2}{*}{  60.6$\; \pm \;$8.4}&+ 0.0\\
&&-- 0.2&&-- 1.1&&-- 0.0&&-- 1.9&&-- 8.2\\
\hline
\multirow{2}{*}{  22.5}&\multirow{2}{*}{ 108.6$\; \pm \;$3.6}&+ 1.6&\multirow{2}{*}{  80.6$\; \pm \;$3.3}&+ 3.6&\multirow{2}{*}{  71.9$\; \pm \;$4.0}&+ 2.6&\multirow{2}{*}{  60.2$\; \pm \;$4.9}&+ 0.0&\multirow{2}{*}{  54.8$\; \pm \;$6.3}&+ 0.0\\
&&-- 0.9&&-- 0.0&&-- 0.6&&-- 3.9&&-- 3.3\\
\hline
\multirow{2}{*}{  37.5}&\multirow{2}{*}{ 100.3$\; \pm \;$3.4}&+ 2.6&\multirow{2}{*}{  70.5$\; \pm \;$3.0}&+ 0.1&\multirow{2}{*}{  58.0$\; \pm \;$3.2}&+ 0.0&\multirow{2}{*}{  42.3$\; \pm \;$3.4}&+ 0.9&\multirow{2}{*}{  41.2$\; \pm \;$4.1}&+ 2.2\\
\hline
\end{tabular}
\caption{\label{tab:Kin3} Unpolarized cross sections in pb$\cdot$GeV$^{-4}$ with their statistical and asymmetric point-to-point systematic uncertainties for the Kin3 setting, for each bin in $\phi$ (vertical) and $-t$ (horizontal).}
\end{table*}

\begin{table*}
\footnotesize
\centering
\begin{tabular}{|c||rr||rr||rr||rr||rr|}
\hline
\multirow{3}{*}{$\phi$ (deg)} & \multicolumn{2}{|c||}{$x_{B}$ = 0.345} & \multicolumn{2}{|c||}{$x_{B}$ = 0.374} &\multicolumn{2}{|c||}{ $x_{B}$ = 0.385} &\multicolumn{2}{|c||}{$x_{B}$ = 0.391} &\multicolumn{2}{|c|}{ $x_{B}$ = 0.399} \\
 &\multicolumn{2}{|c||}{ $Q^{2}$ = 1.453 GeV$^{2}$} &\multicolumn{2}{|c||}{ $Q^{2}$ = 1.552 GeV$^{2}$} &\multicolumn{2}{|c||}{ $Q^{2}$ = 1.589 GeV$^{2}$ } &\multicolumn{2}{|c||}{ $Q^{2}$ = 1.608 GeV$^{2}$} &\multicolumn{2}{|c|}{ $Q^{2}$ = 1.633 GeV$^{2}$}\\
 & \multicolumn{2}{|c||}{$t$ = -0.170 GeV$^{2}$} & \multicolumn{2}{|c||}{$t$ = -0.232 GeV$^{2}$} & \multicolumn{2}{|c||}{$t$ = -0.278 GeV$^{2}$} &\multicolumn{2}{|c||}{ $t$ = -0.323 GeV$^{2}$} &\multicolumn{2}{|c|}{ $t$ = -0.370 GeV$^{2}$} \\
\hline
\multirow{2}{*}{   7.5}&\multirow{2}{*}{   0.8$\; \pm \;$7.9}&+ 1.0&\multirow{2}{*}{   5.7$\; \pm \;$7.9}&+ 0.0&\multirow{2}{*}{  -7.2$\; \pm \;$10.8}&+ 0.0&\multirow{2}{*}{   4.2$\; \pm \;$13.8}&+ 4.5&\multirow{2}{*}{  12.7$\; \pm \;$20.7}&+ 0.0\\
&&-- 3.0&&-- 3.8&&-- 13.5&&-- 3.2&&-- 16.2\\
\hline
\multirow{2}{*}{  22.5}&\multirow{2}{*}{   5.2$\; \pm \;$7.1}&+ 0.0&\multirow{2}{*}{  -2.2$\; \pm \;$7.1}&+ 8.4&\multirow{2}{*}{   3.1$\; \pm \;$8.9}&+ 9.9&\multirow{2}{*}{  10.8$\; \pm \;$13.1}&+ 0.0&\multirow{2}{*}{  25.7$\; \pm \;$16.3}&+ 2.6\\
&&-- 9.4&&-- 0.0&&-- 1.8&&-- 11.6&&-- 1.3\\
\hline
\multirow{2}{*}{  37.5}&\multirow{2}{*}{  16.5$\; \pm \;$6.3}&+ 0.0&\multirow{2}{*}{   4.3$\; \pm \;$6.1}&+ 1.9&\multirow{2}{*}{  20.3$\; \pm \;$6.9}&+ 0.0&\multirow{2}{*}{  18.0$\; \pm \;$8.7}&+ 3.8&\multirow{2}{*}{  25.2$\; \pm \;$11.2}&+ 0.0\\
\hline
\end{tabular}
\caption{\label{tab:diffKin1} Cross-section differences for opposite beam helicities in pb$\cdot$GeV$^{-4}$ with their statistical and asymmetric point-to-point systematic uncertainties for the Kin1 setting, for each bin in $\phi$ (vertical) and $-t$ (horizontal).}
\end{table*}

\begin{table*}
\footnotesize
\centering
\begin{tabular}{|c||rr||rr||rr||rr||rr|}
\hline
\multirow{3}{*}{$\phi$ (deg)} & \multicolumn{2}{|c||}{$x_{B}$ = 0.343} & \multicolumn{2}{|c||}{$x_{B}$ = 0.368} &\multicolumn{2}{|c||}{ $x_{B}$ = 0.375} &\multicolumn{2}{|c||}{$x_{B}$ = 0.379} &\multicolumn{2}{|c|}{ $x_{B}$ = 0.381} \\
 &\multicolumn{2}{|c||}{ $Q^{2}$ = 1.820 GeV$^{2}$} &\multicolumn{2}{|c||}{ $Q^{2}$ = 1.933 GeV$^{2}$} &\multicolumn{2}{|c||}{ $Q^{2}$ = 1.964 GeV$^{2}$ } &\multicolumn{2}{|c||}{ $Q^{2}$ = 1.986 GeV$^{2}$} &\multicolumn{2}{|c|}{ $Q^{2}$ = 1.999 GeV$^{2}$}\\
 & \multicolumn{2}{|c||}{$t$ = -0.172 GeV$^{2}$} & \multicolumn{2}{|c||}{$t$ = -0.232 GeV$^{2}$} & \multicolumn{2}{|c||}{$t$ = -0.278 GeV$^{2}$} &\multicolumn{2}{|c||}{ $t$ = -0.323 GeV$^{2}$} &\multicolumn{2}{|c|}{ $t$ = -0.371 GeV$^{2}$} \\
\hline
\multirow{2}{*}{   7.5}&\multirow{2}{*}{  -1.0$\; \pm \;$6.0}&+ 1.0&\multirow{2}{*}{   3.1$\; \pm \;$6.1}&+ 0.0&\multirow{2}{*}{   1.7$\; \pm \;$7.7}&+ 0.9&\multirow{2}{*}{   7.3$\; \pm \;$11.1}&+ 10.3&\multirow{2}{*}{ -12.1$\; \pm \;$16.0}&+ 1.4\\
&&-- 1.1&&-- 4.8&&-- 2.5&&-- 0.0&&-- 5.2\\
\hline
\multirow{2}{*}{  22.5}&\multirow{2}{*}{  12.2$\; \pm \;$5.7}&+ 2.3&\multirow{2}{*}{  -1.9$\; \pm \;$5.4}&+ 1.9&\multirow{2}{*}{   4.1$\; \pm \;$6.2}&+ 0.0&\multirow{2}{*}{  -0.1$\; \pm \;$9.0}&+ 0.0&\multirow{2}{*}{   1.1$\; \pm \;$12.0}&+ 3.6\\
&&-- 0.0&&-- 1.4&&-- 7.5&&-- 3.7&&-- 1.4\\
\hline
\multirow{2}{*}{  37.5}&\multirow{2}{*}{  16.8$\; \pm \;$4.9}&+ 2.6&\multirow{2}{*}{  12.0$\; \pm \;$4.5}&+ 1.6&\multirow{2}{*}{  16.1$\; \pm \;$5.2}&+ 4.2&\multirow{2}{*}{   7.9$\; \pm \;$6.1}&+ 3.2&\multirow{2}{*}{   4.1$\; \pm \;$7.5}&+ 4.0\\
\hline
\end{tabular}
\caption{\label{tab:diffKin2} Cross-section differences for opposite beam helicities in pb$\cdot$GeV$^{-4}$ with their statistical and asymmetric point-to-point systematic uncertainties for the Kin2 setting, for each bin in $\phi$ (vertical) and $-t$ (horizontal).}
\end{table*}

\begin{table*}
\footnotesize
\centering
\begin{tabular}{|c||rr||rr||rr||rr||rr|}
\hline
\multirow{3}{*}{$\phi$ (deg)} & \multicolumn{2}{|c||}{$x_{B}$ = 0.345} & \multicolumn{2}{|c||}{$x_{B}$ = 0.363} &\multicolumn{2}{|c||}{ $x_{B}$ = 0.368} &\multicolumn{2}{|c||}{$x_{B}$ = 0.371} &\multicolumn{2}{|c|}{ $x_{B}$ = 0.373} \\
 &\multicolumn{2}{|c||}{ $Q^{2}$ = 2.218 GeV$^{2}$} &\multicolumn{2}{|c||}{ $Q^{2}$ = 2.318 GeV$^{2}$} &\multicolumn{2}{|c||}{ $Q^{2}$ = 2.348 GeV$^{2}$ } &\multicolumn{2}{|c||}{ $Q^{2}$ = 2.360 GeV$^{2}$} &\multicolumn{2}{|c|}{ $Q^{2}$ = 2.375 GeV$^{2}$}\\
 & \multicolumn{2}{|c||}{$t$ = -0.176 GeV$^{2}$} & \multicolumn{2}{|c||}{$t$ = -0.232 GeV$^{2}$} & \multicolumn{2}{|c||}{$t$ = -0.279 GeV$^{2}$} &\multicolumn{2}{|c||}{ $t$ = -0.325 GeV$^{2}$} &\multicolumn{2}{|c|}{ $t$ = -0.372 GeV$^{2}$} \\
\hline
\multirow{2}{*}{   7.5}&\multirow{2}{*}{   5.2$\; \pm \;$5.1}&+ 0.0&\multirow{2}{*}{   7.4$\; \pm \;$4.9}&+ 0.2&\multirow{2}{*}{  11.5$\; \pm \;$5.9}&+ 3.9&\multirow{2}{*}{   1.7$\; \pm \;$7.3}&+ 2.9&\multirow{2}{*}{   9.4$\; \pm \;$10.9}&+ 2.0\\
&&-- 5.4&&-- 1.8&&-- 0.1&&-- 0.0&&-- 2.5\\
\hline
\multirow{2}{*}{  22.5}&\multirow{2}{*}{   1.8$\; \pm \;$5.1}&+ 3.6&\multirow{2}{*}{  15.4$\; \pm \;$4.6}&+ 0.9&\multirow{2}{*}{   8.6$\; \pm \;$5.3}&+ 1.6&\multirow{2}{*}{   4.4$\; \pm \;$6.2}&+ 0.0&\multirow{2}{*}{   6.4$\; \pm \;$8.2}&+ 4.6\\
&&-- 0.0&&-- 1.3&&-- 0.4&&-- 4.3&&-- 1.9\\
\hline
\multirow{2}{*}{  37.5}&\multirow{2}{*}{   8.0$\; \pm \;$4.7}&+ 1.5&\multirow{2}{*}{   8.3$\; \pm \;$4.1}&+ 1.8&\multirow{2}{*}{  15.4$\; \pm \;$4.2}&+ 0.0&\multirow{2}{*}{   1.9$\; \pm \;$4.4}&+ 0.0&\multirow{2}{*}{  10.0$\; \pm \;$5.3}&+ 0.0\\
\hline
\end{tabular}
\caption{\label{tab:diffKin3} Cross-section differences for opposite beam helicities in pb$\cdot$GeV$^{-4}$ with their statistical and asymmetric point-to-point systematic uncertainties for the Kin3 setting, for each bin in $\phi$ (vertical) and $-t$ (horizontal).}
\end{table*}

\begin{table*}
\footnotesize
\centering
\begin{tabular}{|c||rr||rr||rr||rr||rr|}
\hline
\multirow{3}{*}{$\phi$ (deg)} & \multicolumn{2}{|c||}{$x_{B}$ = 0.378} & \multicolumn{2}{|c||}{$x_{B}$ = 0.392} &\multicolumn{2}{|c||}{ $x_{B}$ = 0.398} &\multicolumn{2}{|c||}{$x_{B}$ = 0.400} &\multicolumn{2}{|c|}{ $x_{B}$ = 0.401} \\
 &\multicolumn{2}{|c||}{ $Q^{2}$ = 2.012 GeV$^{2}$} &\multicolumn{2}{|c||}{ $Q^{2}$ = 2.054 GeV$^{2}$} &\multicolumn{2}{|c||}{ $Q^{2}$ = 2.074 GeV$^{2}$ } &\multicolumn{2}{|c||}{ $Q^{2}$ = 2.084 GeV$^{2}$} &\multicolumn{2}{|c|}{ $Q^{2}$ = 2.091 GeV$^{2}$}\\
 & \multicolumn{2}{|c||}{$t$ = -0.192 GeV$^{2}$} & \multicolumn{2}{|c||}{$t$ = -0.233 GeV$^{2}$} & \multicolumn{2}{|c||}{$t$ = -0.279 GeV$^{2}$} &\multicolumn{2}{|c||}{ $t$ = -0.324 GeV$^{2}$} &\multicolumn{2}{|c|}{ $t$ = -0.371 GeV$^{2}$} \\
\hline
\multirow{2}{*}{   7.5}&\multirow{2}{*}{  64.0$\; \pm \;$6.3}&+ 1.1&\multirow{2}{*}{  57.0$\; \pm \;$3.8}&+ 0.1&\multirow{2}{*}{  56.3$\; \pm \;$4.4}&+ 1.4&\multirow{2}{*}{  50.5$\; \pm \;$5.9}&+ 2.1&\multirow{2}{*}{  53.3$\; \pm \;$9.0}&+ 8.9\\
&&-- 0.9&&-- 1.2&&-- 1.7&&-- 1.3&&-- 0.0\\
\hline
\multirow{2}{*}{  22.5}&\multirow{2}{*}{  68.2$\; \pm \;$6.2}&+ 1.8&\multirow{2}{*}{  55.1$\; \pm \;$3.6}&+ 0.0&\multirow{2}{*}{  45.2$\; \pm \;$3.8}&+ 0.1&\multirow{2}{*}{  51.7$\; \pm \;$5.1}&+ 0.8&\multirow{2}{*}{  52.0$\; \pm \;$6.9}&+ 3.4\\
&&-- 2.0&&-- 2.5&&-- 0.9&&-- 1.7&&-- 0.3\\
\hline
\multirow{2}{*}{  37.5}&\multirow{2}{*}{  52.8$\; \pm \;$5.6}&+ 0.0&\multirow{2}{*}{  48.2$\; \pm \;$3.3}&+ 0.5&\multirow{2}{*}{  44.9$\; \pm \;$3.5}&+ 1.4&\multirow{2}{*}{  36.2$\; \pm \;$4.0}&+ 1.0&\multirow{2}{*}{  36.9$\; \pm \;$4.8}&+ 0.0\\
\hline
\end{tabular}
\caption{\label{tab:KinX2} Unpolarized cross sections in pb$\cdot$GeV$^{-4}$ with their statistical and asymmetric point-to-point systematic uncertainties for the KinX2 setting, for each bin in $\phi$ (vertical) and $-t$ (horizontal).}
\end{table*}

\begin{table*}
\footnotesize
\centering
\begin{tabular}{|c||rr||rr||rr||rr||rr|}
\hline
\multirow{3}{*}{$\phi$ (deg)} & \multicolumn{2}{|c||}{$x_{B}$ = 0.336} & \multicolumn{2}{|c||}{$x_{B}$ = 0.342} &\multicolumn{2}{|c||}{ $x_{B}$ = 0.343} &\multicolumn{2}{|c||}{$x_{B}$ = 0.342} &\multicolumn{2}{|c|}{ $x_{B}$ = 0.342} \\
 &\multicolumn{2}{|c||}{ $Q^{2}$ = 2.161 GeV$^{2}$} &\multicolumn{2}{|c||}{ $Q^{2}$ = 2.190 GeV$^{2}$} &\multicolumn{2}{|c||}{ $Q^{2}$ = 2.194 GeV$^{2}$ } &\multicolumn{2}{|c||}{ $Q^{2}$ = 2.191 GeV$^{2}$} &\multicolumn{2}{|c|}{ $Q^{2}$ = 2.193 GeV$^{2}$}\\
 & \multicolumn{2}{|c||}{$t$ = -0.171 GeV$^{2}$} & \multicolumn{2}{|c||}{$t$ = -0.231 GeV$^{2}$} & \multicolumn{2}{|c||}{$t$ = -0.278 GeV$^{2}$} &\multicolumn{2}{|c||}{ $t$ = -0.324 GeV$^{2}$} &\multicolumn{2}{|c|}{ $t$ = -0.371 GeV$^{2}$} \\
\hline
\multirow{2}{*}{   7.5}&\multirow{2}{*}{ 120.7$\; \pm \;$4.7}&+ 0.9&\multirow{2}{*}{ 100.1$\; \pm \;$7.6}&+ 2.1&\multirow{2}{*}{--}&&\multirow{2}{*}{--}&&\multirow{2}{*}{--}&\\
&&-- 0.4&&-- 2.5&&&&&&\\
\hline
\multirow{2}{*}{  22.5}&\multirow{2}{*}{ 132.5$\; \pm \;$4.9}&+ 3.7&\multirow{2}{*}{ 106.9$\; \pm \;$6.8}&+ 4.5&\multirow{2}{*}{--}&&\multirow{2}{*}{--}&&\multirow{2}{*}{--}&\\
&&-- 0.0&&-- 0.5&&&&&&\\
\hline
\multirow{2}{*}{  37.5}&\multirow{2}{*}{ 118.7$\; \pm \;$4.5}&+ 2.8&\multirow{2}{*}{  93.9$\; \pm \;$6.1}&+ 0.2&\multirow{2}{*}{  58.7$\; \pm \;$7.0}&+ 1.1&\multirow{2}{*}{  53.8$\; \pm \;$10.1}&+ 0.7&\multirow{2}{*}{  68.8$\; \pm \;$21.3}&+ 0.9\\
\hline
\end{tabular}
\caption{\label{tab:KinX3} Unpolarized cross sections in pb$\cdot$GeV$^{-4}$ with their statistical and asymmetric point-to-point systematic uncertainties for the KinX3 setting, for each bin in $\phi$ (vertical) and $-t$ (horizontal).}
\end{table*}

\begin{table*}
\footnotesize
\centering
\begin{tabular}{|c||rr||rr||rr||rr||rr|}
\hline
\multirow{3}{*}{$\phi$ (deg)} & \multicolumn{2}{|c||}{$x_{B}$ = 0.378} & \multicolumn{2}{|c||}{$x_{B}$ = 0.392} &\multicolumn{2}{|c||}{ $x_{B}$ = 0.398} &\multicolumn{2}{|c||}{$x_{B}$ = 0.400} &\multicolumn{2}{|c|}{ $x_{B}$ = 0.401} \\
 &\multicolumn{2}{|c||}{ $Q^{2}$ = 2.012 GeV$^{2}$} &\multicolumn{2}{|c||}{ $Q^{2}$ = 2.054 GeV$^{2}$} &\multicolumn{2}{|c||}{ $Q^{2}$ = 2.074 GeV$^{2}$ } &\multicolumn{2}{|c||}{ $Q^{2}$ = 2.084 GeV$^{2}$} &\multicolumn{2}{|c|}{ $Q^{2}$ = 2.091 GeV$^{2}$}\\
 & \multicolumn{2}{|c||}{$t$ = -0.192 GeV$^{2}$} & \multicolumn{2}{|c||}{$t$ = -0.233 GeV$^{2}$} & \multicolumn{2}{|c||}{$t$ = -0.279 GeV$^{2}$} &\multicolumn{2}{|c||}{ $t$ = -0.324 GeV$^{2}$} &\multicolumn{2}{|c|}{ $t$ = -0.371 GeV$^{2}$} \\
\hline
\multirow{2}{*}{   7.5}&\multirow{2}{*}{  -2.8$\; \pm \;$9.1}&+ 4.3&\multirow{2}{*}{  -2.4$\; \pm \;$5.1}&+ 0.0&\multirow{2}{*}{  -0.2$\; \pm \;$5.9}&+ 0.8&\multirow{2}{*}{   3.8$\; \pm \;$7.7}&+ 6.9&\multirow{2}{*}{  -9.0$\; \pm \;$11.9}&+ 1.1\\
&&-- 2.5&&-- 3.5&&-- 1.0&&-- 0.0&&-- 3.2\\
\hline
\multirow{2}{*}{  22.5}&\multirow{2}{*}{   2.7$\; \pm \;$8.8}&+ 0.7&\multirow{2}{*}{  -0.6$\; \pm \;$4.8}&+ 0.5&\multirow{2}{*}{   1.4$\; \pm \;$5.0}&+ 0.0&\multirow{2}{*}{  -1.2$\; \pm \;$6.8}&+ 0.0&\multirow{2}{*}{   0.9$\; \pm \;$9.2}&+ 2.7\\
&&-- 5.6&&-- 2.4&&-- 4.0&&-- 3.5&&-- 1.1\\
\hline
\multirow{2}{*}{  37.5}&\multirow{2}{*}{  -6.0$\; \pm \;$7.9}&+ 3.6&\multirow{2}{*}{   4.1$\; \pm \;$4.4}&+ 1.6&\multirow{2}{*}{  13.3$\; \pm \;$4.6}&+ 4.3&\multirow{2}{*}{   5.5$\; \pm \;$5.2}&+ 2.0&\multirow{2}{*}{   0.4$\; \pm \;$6.2}&+ 1.4\\
\hline
\end{tabular}
\caption{\label{tab:diffKinX2} Cross-section differences for opposite beam helicities in pb$\cdot$GeV$^{-4}$ with their statistical and asymmetric point-to-point systematic uncertainties for the KinX2 setting, for each bin in $\phi$ (vertical) and $-t$ (horizontal).}
\end{table*}

\begin{table*}
\footnotesize
\centering
\begin{tabular}{|c||rr||rr||rr||rr||rr|}
\hline
\multirow{3}{*}{$\phi$ (deg)} & \multicolumn{2}{|c||}{$x_{B}$ = 0.336} & \multicolumn{2}{|c||}{$x_{B}$ = 0.342} &\multicolumn{2}{|c||}{ $x_{B}$ = 0.343} &\multicolumn{2}{|c||}{$x_{B}$ = 0.342} &\multicolumn{2}{|c|}{ $x_{B}$ = 0.342} \\
 &\multicolumn{2}{|c||}{ $Q^{2}$ = 2.161 GeV$^{2}$} &\multicolumn{2}{|c||}{ $Q^{2}$ = 2.190 GeV$^{2}$} &\multicolumn{2}{|c||}{ $Q^{2}$ = 2.194 GeV$^{2}$ } &\multicolumn{2}{|c||}{ $Q^{2}$ = 2.191 GeV$^{2}$} &\multicolumn{2}{|c|}{ $Q^{2}$ = 2.193 GeV$^{2}$}\\
 & \multicolumn{2}{|c||}{$t$ = -0.171 GeV$^{2}$} & \multicolumn{2}{|c||}{$t$ = -0.231 GeV$^{2}$} & \multicolumn{2}{|c||}{$t$ = -0.278 GeV$^{2}$} &\multicolumn{2}{|c||}{ $t$ = -0.324 GeV$^{2}$} &\multicolumn{2}{|c|}{ $t$ = -0.371 GeV$^{2}$} \\
\hline
\multirow{2}{*}{   7.5}&\multirow{2}{*}{   7.3$\; \pm \;$6.5}&+ 0.0&\multirow{2}{*}{  14.4$\; \pm \;$10.1}&+ 0.3&\multirow{2}{*}{--}&&\multirow{2}{*}{--}&&\multirow{2}{*}{--}&\\
&&-- 8.0&&-- 6.4&&&&&&\\
\hline
\multirow{2}{*}{  22.5}&\multirow{2}{*}{   1.7$\; \pm \;$6.7}&+ 2.8&\multirow{2}{*}{  32.4$\; \pm \;$8.9}&+ 0.3&\multirow{2}{*}{--}&&\multirow{2}{*}{--}&&\multirow{2}{*}{--}&\\
&&-- 1.8&&-- 3.2&&&&&&\\
\hline
\multirow{2}{*}{  37.5}&\multirow{2}{*}{   7.6$\; \pm \;$6.1}&+ 2.8&\multirow{2}{*}{  16.2$\; \pm \;$7.9}&+ 8.5&\multirow{2}{*}{  16.8$\; \pm \;$8.7}&+ 0.0&\multirow{2}{*}{   5.3$\; \pm \;$11.9}&+ 2.6&\multirow{2}{*}{ -23.9$\; \pm \;$23.3}&+ 5.6\\
\hline
\end{tabular}
\caption{\label{tab:diffKinX3} Cross-section differences for opposite beam helicities in pb$\cdot$GeV$^{-4}$ with their statistical and asymmetric point-to-point systematic uncertainties for the KinX3 setting, for each bin in $\phi$ (vertical) and $-t$ (horizontal).}
\end{table*}

\bibstyle{apsref} 
\bibliography{Compton2014}

\end{document}